\documentclass[a4paper,fleqn,usenatbib]{mnras}

\usepackage[T1]{fontenc}
\usepackage{ae,aecompl}

\usepackage{graphicx}	% Including figure files
\usepackage{amsmath}	% Advanced maths commands
\usepackage{amssymb}	% Extra maths symbols
\usepackage{xcolor}

\title[Global Correlations in Dwarf Galaxies]{Global Correlations Between the Radio Continuum, Infrared and CO Emission in Dwarf Galaxies}

\author[Filho, Tabatabaei, S\'anchez Almeida, Mu\~noz-Tu\~n\'on \& Elmegreen]{
Mercedes E. Filho,$^{1,2}$\thanks{E-mail: mfilho@fe.up.pt (MEF)}
Fatemeh S. Tabatabaei$^{3,4}$, Jorge S\'anchez Almeida$^{3,4}$,
\newauthor Casiana Mu\~noz-Tu\~n\'on$^{3,4}$ and Bruce G. Elmegreen$^{5}$ 
\\
$^{1}$CENTRA/SIM, Instituto Superior T\'ecnico, Universidade de Lisboa, Av. Rovisco Pais 1, P-1049-001 Lisbon, Portugal \\
$^{2}$Universidade do Porto, Departamento de Engenharia F\'\i sica da Faculdade de Engenharia, Rua Dr. Roberto Frias, s/n, \\
P-4200-465 Oporto, Portugal\\
$^{3}$Instituto Astrof\'\i sica de Canarias, E-38205 La Laguna, Tenerife, Spain\\
$^{4}$Departamento de Astrof\'\i sica, Universidad La Laguna, E-38206 La Laguna, Tenerife, Spain\\
$^{5}$IBM, T. J. Watson Research Center, Yorktown Heights, NY 10598, USA
}

\date{Accepted XXX. Received YYY; in original form ZZZ}

\pubyear{2015}

\begin{document}
\label{firstpage}
\pagerange{\pageref{firstpage}--\pageref{lastpage}}
\maketitle

\begin{abstract}
Correlations between the radio continuum, infrared and CO emission are known to exist for several types of galaxies and across several orders of magnitude. However, the low-mass, low-luminosity and low-metallicity regime of these correlations is not well known. A sample of metal-rich and metal-poor dwarf galaxies from the literature has been assembled to explore this extreme regime. The results demonstrate that the properties of dwarf galaxies are not simple extensions of those of more massive galaxies; the different correlations reflect different star-forming conditions and different coupling between the star formation and the various quantities. It is found that dwarfs show increasingly weaker CO and infrared emission for their luminosity, as expected for galaxies with a low dust content, slower reaction rates, and a hard ionizing radiation field. In the higher-luminosity dwarf regime (L$_{\rm 1.4 \, GHz} \gtrsim$ 10$^{27}$ W, where L$_{\rm 1.4 \, GHz} \simeq$ 10$^{29}$ W for a Milky Way star formation rate of $\simeq$1 M$_{\odot}$ yr$^{-1}$), the total and non-thermal radio continuum emission appear to adequately trace the star formation rate. A breakdown of the dependence of the (H$\alpha$-based) thermal, non-thermal, and, hence, total radio continuum emission on star formation rate occurs below L$_{\rm 1.4 \, GHz} \simeq$ 10$^{27}$ W, resulting in a steepening or downturn of the relations at extreme low luminosity. Below L$_{\rm FIR} \simeq$ 10$^{36}$ W $\simeq$ 3 $\times$ 10$^{9}$ L$_{\odot}$, the infrared emission ceases to adequately trace the star formation rate. A lack of a correlation between the magnetic field strength and the star formation rate in low star formation rate dwarfs suggests a breakdown of the equipartition assumption. As extremely metal-poor dwarfs mostly populate the low star formation rate and low luminosity regime, they stand out in their infrared, radio continuum and CO properties.
\end{abstract}


%%%%%%%%%%%%%%%%%%%%%%%%%%%%%%%%%%%%%%%%%%%%%%%%%%

%%%%%%%%%%%%%%%%% BODY OF PAPER %%%%%%%%%%%%%%%%%%

%Pearson is a measure of the linear correlation between two variables X and Y. It has a value between +1 and −1, where 1 is total positive linear correlation, 0 is no linear correlation, and −1 is total negative linear correlation. 

%The Spearman correlation between two variables is equal to the Pearson correlation between the rank values of those two variables; while Pearson's correlation assesses linear relationships, Spearman's correlation assesses monotonic relationships (whether linear or not). If there are no repeated data values, a perfect Spearman correlation of +1 or −1 occurs when each of the variables is a perfect monotone function of the other.

\section{Introduction}

Correlations are known to exist between the radio continuum (RC), infrared (IR) and CO emission over several orders of magnitude, across a wide variety of galaxy types, over a range in redshift, and both on local and global scales (e.g., de Jong et al. 1985; Young \& Scoville 1991; Price \& Duric 1992; Niklas \& Beck 1997; Yun, Reddy \& Condon 2001; Bell 2003; Murgia et al. 2002; Appleton et al. 2004; Murgia et al. 2005; Leroy et al. 2005; Seymour et al. 2009; Jarvis et al. 2010; Sargent et al. 2010; Bourne et al. 2011; Pannella et al. 2015). Globally, the IR -- RC correlation has been shown to be nearly linear, although there is some sign of a luminosity (e.g., Bell 2003 and references therein) and redshift (e.g., Magnelli et al. 2015) dependence. A global, nearly-linear, correlation between the CO and IR emission has also been shown to exist, with a potential metallicity dependence (e.g., Shetty et al. 2016 and references therein). The global CO -- RC correlation has been shown to be supra-linear (e.g., Liu \& Gao 2010 and references therein).

In the conventional picture (see also Niklas \& Beck 1997), the bridging factor between these quantities is massive star formation. The fundamental relation connecting the interstellar medium (ISM) properties to star formation is the Kennicutt-Schmidt relation (Schmidt 1959; Kennicutt 1989, 1998; Kennicutt \& Evans 2012). The star formation rate (SFR) is generally quantified using monochromatic, bolometric or a combination of (continuum and line) tracers in the ultraviolet (UV), IR, optical (i.e., H$\alpha$) and radio (i.e., RC) regime, which sample different stellar mass ranges and timescales (e.g., Calzetti 2013; Kennicutt \& Evans 2012). SFR indicators in the UV generally probe direct stellar light, while the optical/near-infrared range traces gas ionized by stellar light, the mid-IR/far-infrared (FIR) probes dust-reprocessed stellar light, and the radio regime indirectly probes stellar light. 

Interpretations of the IR -- RC -- CO relations have been made in terms of the calorimetric model (e.g., V\"olk 1989), the phenomenological model (e.g., Helou \& Bicay 1993), the cosmic ray (CR)-heating model (e.g., Suchkov, Allen \& Heckman 1993), and the chemical evolution model (e.g., Bettens et al. 1993). Models which are independent of conventional star-formation scenarios have been proposed by Murgia et al. (2005), Lacki, Thompson \& Quataert (2010) and Tabatabaei et al. (2013a, b). Murgia et al. (2005) proposed a mechanism based on hydrostatic pressure regulation to explain the correlations. Lacki, Thompson \& Quataert (2010) suggested that, in starbursts, although ionization and Bremsstrahlung losses dominate, the IR -- RC correlation arises due to secondary electrons/positrons, and is not directly related to the star formation. Tabatabaei et al. (2013a, b) provided an observationally and physically-motivated explanation involving the coupling between the gas and the equipartition magnetic field (B) with or without (i.e., CR diffusion losses dominate over injection) massive star formation.

Observationally, and also theoretically, it is unclear whether these relations should hold in low-mass, low-luminosity, low-metallicity galaxies, as the physical conditions in these galaxies are distinct. Indeed, the effectiveness of using the RC, IR and CO observables to trace different ISM components depends on many factors, such as mass, luminosity, star-forming conditions, density and metallicity. 

The RC traces the ionized gas, as well as CRs and magnetic fields. The RC emission, at centimeter wavelengths, is generally produced by a combination of free-free emission from thermal electrons, and non-thermal synchrotron emission from relativistic electrons embedded in magnetic fields (e.g. Condon 1992). Increased ionizing UV photon production, associated with transient starbursts, low metallicity (decreased line-blanketing) and reduced dust absorption, can increase the ionization of the gas, and, hence, the thermal RC component. This effect is counteracted by increased ionizing photon escape from the galaxy, associated with, e.g., low mass and high star formation rates, which can decrease the thermal contribution (e.g. Fernandez \& Shull 2011; Benson, Venkatesan \& Shull 2013; Leitherer et al. 2016). Supernovae (SN) inject turbulence into the medium, which drives magnetic field generation and amplification. However, high velocity dispersion and slow differential rotation can weaken or destroy the magnetic fields (e.g., Beck 1006). In turn, SN explosions, SN remnants and strong ISM fields produce and accelerate a part of the CRs, while another part are secondary electrons (e.g., Condon 1992). Together with the star formation rate, mass determines the CR loss versus injection budget, as the star formation threshold to drive winds and outflows, which can be responsible for CR advection and escape from the galaxy, is lower in smaller gravitational potential wells. However, increased SN rates per unit stellar mass appear to be associated with low-metallicity galaxies (e.g., Kistler et al. 2013).

Different dust components absorb stellar radiation and re-emit the energy in the IR regime. Mid-IR emission includes continuum emission, and is produced by polycyclic aromatic hydrocarbons and small dust grains, while FIR emission is produced by large dust grains (e.g., Draine 2011). At low metallicities, the dust content per unit stellar mass is generally depressed, with the bulk of the dust mass dominated by oxygen-rich silicate grains due to heavy element depletion. As the radiation field is harder due to less line-blanketing, and more intense if the sources are undergoing a starburst, PAHs are destroyed, small grains become overabundant, and the dust temperature increases, as does the FIR -- submm emission per unit dust mass (e.,g. R\'emy-Ruyer et al. 2013; Shi et al. 2014). 

Molecular gas is the fuel from which stars are formed. H$_2$, the most abundant molecule, is mostly undetectable due to its lack of a permanent dipole moment, so that dipole transitions between different vibrational and rotational levels within the electronic ground state are forbidden. Instead, the $J = 1 - 0$ rotational transition of CO, emitting at 115~GHz, is commonly used as a molecular gas tracer, as it is excited at typical densities (n $\simeq$ 100 cm$^{-3}$) and temperatures (T $\simeq$ 10 -- 100 K) found in molecular cloud cores. In low-mass galaxies, the CO (and H$_2$) formation rate is lower due to the presence of less dense material (e.g., Krumholz, McKee \& Tumlinson 2008, 2009a, 2009b; Wolfire, Hollenbach \& McKee 2010; Krumholz 2013). CO has also been shown to be metallicity-dependent (e.g., Leroy et al. 2011; Schruba et al. 2012; Bolatto, Wolfire \& Leroy 2013; Elmegreen et al. 2013; Cormier et al. 2014; Hunt et al. 2015; Shi et al. 2015; Amor\'\i n et al. 2016). At low metallicity, there is a suppression of CO per unit stellar mass due to dissociation by the hard, intense (when there is a starburst) radiation field, a lack of dust shielding and slower chemical reaction rates (e.g., Krumholz, McKee \& Tumlinson 2008, 2009a, 2009b; Wolfire, Hollenbach \& McKee 2010; Shetty et al. 2011; Krumholz 2013), so that CO is no longer an efficient tracer of the molecular gas. In addition, an increased CR ionization rate, induced by, e.g., transient starbursts, has also been shown to destroy CO (e.g., Bisbas et al. 2017).

The main goal of this paper is to sample the less-explored low-mass, low-luminosity and low-metallicity regime of the global IR, RC and CO relations. A significant dataset for low-mass galaxies in general, and low-metallicity sources in particular, has been compiled. This is a gain relative to previous studies, which either did not probe such an extreme regime, or utilized a smaller number of sources in this regime. This poorly-known regime is of particular interest for characterizing star formation at high redshifts, where the galaxies were gas-rich, low-mass, and metal-poor (e.g., Sparre et al. 2015; Katsianis et al. 2017). In addition, datasets and relations for more massive galaxies and surveys have also been included, where possible, for comparison. 

The paper is organized as follows. Section~2 presents the low-mass, low-luminosity and low-metallicity galaxy datasets used in the analysis, plus auxiliary data for more massive galaxies and surveys. Results are presented in Section 3, which includes an analysis of the IR -- RC and B - SFR correlation (Sect.~3.1), the CO -- RC correlation (Sect.~3.2) and the IR -- CO correlation (Sect.~3.3), and an interpretation and discussion of the results. Section~4 contains summary of the results and conclusions.

Throughout, a cosmological model where $H_0$ = 69.6 km s$^{-1}$ Mpc$^{-1}$, $\Omega_{\rm m}$ = 0.286, and $\Omega_{\rm vac}$ = 0.714 (Bennett et al. 2014), has been adopted.

%%%%%%%%%%%%%%%%%%%%%%%%%%%%%%%%%%%%%%%%%%%%%%%%%%%%%%%%%

\section{Data}

For the analysis of the IR, RC and CO relations in dwarf galaxies, particularly in the low-metallicity regime, three galaxy samples have been used: the extremely metal-poor dwarf galaxy (XMP) sample (Morales-Luis et al. 2011), the Dwarf Galaxy Survey (DGS; Madden et al. 2013) and the Local Irregulars That Trace Luminosity Extremes The HI Nearby Galaxy Survey (LITTLE THINGS; Hunter et al. 2012). These samples have been chosen mainly because they possess a suite of multi-wavelength properties. Typical stellar masses for the dwarfs fall in the range from 10$^6$ -- 10$^{9}$ M$_{\odot}$, with several dwarfs possessing stellar masses as large as 10$^{10}$ M$_{\odot}$ (Madden et al. 2013; Filho et al. 2013). The direct comparative analysis has been performed with the Key Insights on Nearby Galaxies: a Far-Infrared Survey with Herschel (KINGFISH; Kennicutt et al. 2011). Relations obtained for several samples of (different) dwarfs, spirals, luminous infrared galaxies (LIRGs) and ultraluminous infrared galaxies (ULIRGs) have also been utilized for comparison (Murgia et al. 2005; Liu \& Gao 2010; Chy\.{z}y et al. 2011).

The XMP sample consists of 140 extremely metal-poor (explicitly, 12+log(O/H) $\lesssim$ 7.69) dwarf galaxies selected from the Sloan Digital Sky Survey (SDSS) data release (DR) 7 (Abazajian et al. 2009) and the literature. The DGS is a multi-wavelength survey of 50 nearby dwarf galaxies with a broad range in metallicity, with the goal to map the dust and gas emission. The objective of LITTLE THINGS (41 sources) is to determine what drives the star formation in dwarf galaxies, using multi-wavelength information. KINGFISH is an imaging and spectroscopic survey of 61 nearby galaxies, chosen, in the present study as a reference, to cover a wide range of galaxy properties and environments.

The XMP sample has 19 sources in common with the DGS, 11 sources in common with LITTLE THINGS, and 2 sources in common with KINGFISH (see also Appendix A). LITTLE THINGS and the DGS have 7 sources in common (one of which is an XMP), while KINGFISH and LITTLE THINGS have 5 sources in common (2 of which are XMPs; see also Appendix A). There are no sources in common between KINGFISH and the DGS.

XMP multi-wavelength data generally come from Filho et al. (2013) or references below, for the sources in common with the DGS (Madden et al. 2013), LITTLE THINGS (Hunter et al. 2012) and KINGFISH samples (Kennicutt et al. 2011; Hunt et al. 2015). SFRs from Filho et al. (2013) are derived from the H$\alpha$ emission. 70 (or 60 for Infrared Astronomical Satellite (IRAS) data; L$_{\rm 70 \, \mu m}$) $\mu$m, 100 (L$_{\rm 100 \, \mu m}$) $\mu$m, 160 (L$_{\rm 160 \, \mu m}$) $\mu$m and FIR luminosities (L$_{\rm FIR}$) are from Lisenfield et al. (2007), Engelbracht et al. (2008), Dale et al. (2009), Dale et al. (2012), Madden et al. (2013) and R\'emy-Ruyer et al. (2013).

%FIR (L$_{\rm FIR}$) luminosities are from Dale et al. (2009), Dale et al. (2012) and Madden et al. (2013).

For the DGS sources, metallicities and distances are from Madden et al. (2013). SFRs are derived from the TIR emission, or from the H$\alpha$ or H$\beta$ emission, when no IR data is available (Madden et al. 2013). Inclination angles for the DGS sources were estimated from the source sizes in Madden et al. (2013), according to the procedure outlined in Filho et al. (2013; their Eq.~[1]). It has been assumed that the DGS galaxies are disks of intrinsic thickness q$_{0}$ = 0.25 (e.g., S\'anchez-Janssen, M\'endez-Abreu \& Aguerri 2010). 70 $\mu$m, 100 $\mu$m, 160 $\mu$m and FIR luminosities are from Madden et al. (2013) and R\'emy-Ruyer et al. (2013). 

%while FIR luminosities are from Madden et al. (2013).} 

%FOR THE DGS SOURCES (MADDEN ET AL. 2013), SFR (TABLE 2) ESTIMATED FROM TIR (BOLOMETRIC), USING FORMULA [3] IN KENNICUTT 1998. THE REMY-RUYER (2013) TIR LUMINOSITY COMES FROM A BB FIT TO THE DATA. TIR IS IN MADDEN ETAL. 2013, OBTAINED USING DALE AND HELOU 2002 EQUATION 4 AND SPITZER DATA.

LITTLE THINGS distances, metallicities, inclination angles, and H$\alpha$ SFRs are from Hunter et al. (2012). 70 (or 60 for IRAS data) $\mu$m, 100 $\mu$m, 160 $\mu$m and FIR luminosities are from Engelbracht et al. (2008), Dale et al. (2009) and Dale et al. (2012). 

%\suri{TIR luminosities are from Dale et al. (2009) and Dale et al. (2012).}}

%FOR THE LITTLE THINGS SOURCES SFR ARE FROM HALPHA. DALE 2009 (TABLE 1) AND DALE 2012 (TABLE 1) IS TIR OBTAINED USING DALE AND HELOU 2002 EQUATION 4.}

Wolf-Lundmark-Melotte (WLM) is included in the LITTLE THINGS sample, and the distance, metallicity, inclination angle, and H$\alpha$ SFR are from Hunter et al. (2012). 70 $\mu$m, 100 $\mu$m, 160 $\mu$m and FIR luminosities are from Rice et al. (1988) and Dale et al. (2009). 

%{TIR luminosities are from Dale et al. (2009).}}
 
KINGFISH 1.4 GHz RC (L$_{\rm 1.4 \, GHz}$) data comes from Tabatabaei et al. (2017; hereinafter T17), while the 70 $\mu$m, 100 $\mu$m and 160 $\mu$m data are from Dale et al. (2012). FIR luminosities are estimated from the IRAS 60 $\mu$m and 100 $\mu$m data, using the formulation in Condon (1992). H$\alpha$ plus 24 $\mu$m SFRs, metallicites and distances are taken from Kennicutt et al. (2011), while inclination angles come from Hunt et al. (2015).

%TIR DATA FROM DALE ET AL. 2012 IS TIR FROM DALE AND HELOU 2002 EQUATION 4 AND HERSCHEL DATA.}

RC data for the DGS, LITTLE THINGS (including WLM) and XMP sources, ranging from several arcseconds to several arcminutes angular resolution, are compiled from (and references therein): Dressel \& Condon (1978; 2.4 GHz), Klein \& Graeve (1986; 4.9 GHz), Zijlstra, Pottasch \& Bignell (1990; 4.9 GHz), Bicay et al. (1995; 4.8 GHz), Faint Images of the Radio Sky at Twenty-Centimeters (FIRST; White et al. 1997), National Radio Astronomy Observatory Very Large Array Sky Survey (NVSS; Condon et al. 1998), Condon, Cotton \& Broderick (2002), Cannon \& Skillman (2004), Lisenfield et al. (2004), Thuan et al. (2004), Leroy et al. (2005), Schmitt et al. (2006; 8.5 GHz), Rosa-Gonzalez et al. (2007; 4.9 GHz), Healey et al. (2007), Roy, Goss \& Anantharamaiah (2008; 8.5 GHz), Heesen et al. (2011; 5 GHz) and Chy\.{z}y et al. (2011; 2.6 GHz). For arcminute RC data, the radio maps were visually inspected for contaminating RC sources; none were found. If not explicit in parentheses, the RC data were obtained at 1.4 GHz. For radio data taken at a frequency different from 1.4 GHz, a steep radio spectrum (F$_{\nu} \propto \nu^{-0.7}$) has been assumed. This choice is justified by the mainly dominant non-thermal steep spectrum component in the total RC emission (see also Sect.~3.1.1).

CO data (L$_{\rm CO}$) for the DGS, LITTLE THINGS (including WLM) and XMP sources are compiled from (and references therein): Sage et al. (1992), Tacconi \& Young (1985), Tacconi \& Young (1987), Israel, Tacconi \& Baas (1995), Kobulnicky et al. (1995), Taylor, Kobulnicky \& Skillman (1998), Helfer et al. (2003), Leroy et al. (2005, 2007, 2009), Bolatto et al. (2008), Schruba et al. (2012), Shi et al. (2014, 2015), Cormier et al. (2014) and Rubio et al. (2016).

% TIR data from R\'emy-Ruyer et al. (2013) are derived by integrating the modelled SED\footnote{spectral energy distribution} curve between 50 and 650 $\mu$m. TIR data from Dale et al. (2012) are based on Herschel PACS\footnote{Photoconductor Array Camera and Spectrometer} and SPIRE\footnote{Spectral and Photometric Imaging Receiver} data.

%%%%%%%%%%%%%%%%%%%%%%%%%%%%%%%%%%%%%%%%%%%%%%%%%%%%%%%%%%%%%%%%%%%%%%%%%%%%%%%%%%%%%%%%%%%%%%%%%%%%%%%%%%%%%%%%%%%%%%%%%%%%%%%%%%%%%%%%%

\section{Results and Interpretation}

Because low-metallicity galaxies are generally also the galaxies that exhibit the lowest mass and lowest luminosity (e.g., Filho et al. 2013, 2016), the effects of mass, luminosity and metallicity on galaxy properties are not always straightforwardly extricable.

Hereinafter, LITTLE THINGS and DGS sources will be referred to simply as dwarfs or dwarf galaxies.

In the following plots (Fig.~1 -- 7), sources with total RC upper limits below 0.1 mJy are suppressed for clarity. Errorbars for the KINGFISH sample, and errorbars on L$_{\rm FIR}$ and L$_{\rm CO}$ are also suppressed for clarity. Because SFR errors for the XMP and DGS sources are unavailable, there are no errorbars in the SFR, L$^{\rm thermal}_{\rm 1.4 \, GHz}$ (the thermal 1.4 GHz RC emission; see also Sect.~3.1.1) and q$^{\rm thermal}_{\rm FIR}$ parameter ($\equiv$ log (L$_{\rm FIR}$/L$^{\rm thermal}_{\rm 1.4 \, GHz}$); see also Sect.~3.1.3). The L$^{\rm non-thermal}_{\rm 1.4 \, GHz}$ (the non-thermal 1.4 GHz RC emission; see also Sect.~3.1.1), q$^{\rm non-thermal}_{\rm FIR}$ parameter ($\equiv$ log (L$_{\rm FIR}$/L$^{\rm non-thermal}_{\rm 1.4 \, GHz}$); see also Sect.~3.1.3) and B (see also Sect.~3.1.4) errorbars in DGS sources and XMPs only include errors in the total RC emission. The errorbars for the L$^{\rm non-thermal}_{\rm 1.4 \, GHz}$, q$^{\rm non-thermal}_{\rm FIR}$ parameter and B in LITTLE THINGS sources include errors in the total RC emission and SFR. It is to be noted that the errorbars in L$^{\rm non-thermal}_{\rm 1.4 \, GHz}$ and L$^{\rm thermal}_{\rm 1.4 \, GHz}$ are highly anti-correlated (see Sect.~3.1.1). Some sources (i.e., KINGFISH sources) and individual data points do not possess radio flux errors, and some errorbars are smaller than the symbols.

Throughout, the XMPs correspond to the red open and solid squares, the DGS sources to the blue open and solid circles, the LITTLE THINGS sources to the purple open and solid triangles, the KINGFISH sources to the green crosses, and WLM to the grey solid and open pentagons.

Table~1 contains the line fits for the data in the following plots (Fig.~1 -- 7), including the Pearson coefficient (r$_{\rm p}$), the Spearman rank (r$_{\rm s}$), and the slope and offset for a linear least squares fit. For data points with more than one L$_{\rm CO}$ value (data points connected by a dotted line), fits have been performed considering only the faintest CO luminosity. Because there are points that do not possess errors, errobars are not used to weight the data for the fits, and upper limits to the data are not incorporated in the fits. Errors in the fits are standard errors, defined as the square root of the estimated error variance. Table~2 contains the statistics for the q$_{\rm FIR}$, q$_{\rm FIR}^{\rm thermal}$ and q$_{\rm FIR}^{\rm non-thermal}$ parameter (see also Sect.~3.1.3). Fits and statistics are only performed for the 'low' angular resolution (several arcmin resolution) radio data.

%%%%%%%%%%%%%%%%%%%%%%%%%%%%%%%%%%%%%%%%%%%%%%%%%%%%%%%%%%%%%%%%%%%%%%%%%%%%%%%%%%%%%%%%%%%%%%%%%%%%%%%%%%%%%%%%%%%%%%%%%%%%%%%%%%%%%%%%%%%%%%%%%%%

\subsection{IR and RC Emission}

\subsubsection{Thermal versus Non-thermal RC Emission}

In the frequency range 1 $<$ $\nu$ $<$ 10 GHz, two mechanisms contribute to the RC: free-free emission from thermal electrons and non-thermal emission from relativistic electrons embedded in magnetic fields (e.g., Condon 1992; T17). In spirals, the average thermal fraction is typically 10\% at 1.4 GHz, increasing to 50\% at 5 GHz (e.g., Beck 2015). For dwarfs, the thermal fraction is generally no more than 30\% at 1.4 GHz (e.g., Niklas \& Beck 1997; Hunt et al. 2004; Roychowdhury \& Chengalur 2012).

Ideally, the thermal -- non-thermal decomposition should be performed through a complete mid-RC spectral energy distribution (SED) analysis, as done by T17 for the KINGFISH galaxies. As such a precise separation is impossible for the dwarf galaxies and XMPs in this study due to the lack of RC data, and for the sake of consistency, the KINGFISH thermal and non-thermal fluxes have been estimated using the same method applied for the dwarf galaxies and XMPs, detailed below. 

A relation between the thermal RC emission and other SFR indicators stems from the proportionality between the thermal RC emission and the total photoizonization rate. In low optical extinction galaxies, such as dwarf galaxies and XMPs (e.g., R\'emy-Ruyer et al. 2013; Shi et al. 2014; see also Sect.~3.1.2), the H$\alpha$ emission is assumed to trace the bulk of the star formation (see Sect.~3.2.1, however). In such systems, assuming case B (Osterbrock \& Ferland 2006) and an electron density (n$_{\rm e}$) of 100 cm$^{-3}$, and considering a pure thermal RC source, the H$\alpha$-to-thermal RC emission ratio is given by (Caplan \& Deharveng 1986)

\begin{equation}
{\rm \frac{F_{H\alpha} \, [erg \, s^{-1} \, cm^{-2}]}{F^{thermal}_{\nu} \, [Jy]} = \frac{8.67 \times 10^{-9} \, \left(\frac{T_{e} \, [K]}{10^4}\right)^{-0.44}} {10.811 + 1.5 \, ln \left(\frac{T_{e} \, [K]}{10^4}\right) - ln (\nu \, [GHz]) }}
\end{equation}

\noindent where $\nu$ is the observed frequency, F$_{\rm H\alpha}$ is the H$\alpha$ flux (Eq.~[2]), F$^{\rm thermal}_{\nu}$ is the thermal flux density at $\nu$, and T$_{\rm e}$ is the electron temperature. 

In order to recover the H$\alpha$ emission from the SFRs provided for the sample sources (Sect.~2), a Kroupa initial mass function (IMF), T$_{\rm e}$ = 10~000K and n$_{\rm e}$ = 100 cm$^{-3}$ have been assumed (Calzetti 2013)

\begin{equation}
{\rm SFR_{H\alpha} \, [M_{\odot} \, yr^{-1}] = 5.5 \times 10^{-42} L_{H\alpha} \, [erg \, s^{-1}]}
\end{equation}

\noindent where L$_{\rm H\alpha}$ is the H$\alpha$ luminosity. Setting $\nu$ to 1.4 GHz and T$_{\rm e}$ to 10~000 K (e.g., Nicolls et al. 2014) allows to determine the thermal 1.4 GHz RC emission. The non-thermal 1.4 GHz RC emission is then obtained by subtracting the thermal contribution from the total 1.4 GHz RC emission.

The thermal and non-thermal 1.4 GHz RC emission for the KINGFISH galaxies obtained with the present method are consistent with the values obtained with the more precise procedure of SED fitting (T17).
 
The procedure entails several caveats. For many of the DGS sources the SFRs are derived from the TIR emission (Sect.~2); in these cases, using the SFR$_{\rm TIR}$ as a proxy for SFR$_{\rm H\alpha}$ provides a lower limit to the true SFR (and hence, L$_{\rm H\alpha}$ and F$_{\nu}$), as the TIR SFR indicator depends strongly on dust content, and on dust absorption properties and timescales (e.g., Kennicutt \& Evans 2012; Calzetti 2013). Because the KINGFISH sample may contain galaxies of high optical extinction, and because the SFRs contain both a dust unobscured (H$\alpha$) and dust-obscured (24 $\mu m$) indicator (Sect.~2), the true SFR (and hence, L$_{\rm H\alpha}$ and F$_{\nu}$) may be overestimated. The true thermal RC contribution may be further overestimated, as continuous star formation is assumed, while most dwarfs and XMPs are characterized by bursty or stochastic star formation episodes (e.g., Weisz et al. 2012). In addition, in low-metallicity systems, where a larger number of UV photons are produced, the thermal RC estimation, based on solar metallicity populations, will tend to overestimate the true thermal RC emission (e.g., Lee et al. 2009). On the other hand, as it is assumed that all emitted Lyman continuum photons result in the ionization of a hydrogen atom, the true thermal RC emission may be underestimated. However, the effect should be small or negligible; low-mass dwarfs are known to possess a modest 2 -- 10\% ionizing photon escape fraction (e.g., Rutkowski et al. 2016).

Figure~1 contains the thermal fraction (thermal-to-total 1.4 GHz RC luminosity) as a function of the total 1.4 GHz RC luminosity (a) and the relation between the thermal and non-thermal 1.4 GHz RC luminosity (b).

%Because the thermal RC luminosity was directly recovered from the H$\alpha$ (or H$\alpha$ plus 24 $\mu$m for the KINGFISH sources) SFR (Eq.~[1]), this plot is equivalent to a plot of the non-thermal 1.4 GHz RC luminosity as a function of the SFR$_{\rm H\alpha}$ (or SFR$_{\rm H\alpha+24\, \mu m}$ for the KINGFISH sources). 
 
%%%%%%%%%%%%%%%%%%%%%%%%%%%%%%%%%%%%%%%%%%%%%%%%%%%%%%%%%%%%%%%%%%%%%%%%%%

% Figure 1 - THERMAL-NON-THERMAL

\begin{figure}
	\includegraphics[width=8cm]{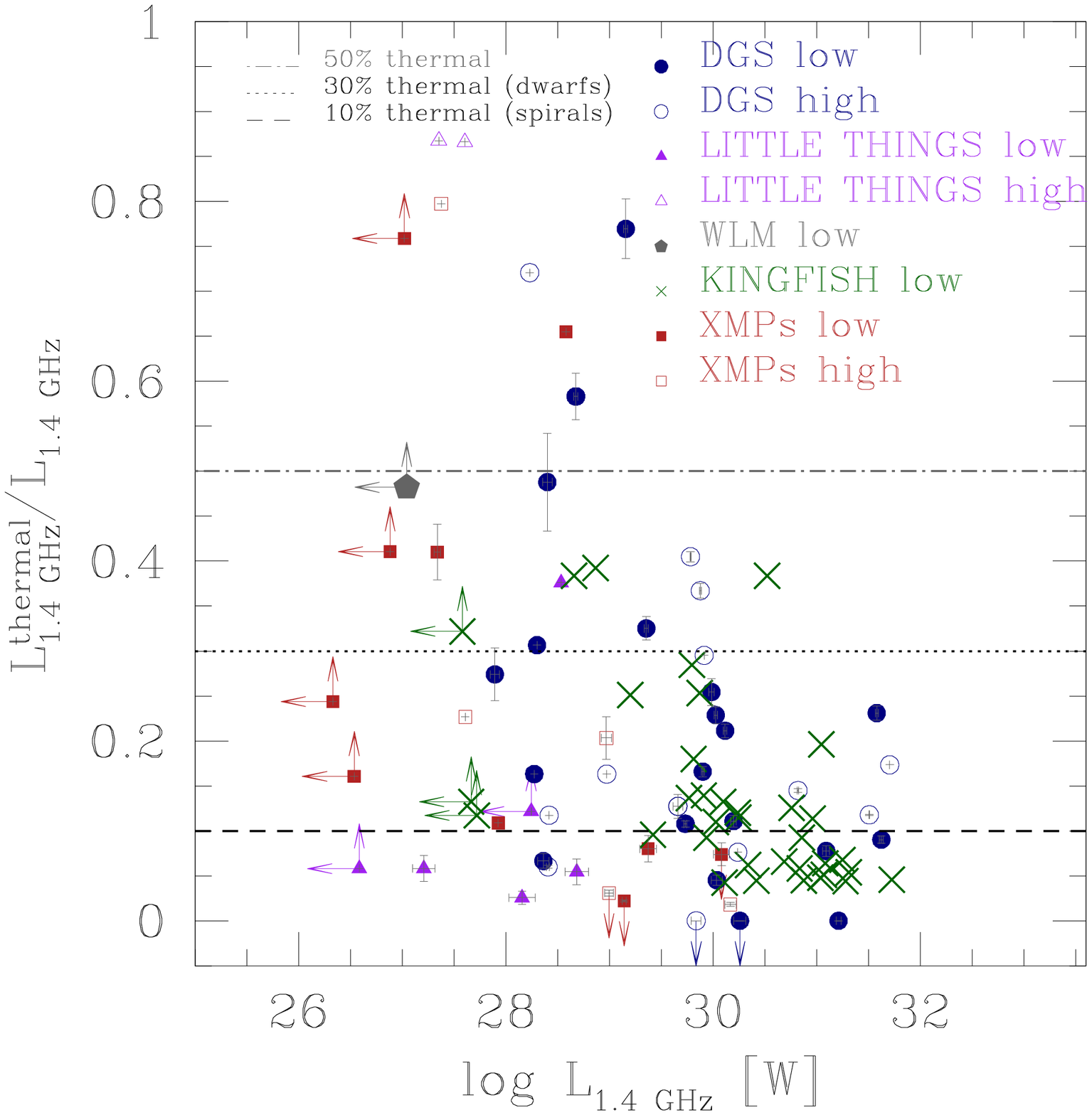}
	(a)
	\\
	\includegraphics[width=8cm]{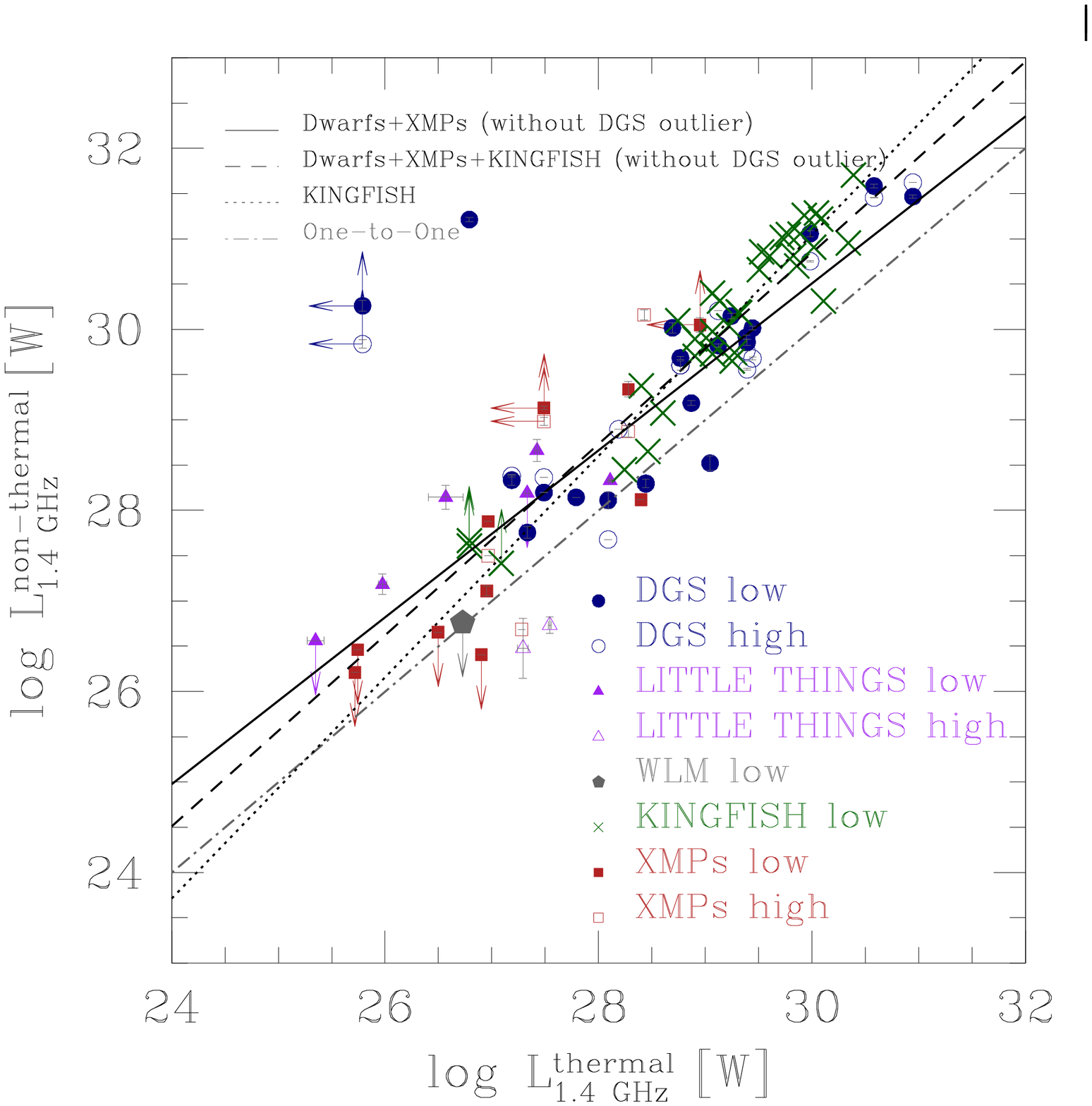}
	(b)
	\caption{{\it Top (a):} Thermal-to-total 1.4 GHz RC luminosity as a function of the logarithm of the total 1.4 GHz RC luminosity. The grey dotted -- dashed line line marks the 50\% thermal contribution, the black dotted line marks the 30\% thermal contribution, typical of dwarf galaxies, and the black dashed line marks the 10\% thermal contribution, typical of spiral galaxies. {\it Bottom (b):} Logarithm of the non-thermal 1.4 GHz RC luminosity as a function of the logarithm of the thermal 1.4 GHz RC luminosity. L$^{\rm thermal}_{\rm 1.4 \, GHz}$ was derived from the H$\alpha$ emission for the LITTLE THINGS sources and XMPs, from the TIR, H$\beta$ or H$\alpha$ emission for the DGS sources, and from the H$\alpha+24\, \mu m$ emission for the KINGFISH sources (Eq.~[1] and [2]). 'High' and 'low' refer to a RC angular resolution of several acseconds to several arminutes, respectively. The fit to the 'low' dwarfs and XMPs is plotted as the black solid line (without the DGS Mrk1089 outlier), to the 'low' dwarfs and XMPs plus KINGFISH galaxies as the black dashed line (without the DGS Mrk1089 outlier), and to the KINGFISH galaxies as the black dotted line (see also Table~1). The grey dotted -- dashed line marks the one-to-one relation. It is to be noted that the thermal and non-thermal 1.4 GHz RC errorbars are highly anti-correlated. XMPs (red open and solid squares), DGS sources (blue open and solid circles), LITTLE THINGS sources (purple open and solid triangles), WLM (grey open and solid pentagons) and KINGFISH sources (green crosses) are plotted.}
\end{figure}
 
%%%%%%%%%%%%%%%%%%%%%%%%%%%%%%%%%%%%%%%%%%%%%%%%%%%%%%%%%%%%%%%%%%%%%%%%%%%

At high luminosity, the thermal fraction for the KINGFISH sources (green crosses; Fig.~1a) is similar to the thermal fraction observed in spiral galaxies ($\sim$10\%; black dashed line; e.g., Beck 2015). The non-thermal-to-thermal 1.4 GHz RC ratio decreases towards lower luminosity (Fig.~1b), as predicted by the standard model (e.g., Condon 1992). The supra-linear slope of the non-thermal-to-thermal 1.4 GHz relation (black dotted line; Fig.~1b) is 1.22$\pm$0.11 (see also Table~1), which is consistent with the (5 GHz) relation L$^{\rm non-thermal}_{\rm 5 \, GHz} \propto$ L$^{\rm thermal \, 1.22\pm0.18}_{\rm 5 \, GHz}$ found by Price \& Duric (1992) for a bright galaxy sample.

Although at lower luminosities the scatter increases, the dwarfs (blue and purple symbols) and XMPs (red symbols) generally show a larger thermal fraction than the KINGFISH sources (Fig.~1a). The dwarfs and XMPs show a non-thermal-to-thermal 1.4 GHz RC relation (black solid line; Fig.~1b) consistent with a slightly sub-linear slope (0.92$\pm$0.09), with a Pearson coefficient of 0.91 (see also Table~1). Notwithstanding the overall behaviour of the non-thermal-to-thermal RC ratio being mainly driven by mass scaling, the ratio for the dwarfs and XMPs is shown to increase slightly with decreasing luminosity, with values ranging from $\approx$2 at high luminosity to $\approx$10 at low luminosity. Lee et al. (2009) find that the H$\alpha$ emission increasingly underestimates the SFR, relative to the UV, towards lower luminosity, from a factor of $\approx$2, to an order of magnitude at the lowest SFRs (see also Sect.~3.1.2 and 3.1.3). Explanations for this variation include increased ionizing photon escape from the galaxy, variations in the initial mass function, increased stochastic star formation (e.g., Weisz et al. 2012), amongst others (see discussion in Lee et al. 2009). As the H$\alpha$ emission is used to derive the thermal RC emission in the majority of the dwarfs and XMPs (Eq.~[2]), this is equivalent to a decreasing thermal RC contribution with decreasing luminosity, which may cause the slightly sub-linear slope observed in the dwarf and XMP correlation (Fig.~1b). It is to be noted, however, that if the XMP and LITTLE THINGS (purple symbols) upper limits are considered below L$_{\rm 1.4 \, GHz}^{\rm non-thermal} \simeq$ L$_{\rm 1.4 \, GHz}^{\rm thermal} \simeq$ 10$^{27}$ W (Fig.~1b), these could suggest a steeper slope, or even a downturn, in the relation at extreme low luminosity. The possible steepening or downturn of the relation below L$_{\rm 1.4 \, GHz}^{\rm non-thermal} \simeq$ L$_{\rm 1.4 \, GHz}^{\rm thermal} \simeq$ 10$^{27}$ W could signal an additional effect: that the non-thermal RC emission ceases to trace the SFR at extreme low luminosity (see also Sect.~3.1.2). 

It is noteworthy that, even at low luminosity, most of the dwarfs and XMPs show a significant non-thermal RC contribution, over 50\% (grey dotted -- dashed line; Fig.~1a; e.g., Niklas \& Beck 1997; Hunt et al. 2004; Roychowdhury \& Chengalur 2012). The still high non-thermal contribution at low luminosity justifies the assumption of a steep radio spectrum for the dwarfs and XMPs (see also Sect.~2). 

The dwarfs and XMPs plus KINGFISH galaxies taken together (black dashed line) suggest that the overall non-thermal-to-thermal RC fraction decreases slightly towards lower luminosity (Fig.~1b), from $\approx$10 at high luminosity to $\approx$3 at low luminosity. The overall relation has a slope of 1.06$\pm$0.06, with a Pearson coefficient of 0.92 (see also Table~1). Overall, the relation is mainly driven by the decreased SFR and luminosity due to mass-scaling, plus the effects of the thermal and non-thermal RC emission ceasing to trace the SFR in low-luminosity, low-SFR dwarfs and XMPs (see also Sect.~3.1.2).

\subsubsection{IR -- RC Relation}

% BIMA SONG range in CO is 30.153 to 31.636 centers and disks of 44 nearby spiral galaxies (Helfer et al. 2003)

A relation between the global IR and RC luminosity was first registered by van der Kruit (1973a, b, c), Dickey \& Salpeter (1984), Helou, Soifer \& Rowan-Robinson (1985) and de Jong et al. (1985). The relation has since been found to be nearly linear (slope = 1.0 -- 1.1), within a factor of two over five orders of magnitude in luminosity, over a broad range of galaxy types, and up to a redshift of $\approx$3 (e.g., Jong et al. 1985; Niklas \& Beck 1997; Yun, Reddy \& Condon 2001; Bell 2003; Appleton et al. 2004; Seymour et al. 2009; Jarvis et al. 2010; Sargent et al. 2010; Bourne et al. 2011; Pannella et al. 2015; however, see also, e.g., Magnelli et al. 2015). 

Bell (2003) found a nearly-linear (T)IR -- RC relation (slope = 1.10$\pm$0.04) down to L$_{\rm TIR} \simeq$ 4 $\times$ 10$^{33}$ W. The nearly-linear relation down to lower luminosities has been attributed to a 'conspiracy', whereby, at low luminosities, both the (T)IR and total 1.4 GHz RC emission underestimate, in the same manner, the SFR (Bell 2003). The effect at low luminosity results from a combination of a non-linear dependence of the non-thermal RC emission on SFR, non-linear effects of dust opacity, and an increased contribution from old stellar populations (see also Sect.~3.1.3). Bell (2003) provided SFR calibrations that take these factors into account at low luminosity (Fig.~2d; blue curved dotted line). However, there is also evidence that the (T)IR -- RC correlation is luminosity-dependent (e.g., Bell 2003 and references therein), with slightly steeper slopes observed for samples weighed by many low-luminosity galaxies. 

% conspiracy, both indicators underestimate the SFR at low luminosity; TIR becomes optically thin to UV and radio because of CR escape. 

Schleicher \& Beck (2016) derive, for dwarfs, three SFR critical surface density ($\Sigma^{\rm crit}_{\rm SFR}$) conditions that can induce a change in the (F)IR -- RC correlation, and can explain the decreasing non-thermal fraction in low-luminosity sources (see also Fig.~1). These critical surface densities depend on the gas density, ionization factor, scale height, electron temperature and filling factor. The first occurs at $\Sigma^{\rm crit}_{\rm SFR} \simeq$ 10$^{-6}$ M$_{\odot}$ yr$^{-1}$ kpc$^{-2}$, and is related to the continuous thermal emission supply, which results from the requirement that the lifetime of massive stars be longer than the timescale of massive star formation. The second occurs at $\Sigma^{\rm crit}_{\rm SFR} \simeq$ 10$^{-6}$ -- 10$^{-5}$ M$_{\odot}$ yr$^{-1}$ kpc$^{-2}$, and is related to the maintenance of the SFR -- magnetic field relation through the continuous injection of turbulent energy via SN explosions, which ensures that the magnetic field is amplified via small-scale dynamos (e.g., Schleicher \& Beck 2013; see also Sect.~3.1.4). The third occurs at $\Sigma^{\rm crit}_{\rm SFR} \simeq$ 10$^{-5}$ -- 10$^{-4}$ M$_{\odot}$ yr$^{-1}$ kpc$^{-2}$, and is related to the dominance of SN explosion injection of CRs over CR diffusion losses, which directly impacts the non-thermal component of the RC emission, and its dependence on the SFR. As long as the above conditions are maintained, and for high $\Sigma_{\rm SFR}$, the characteristic scaling is L$_{\rm 1.4 \, GHz}^{\rm non-thermal} \propto$ L$_{\rm FIR}^{4/3}$ (CR quantity is dependent on the magnetic field strength), while for sources with long rotation periods and low $\Sigma_{\rm SFR}$, the scaling is L$_{\rm 1.4 \, GHz}^{\rm non-thermal} \propto$ L$_{\rm FIR}^{5/3}$ (CR quantity is dependent on the injection rate).

Figure~2 contains plots of the total 1.4 GHz RC luminosity as a function of the 70 $\mu$m (a), 100 $\mu$m (b), 160 (c) $\mu$m and FIR (d) luminosity. 

%%%%%%%%%%%%%%%%%%%%%%%%%%%%%%%%%%%%%%%%%%%%%%%%%%%%%%%%%%%%%%%%%%%%%%%%%%

% Figure 2 - IR-RC

\begin{figure*}
	\includegraphics[width=8cm]{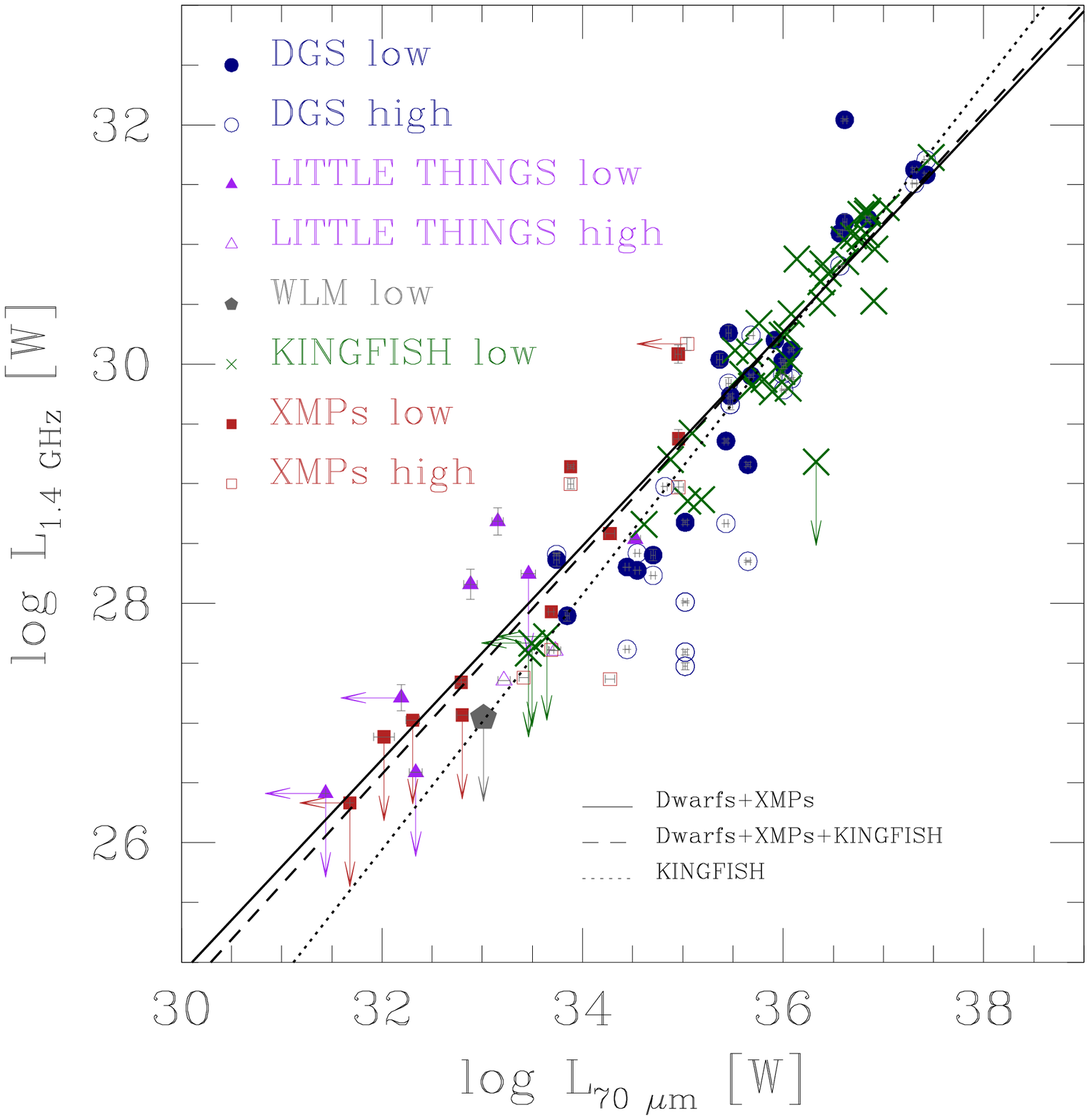}
	(a)
	\includegraphics[width=8cm]{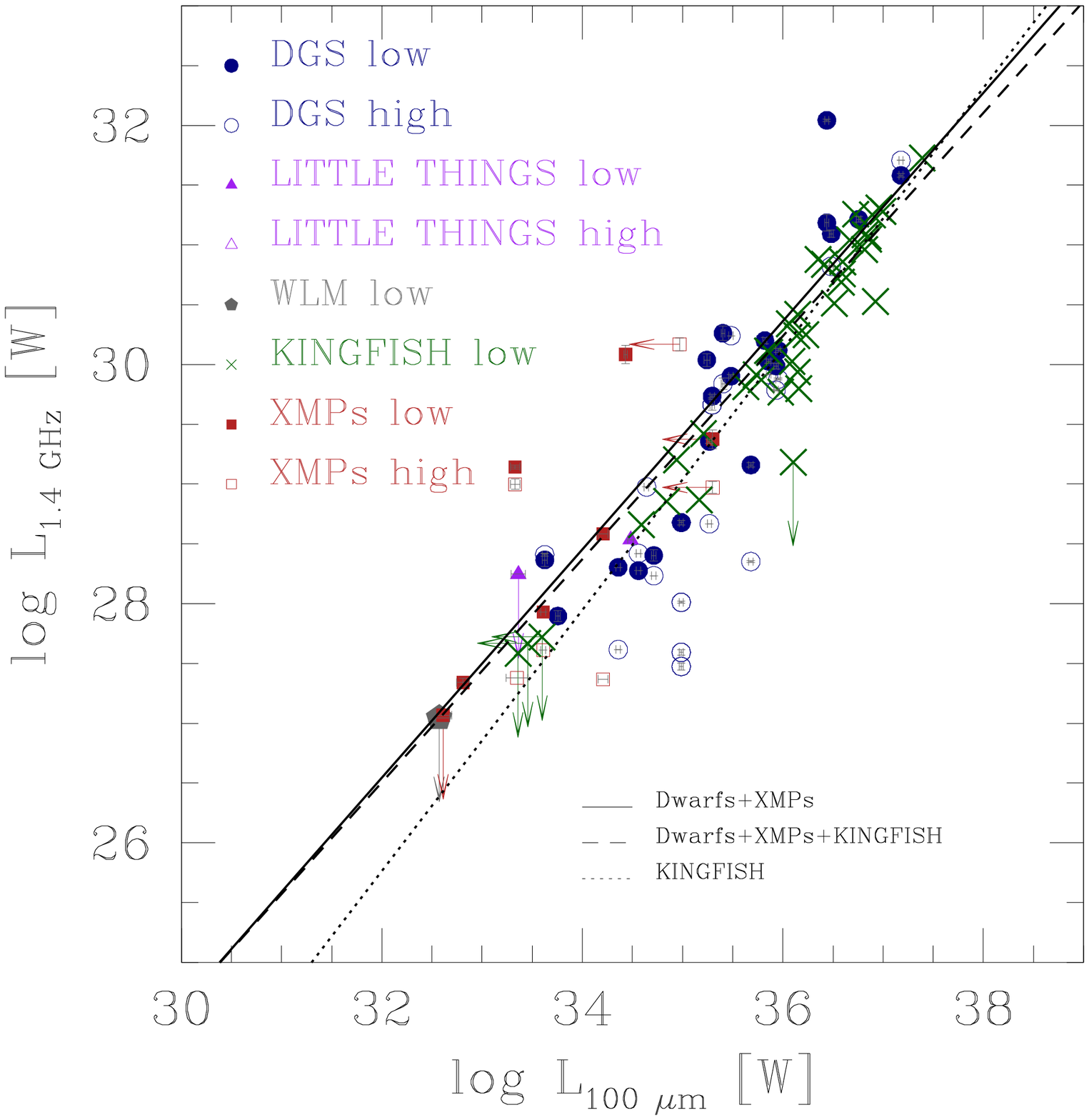}
	(b)
	\\
	\includegraphics[width=8cm]{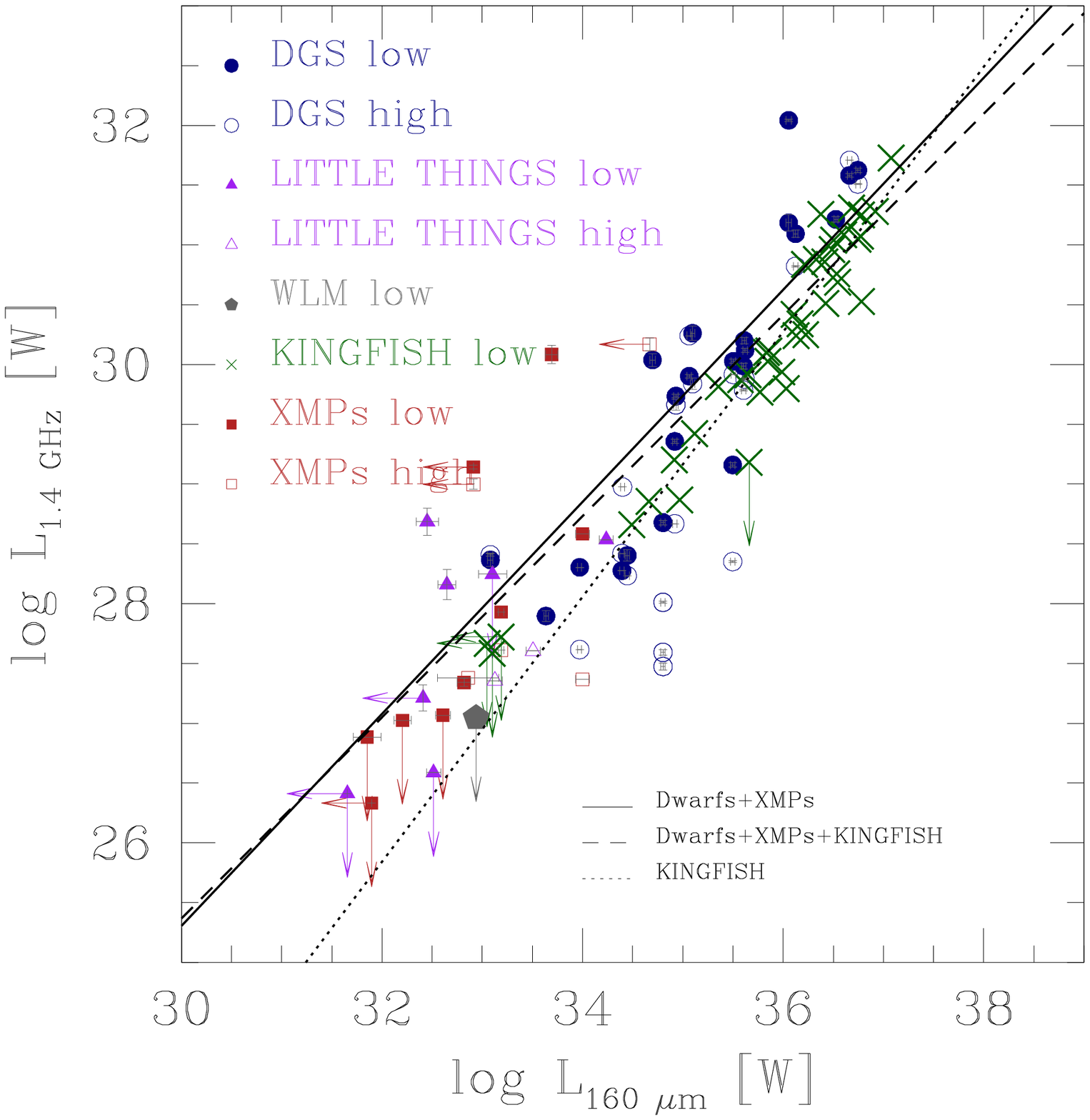}
	(c)
	\includegraphics[width=8cm]{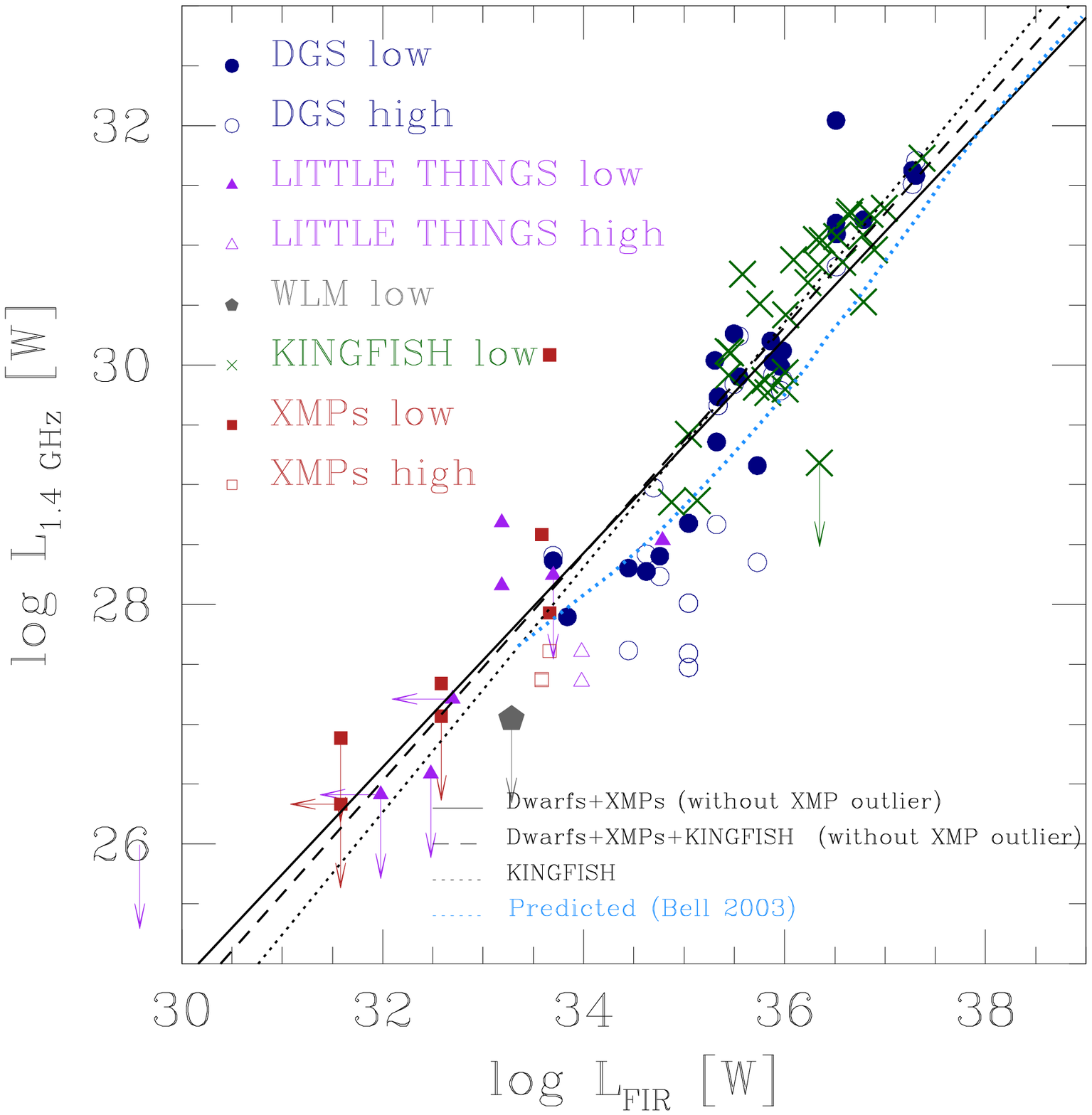}
	(d)
	\caption{{\it Top left (a):} Logarithm of the 1.4 GHz RC luminosity as a function of the logarithm of the 70 $\mu$m luminosity. {\it Top right (b):} Logarithm of the 1.4 GHz RC luminosity as a function of the logarithm of the 100 $\mu$m luminosity. {\it Bottom left (c):} Logarithm of the 1.4 GHz RC luminosity as a function of the logarithm of the 160 $\mu$m luminosity. {\it Bottom right (d):} Logarithm of the 1.4 GHz RC luminosity as a function of the logarithm of the FIR luminosity. The fit to the 'low' dwarfs and XMPs is plotted as the black solid line (for {\it (d)}, without the XMP SBS0335-052 outlier), to the 'low' dwarfs and XMPs plus KINGFISH galaxies as the black dashed line (for {\it (d)}, without the XMP SBS0335-052 outlier), and to the KINGFISH galaxies as the black dotted line (see also Table~1). Plotted also is the predicted (T)IR -- RC relation (blue curved dotted line; {\it (d)}) from the SFR calibrations of Bell (2003). Symbols are the same convention as Figure~1.}
%, and the (60 $\mu$m) IR -- RC  relation (grey dotted -- dashed line; {\it (a)}) for nearby spirals from BIMA SONG (Murgia et al. 2005).
 \end{figure*}

%%%%%%%%%%%%%%%%%%%%%%%%%%%%%%%%%%%%%%%%%%%%%%%%%%%%%%%%%%%%%%%%%%%%%%%%%%%%%

% if the RC luminosity is proportional to the SFR, while the IR only traces a small fraction of the star formation, than the IR -- RC correlation should show a negative curvature, with an additional positive curvature at the high-luminosity end due to the non-linear behavior of the SFR with the dust opacity (bottom right; black dotted line).

The (60 $\mu$m) IR -- RC correlation for the Berkeley Illinois Maryland Association (BIMA) Survey of Nearby Galaxies (SONG; Murgia et al. 2005) and more massive galaxies (black dotted line; Fig.~2a) have nearly linear slopes of 1.05$\pm$0.14 and 1.07$\pm$0.07, respectively. The slope for the dwarf galaxies (blue and purple symbols) and XMPs (red symbols), which are represented by the black solid line fit (Fig.~2a) is sub-linear (0.89$\pm$0.08; see also Table~1).

For all IR luminosities (Fig.~2a, b and c) there is a tendency for the dwarfs and XMPs to be underluminous in the IR for their luminosity, with the IR deficit increasing towards lower luminosity (see also Fig.~3a and 4); the underluminous IR emission drives the shallow slopes of the black solid lines in the figures. The shallow slopes are not related to any violation of the $\Sigma_{\rm SFR}^{\rm crit}$ conditions (Schleicher \& Beck 2016), as the SFR surface density of the dwarfs and XMPs are above the upper threshold of $\Sigma_{\rm SFR}^{\rm crit} \simeq$ 10$^{-4}$ M$_{\odot}$ yr$^{-1}$ kpc$^{-2}$ (Filho et al. 2016). However, if the XMP and LITTLE THINGS (purple symbols) upper limits are considered, these could suggest a steeper slope, or even a downturn, in the relation at extreme low luminosity (L$_{\rm 1.4 \, GHz} \lesssim$ 10$^{27}$ W and L$_{\rm IR} \lesssim$ 10$^{33}$ W; see also Fig.~2d). Although correlations in the literature generally do not probe such low luminosities, this possible steepening of the relation would be consistent with findings that show that lower luminosity galaxies tend to be underluminous in the radio (e.g., Bell 2003).

In the (F)IR -- RC plot (Fig.~2d), the KINGFISH sources (green crosses) show a relation (black dotted line) that is nearly linear, with a slope of 1.02$\pm$0.11 (see also Table~1). The result can be compared to the slopes found by Price \& Duric (1992; slope = 1.12), Yun et al. (2001; slope = 0.99$\pm$0.01) and Bell (2003; slope = 1.10$\pm$0.04) for bright galaxies.

For the dwarfs and XMPs, the (F)IR -- RC correlation (black solid line; Fig.~2d) is consistent with a slightly sub-linear slope (0.89$\pm$ 0.07 and r$_{\rm p}$ = 0.93; see also Table~1). This slope can be compared to the slope of 1.09$\pm$0.07 given in Wu et al. (2008) for a sample of brighter dwarfs. Image stacking of faint dwarf Irregular galaxies (Roychowdury \& Chengalur 2012) also demonstrates that the IR-to-RC ratios in dwarf Irregular galaxies are similar to those of large galaxies, signaling a continuity in IR and RC properties. However, although less pronunced than at other IR luminosities (Fig.~2a, b and c), dwarfs and XMPs appear slightly underluminous in the (F)IR for their luminosity, and the upper limits suggest a possible downturn of the relation at extreme low lumimosities.

Bell (2003) demonstrated that if the total RC is a perfect SFR tracer, than there should be a detectable negative curvature in the (T)IR -- RC correlation towards lower luminosities, as the result of the IR emission tracing only a small fraction of the star formation in fainter galaxies (Fig.~2d). Indeed, there may be some indication that the (F)IR -- RC relation shows an upturn (denoted by the slope flattening) approaching lower luminosity (L$_{\rm 1.4 \, GHz} \simeq$ 10$^{30}$ W and L$_{\rm FIR} \simeq$ 10$^{36}$ W). Nonetheless, below L$_{\rm 1.4 \, GHz} \simeq$ 10$^{27}$ W and L$_{\rm FIR} \simeq$ 10$^{33}$ W, the upper limits in the XMP and LITTLE THINGS data points could suggest a steeper slope, or even a downturn, in the relation at extreme low luminosity, consistent with findings for samples weighed by many faint galaxies (e.g., Bell 2003; see also Fig.~2a, b and c). Together, this could signal that, below L$_{\rm 1.4 \, GHz} \simeq$ 10$^{30}$ W and L$_{\rm IR} \simeq$ 10$^{36}$ W, the IR emission ceases to aqequately trace the SFR, while the total RC emission is still an adequate SFR tracer. Below L$_{\rm 1.4 \, GHz} \simeq$ 10$^{27}$ W and L$_{\rm IR} \simeq$ 10$^{33}$ W, both the total RC and IR emission cease to adequately trace the SFR (see also Sect.~3.1.3). However, it is unclear from the present data if the total RC and IR emission underestimate the SFR in the same manner, i.e., it is unclear if the 'conspiracy' theory of Bell (2003) is valid at extreme low luminosities (see also Fig.~3b and 4). It is to be stressed that the Bell (2003) and Yun et al. (2001) samples do not probe the extreme mass, luminosity and metallicity regime probed by the present sample, which could partly explain why a negative curvature and a downturn in their data is not observed. 

The general behaviour of the IR -- RC relation at low luminosities can be explained as a combination of the following factors (see also Sect.~3.1.1). Mass-scaling is the main driver of the relation. In low-mass, non-star-forming dwarfs, ionizing UV photon production and CR injection is overall suppressed, resulting in weaker thermal and non-thermal contributions. The lower potential wells of low-mass, star-forming dwarfs and XMPs, may imply significant winds and outflows (e.g., Olmo-Garc\'\i a et al. 2017 and references therein), which can contribute to CR advection and escape from the galaxy, and, hence, weaker synchrotron emission. Transient starbursts, associated with low mass and low dust content, can promote increased ionizing photon escape, decreasing the thermal contribution (e.g., Fernandez \& Shull 2011; Benson, Venkatasen \& Shull 2013; Leitherer et al. 2016); however, star-forming dwarfs typically demonstrate a modest ionizing photon escape fraction of 2 -- 10\% (e.g., Rutkowski et al. 2016). Nonetheless, these effects may be counteracted by the increased thermal contribution resulting from the increased ionizing UV photon production associated with transient starbursts, decreased line-blanketing (low metallicity) and low dust content (e.g., Qiu et al. 2017), as well as a change in the non-thermal contribution, as a result of secondary CRs and a change in the CR loss -- CR injection budget due to transient starbursts. In addition, the behaviour of the thermal (e.g., Lee et al. 2009) and non-thermal (e.g., Bell 2003) RC components with SFR play a significant role at extreme low luminosities. Finally, decreased dust content and dust opacity, and, hence, decreased IR emission, is associated with low-luminosity and low-metallicity galaxies (e.g., Klein, Weiland \& Brinks 1991; R\'emy-Ruyer et al. 2013; Shi et al. 2014). 

%It is known that the dust properties and, hence, the IR dust SED, change with metallicity (e.g., R\'emy-Ruyer et al. 2013; Shi et al. 2014); the dust temperature increases with decreasing metallicity, so that the SED of lower-metallicity systems peaks at shorter wavelengths (50 -- 100 $\mu$m) than their higher-metallicity counterparts. For the KINGFISH sample (green crosses), the cooler dust temperatures and the IR SED peaking at longer wavelengths is apparent with the increasing IR luminosity towards longer IR wavelengths (Fig.~2a, b and c). For the dwarfs (blue and purple symbols) and XMPs (red symbols), the shallower (160 $\mu$m) IR -- RC relation (black solid line; slope = 0.89$\pm$0.11; see also Table~1) in Figure~2c signals an increase in the number of sources with decreased IR luminosity towards longer wavelengths, consistent with the expected higher dust temperatures and IR SED characteristics. 

The (F)IR -- RC correlation is also known to hold for both thermal and non-thermal RC emission, but with different slopes (e.g., Price \& Duric 1992); for the (5 GHz) thermal emission it is nearly linear (L$^{\rm thermal}_{\rm 5 \, GHz} \propto$ L$_{\rm FIR}^{0.97\pm0.02}$), while for the (5 GHz) non-thermal emission the relation is steeper (L$^{\rm non-thermal}_{\rm 5 \, GHz} \propto$ L$_{\rm FIR}^{1.33\pm0.10}$). Because the thermal 1.4 GHz RC emission was derived from the SFR (Eq.~[2]), only the non-thermal 1.4 GHz RC component is plotted as a function of the FIR luminosity in Figure~3a. 

%%%%%%%%%%%%%%%%%%%%%%%%%%%%%%%%%%%%%%%%%%%%%%%%%%%%%%%%%%%%%%%%%%%%%%%%%%

% Figure 3 - IR-non-thermal RC

\begin{figure}
	\includegraphics[width=8cm]{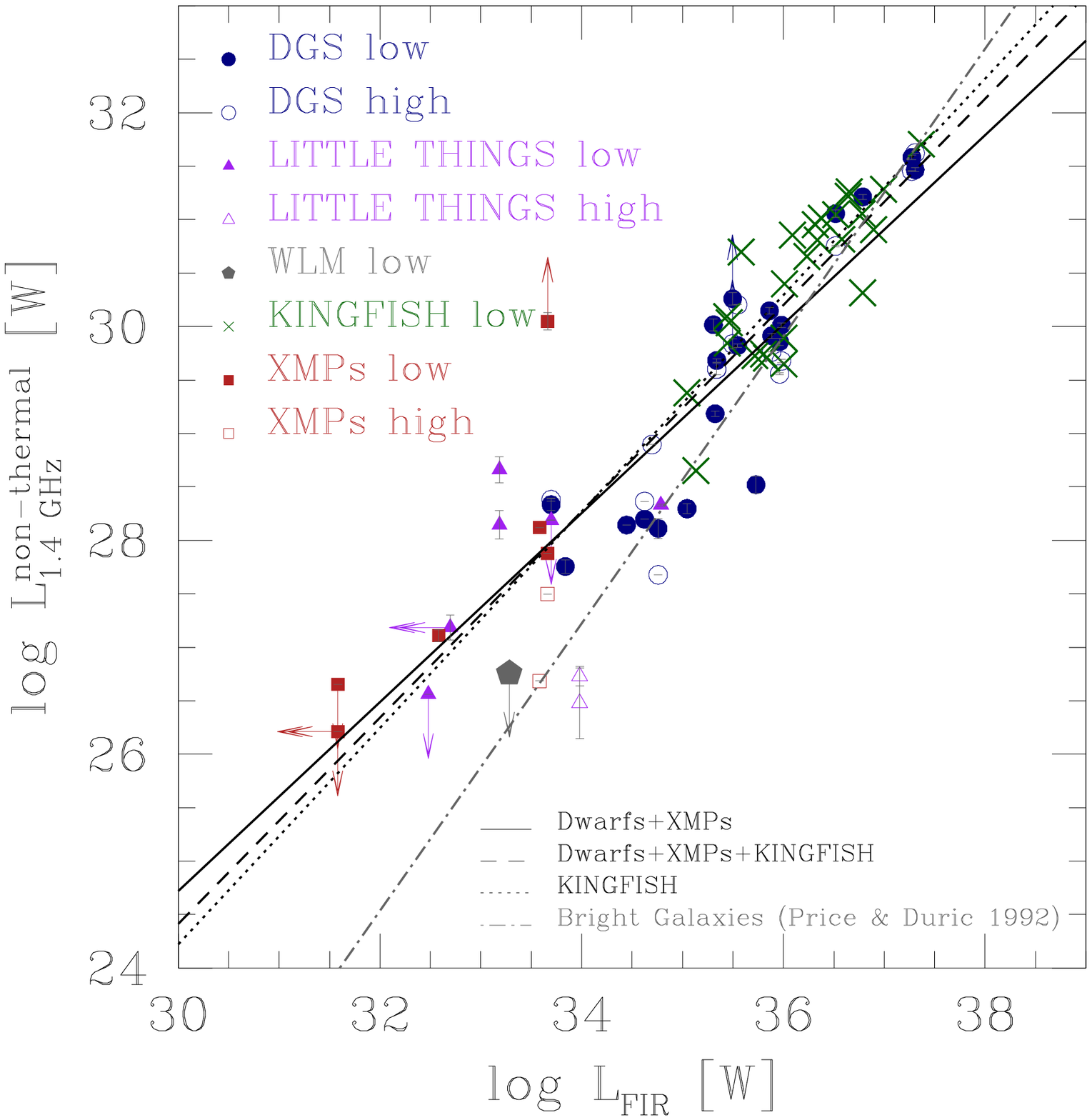}
	(a)
	\\
	\includegraphics[width=8cm]{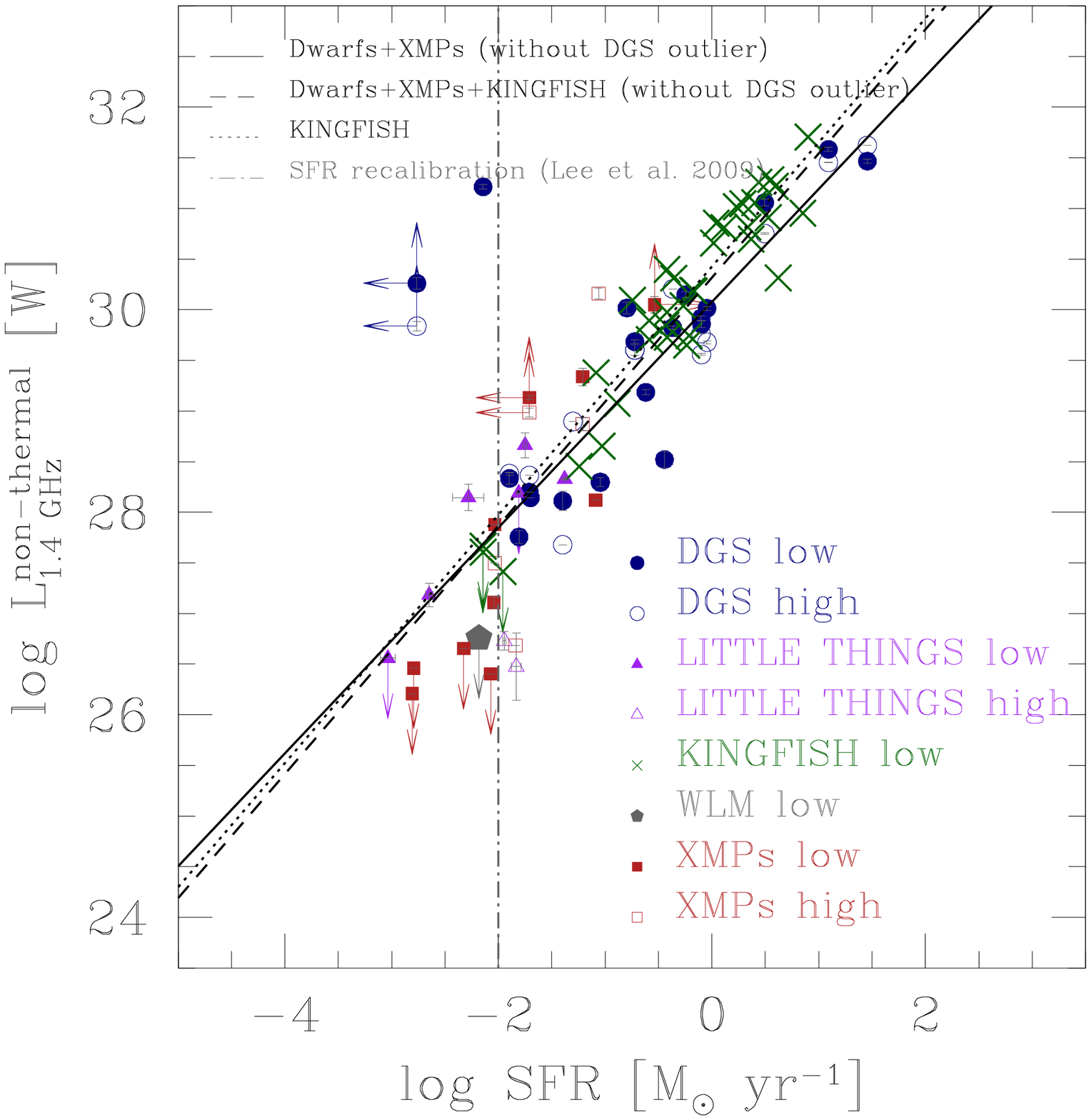}
	(b)
	\caption{{\it Top (a):} Logarithm of the non-thermal 1.4 GHz RC luminosity as a function of the logarithm of the FIR luminosity. The fit to the 'low' dwarfs and XMPs is plotted as the black solid line, to the 'low' dwarfs and XMPs plus KINGFISH galaxies as the black dashed line, and to the KINGFISH galaxies as the black dotted line (see also Table~1). Plotted also is the (F)IR -- (5 GHz) non-thermal RC relation (grey dotted -- dashed line) from Price \& Duric (1992). {\it Bottom (b):} Logarithm of the non-thermal 1.4 GHz RC luminosity as a function of the logarithm of the SFR. For the KINGFISH sources, SFR $\equiv$ SFR$_{\rm H\alpha+24 \, \mu m}$. For dwarfs and XMPs, if SFR $\lesssim$ 0.01 M$_{\odot}$ yr$^{-1}$ (grey dotted -- dashed line), then SFR $\equiv$ SFR$_{\rm UV}$ from the empirical recalibration of Lee et al. (2009), and if SFR $\gtrsim$ 0.01 M$_{\odot}$ yr$^{-1}$, then, generally, SFR $\equiv$ SFR$_{\rm H\alpha}$ (see text for details). The fit to the 'low' dwarfs and XMPs (without the DGS Mrk1089 outlier) is plotted as the black solid line, to the 'low' dwarfs and XMPs plus KINGFISH galaxies (without the DGS Mrk1089 outlier) as the black dashed line, and to the KINGFISH galaxies as the black dotted line (see also Table~1). Symbols are the same convention as Figure~1.}
\end{figure}
 
%%%%%%%%%%%%%%%%%%%%%%%%%%%%%%%%%%%%%%%%%%%%%%%%%%%%%%%%%%%%%%%%%%%%%%%%%%%%%

The KINGFISH galaxies (green crosses) show a correlation between the non-thermal RC and (F)IR emission (black dotted line), with a nearly linear slope of 1.01$\pm$0.13 (Fig.~3a; see also Table~1). This is consistent with the generally low thermal 1.4 RC fraction ($\sim$10\%; Fig.~1a) of the (higher luminosity) KINGFISH sources. The slope is shallower than the (5 GHz) slope (1.33$\pm$0.10) that Price \& Duric (1992) found for a bright galaxy sample.

The dwarf and XMP relation (black solid line) possesses a sub-linear slope of 0.88$\pm$0.09 (Fig.~3a; see also Table~1). The correlation is significantly shallower than the dwarf scaling relations provided by Schleicher \& Beck (2016). Similary to the possible trend seen in Figure~2d, there is a hint, from the shallower slope, that the relation may show an upturn at lower luminosity, approaching L$^{\rm non-thermal}_{\rm 1.4 \, GHz} \simeq$ 10$^{30}$ W and L$_{\rm FIR} \simeq$ 10$^{36}$ W, and possibly a downturn for even lower luminosities (L$^{\rm non-thermal}_{\rm 1.4 \, GHz} \lesssim$ 10$^{27}$ W and L$_{\rm FIR} \lesssim$ 10$^{33}$ W). The overall behaviour signals that the non-thermal 1.4 GHz RC emission may be an adequate SFR indicator in the L$^{\rm non-thermal}_{\rm 1.4 \, GHz} \gtrsim$ 10$^{27}$ W and L$_{\rm FIR} \gtrsim$ 10$^{33}$ W regime (see also Sect.~3.1.3). 

In order to further assess the efficacy of using the non-thermal RC emission to trace SFR at low luminosity, Figure ~3b contains the relation between the non-thermal 1.4 GHz RC luminosity and the SFR. For the KINGFISH sources, the SFR is from H$\alpha$ plus 24 $\mu$m emission, and for the dwarfs and XMPs with SFR $\gtrsim$ 0.01 M$_{\odot}$ yr$^{-1}$, the SFR is, generally, from the H$\alpha$ emission (see also Sect.~2 and 3.1.1). The empirical recalibration of the SFR using the UV SFR as reference (Lee et al. 2009; their Eq.~[10]) has been applied to the dwarfs and XMPs with SFR $\lesssim$ 0.01 M$_{\odot}$ yr$^{-1}$; the recalibration of the SFR takes into account the fact that the H$\alpha$ emission underestimates the SFR, while the UV is a robust tracer of the SFR, in this regime. 

The slope of the KINGFISH relation (black dotted line) is 1.22$\pm$0.11 (see also Table~1), consistent with the (5 GHz) Price \& Duric (1992) slope of 1.2 (Fig.~3b). The dwarfs (blue and purple symbols) and XMPs (red symbols) show a relation (black solid line) with a slope of 1.11$\pm$ 0.09 and r$_{\rm p}$ = 0.90 (see also Table~1). However, below SFR $\simeq$ 0.01 M$_{\odot}$ yr$^{-1}$, the few upper limit XMP and LITTLE THINGS (purple points) data points suggest a possible downturn or steepening of the relation (e.g., Lee et al. 2009). This result signals that for SFR $\gtrsim$ 0.01 M$_{\odot}$ yr$^{-1}$, the non-thermal RC emission adequately traces the SFR, while below this SFR value, the non-thermal RC emission is no longer an adequate SFR tracer. Although the lower SFR and luminosity regime is generally not adequately sampled in the literature, this result is consistent with empirical findings that there is a breakdown in the dependence of the non-thermal RC emission on SFR in low-luminosity sources; however, in contrast, in the literature, it is assumed that this breakdown occurs at slightly higher luminosities, in almost synchroneity with, and with a similar magnitude as, the IR -- SFR breakdown, resulting in a nearly-linear IR -- RC relation ('conspiracy' theory; Bell 2003). 

%%%%%%%%%%%%%%%%%%%%%%%%%%%%%%%%%%%%%%%%%%%%%%%%%%%%%%%%%%%%%%%%%%%%%%%%%%%%%%%%%%%%%%%%%%%%%%%%%%%%%%

\subsubsection{IR -- RC Ratio}

Effects of non-linearity, or an increase in the dispersion of the IR -- RC relation, can be better examined with the IR -- RC ratio, the so-called q$_{\rm IR}$ parameter (e.g., Helou, Soifer \& Rowan-Robinson 1985; Condon, Anderson \& Helou 1991). The IR -- RC ratio can be defined as

\begin{equation}
{\rm q_{IR} = log\Big(\frac{F_{IR}}{F_{radio}}\Big)}
\end{equation}

\noindent where IR refers to the 70, 100 or 160 $\mu$m, or the FIR flux, and F$_{\rm radio}$ refers to the total, non-thermal or thermal 1.4 GHz RC flux. Figure 4 contains plots of the q$_{\rm FIR}$ parameter (IR $\equiv$ FIR), for the total (a), non-thermal (c) and thermal (d) contribution, as a function of the, generally,  H$\alpha$ (or H$\alpha$ plus 24 $\mu$m for the KINGFISH sources) SFR (see also Sect.~2 and 3.1.1), as well as the total q$_{\rm FIR}$ parameter as a function of FIR luminosity (b).

Assuming that the UV emission does trace the SFR in the low-SFR regime (SFR $\lesssim$ 0.01 M$_{\odot}$ yr$^{-1}$), applying the SFR recalibration formula of Lee et al. (2009; their Eq.~[10]) would result in moving low-SFR dwarfs and XMPs to the right in Figure~4a and 4c (see also Fig.~4d), without changing the mean values. The use of SFR$_{\rm H\alpha+24 \, \mu m}$ for the KINGFISH sources (instead of SFR$_{\rm H\alpha}$ generally used for the dwarfs and XMPs; see also Sect.~2 and 3.1.1), places these sources towards the right in Figure~4a and 4c (see also Fig.~4d), without changing the mean values.

%%%%%%%%%%%%%%%%%%%%%%%%%%%%%%%%%%%%%%%%%%%%%%%%%%%%%%%%%%%%%%%%%%%%%%%%%%%%%

% Figure 4 - Q-SFR

\begin{figure*}
	\includegraphics[width=8cm]{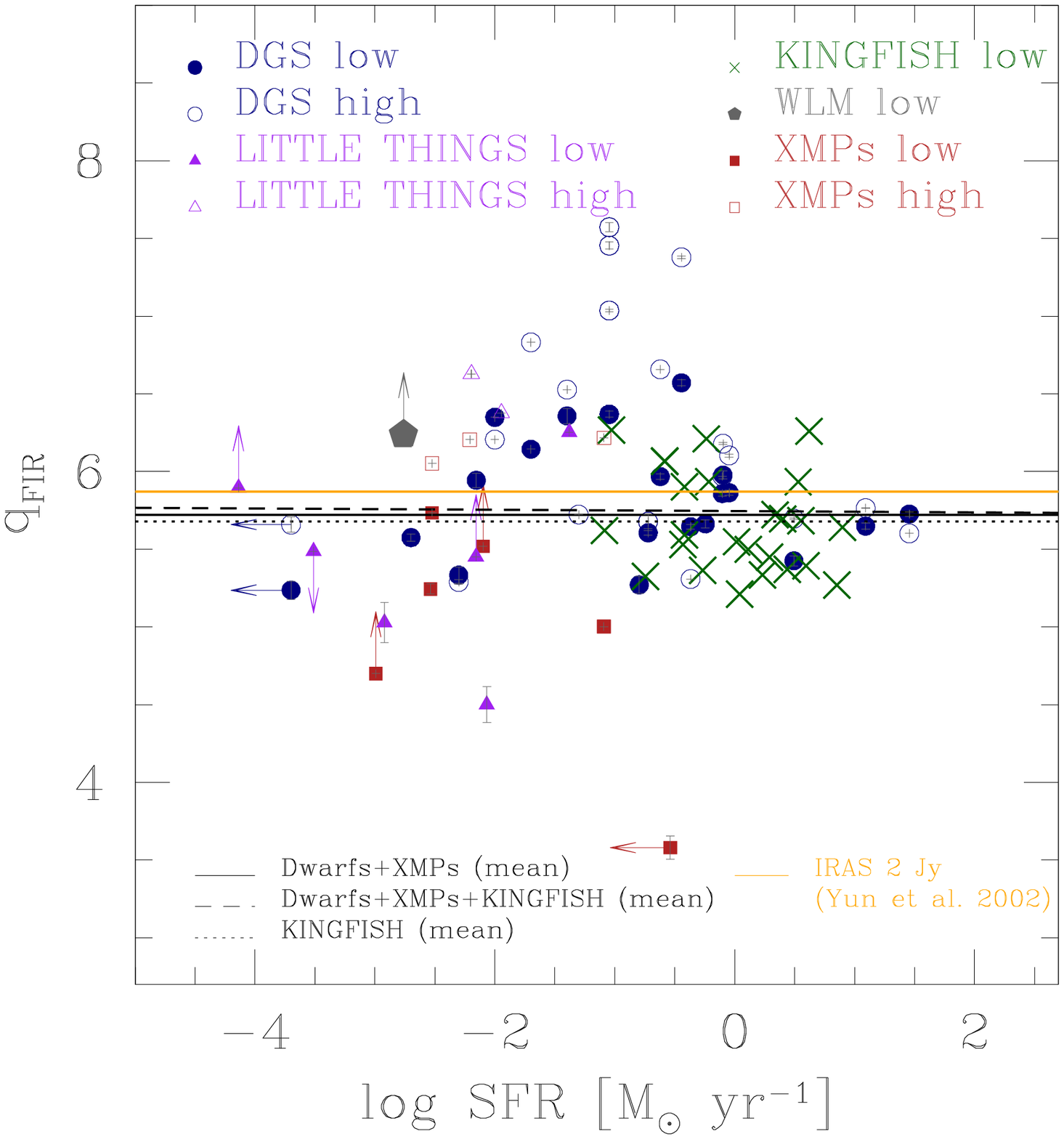}
	(a)
	\includegraphics[width=8cm]{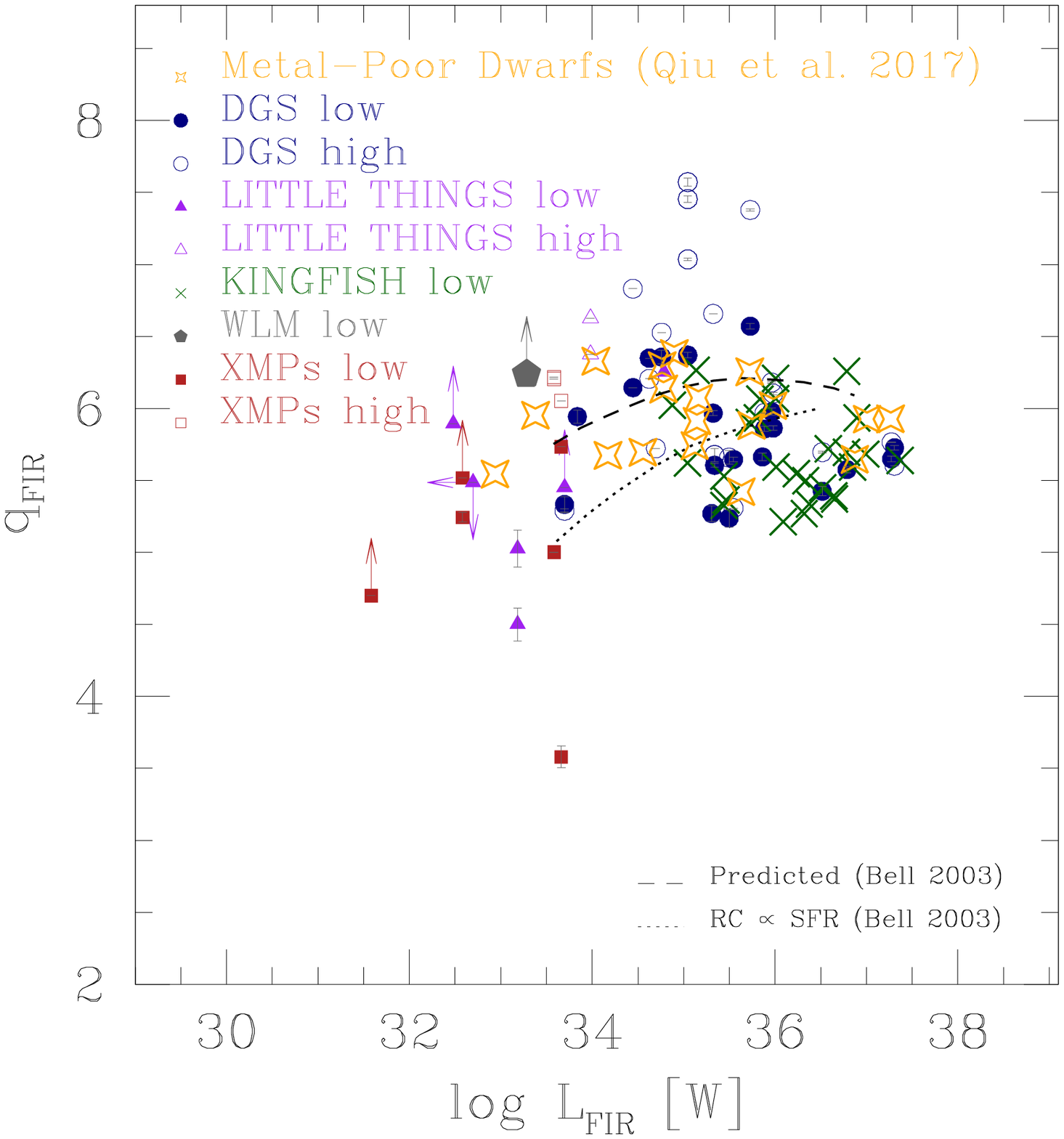}
	(b)
	\\
	\includegraphics[width=8cm]{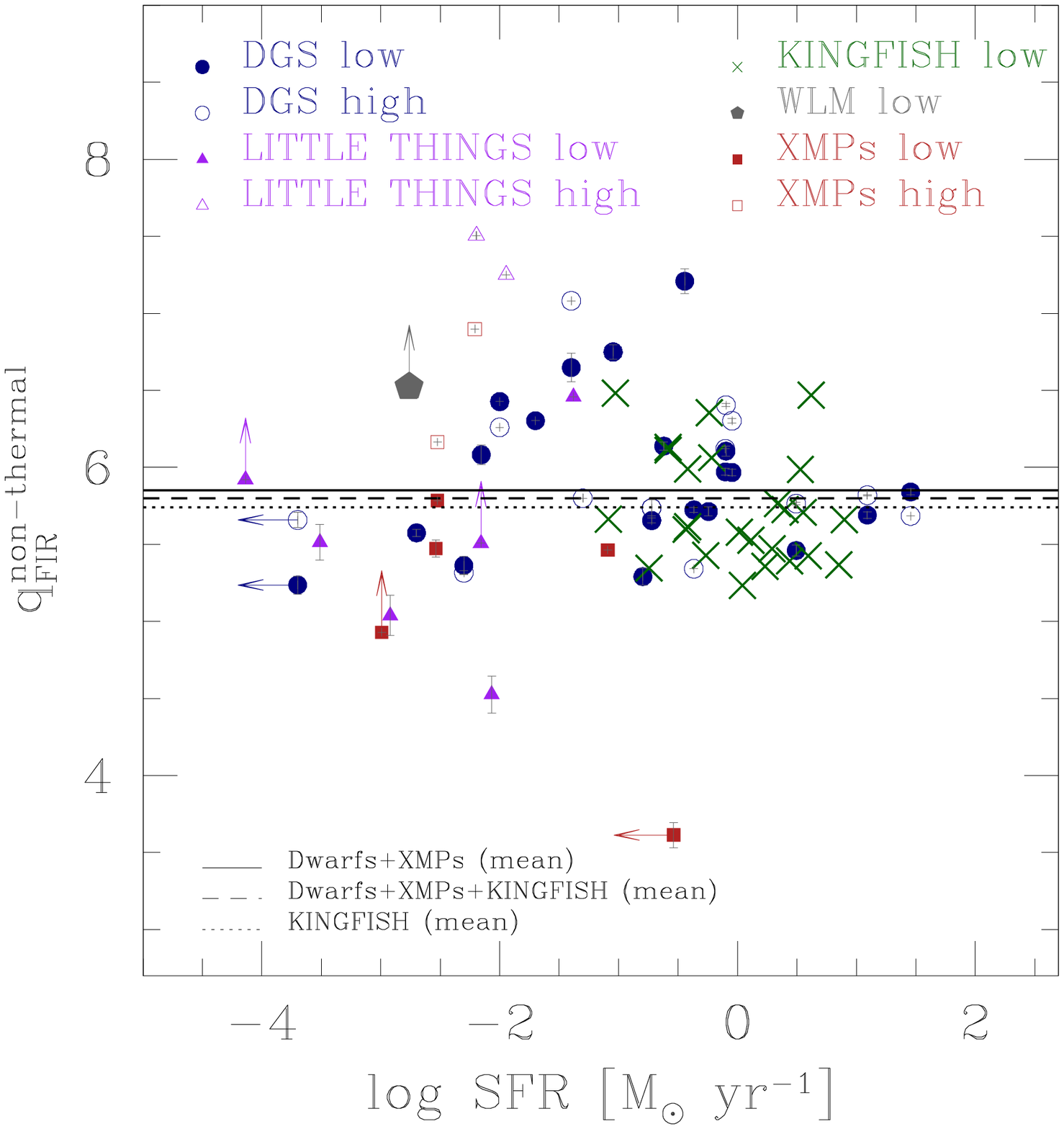}
	(c)
	\includegraphics[width=8cm]{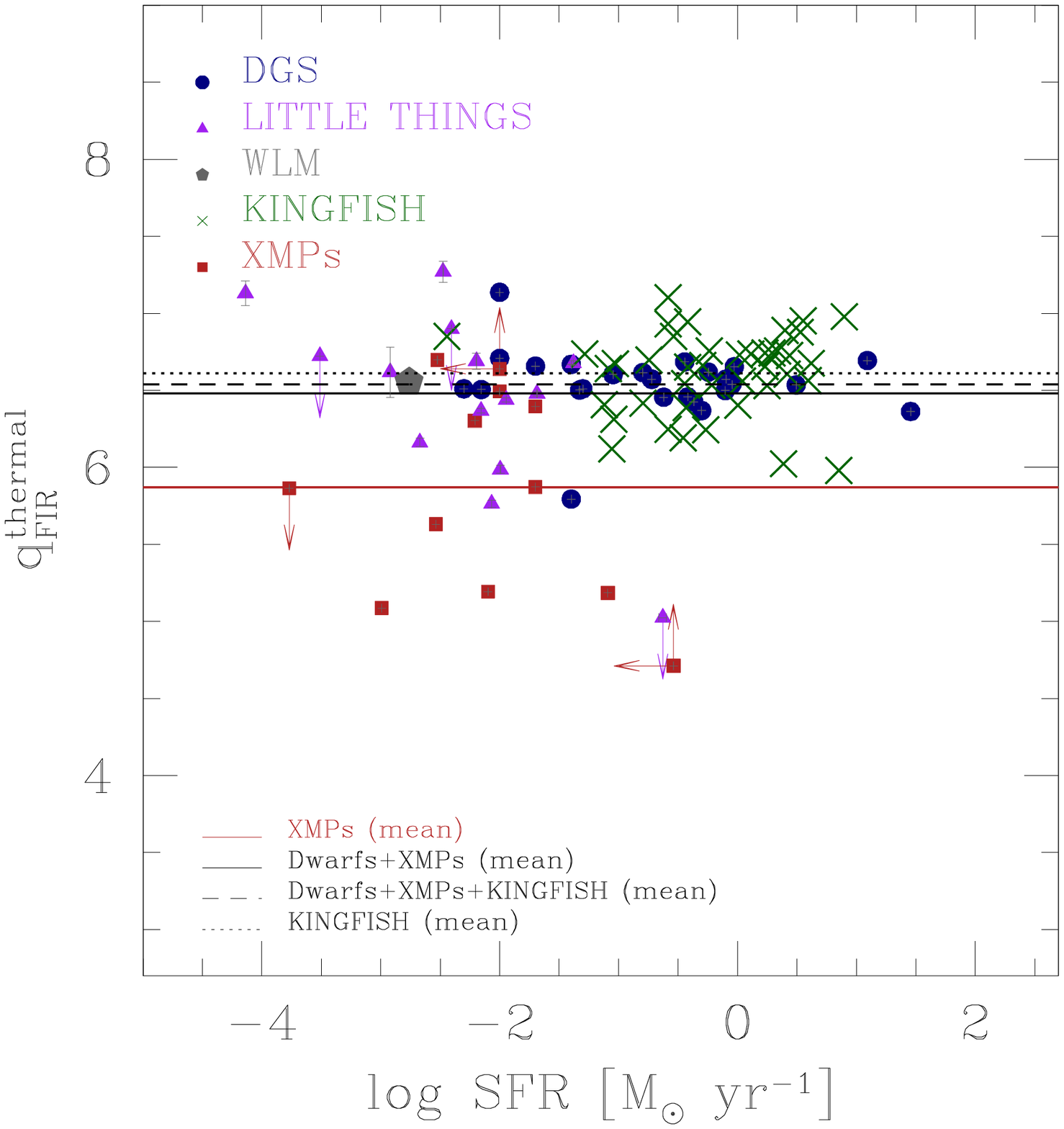}
	(d)

	\caption{{\it Top Left (a):} Logarithm of the FIR-to-total 1.4 GHz RC luminosity as a function of the logarithm of the SFR. {\it Top Right (b):} Logarithm of the FIR-to-total 1.4 GHz RC luminosity as a function of the logarithm of the FIR luminosity. {\it Bottom Left (c):} Logarithm of the FIR-to-non-thermal 1.4 GHz RC luminosity as a function of the logarithm of the SFR. {\it Bottom Right (d):} Logarithm of the FIR-to-thermal 1.4 GHz RC flux as a function of the logarithm of the SFR. For the KINGFISH sources, SFR $\equiv$ SFR$_{\rm H\alpha+24 \, \mu m}$, for the LITTLE THINGS sources and XMPs, SFR $\equiv$ SFR$_{\rm H\alpha}$ and for the DGS sources, SFR $\equiv$ SFR$_{\rm TIR}$, SFR$_{\rm H\beta}$ or SFR$_{\rm H\alpha}$. The mean value for the 'low' XMPs is plotted as the red straight solid line ({\it (d)}), for the 'low' dwarfs and XMPs as the black straight solid line, for the 'low' dwarfs and XMPs plus KINGFISH galaxies as the black straight dashed line, and for the KINGFISH galaxies as the black straight dotted line (see also Table~2). The mean value for the IRAS 2 Jy sample (Yun et al. 2002) is plotted as the orange straight solid line ({\it (a)}). Plotted also is the predicted relation (black curved dashed line; {\it (b)}) from the SFR calibrations of Bell (2003), and the predicted correlation (black curved dotted line; {\it (b)}) if the 1.4 GHz RC emission is a perfect SFR tracer. Metal-poor dwarfs from Qiu et al. (2017) are plotted as orange open stars ({\it (b)}). Symbols are the same convention as Figure~1.}
\end{figure*}

% linear regression
% only dwarfs low resolution (no upper/lower limits) - log SFR = 0.08154515922 q_total + 5.806279659
% dwarfs plus kingfish - log SFR = 0.1277526021 q_total + 6.033910751

% only dwarfs low resolution (no upper/lower limits) - log SFR = 0.1024775654 q_non + 5.993982315
% dwarfs plus kingfish - log SFR = 0.1165804341 q_non + 6.141004086

% only dwarfs low resolution (no upper/lower limits) - log SFR = -0.03612497821 q_therm + 6.383201599
% dwarfs plus kingfish - log SFR = 0.1152599603 q_therm + 6.824923515

% only dwarfs low resolution (no upper/lower limits) - log L_IR = 0.1702640239 q_total -0.6249479651
% dwarfs plus kingfish - log L_IR = 0.2001236379 q_total -1.245579123
% this has the Bell 2003 curve

%%%%%%%%%%%%%%%%%%%%%%%%%%%%%%%%%%%%%%%%%%%%%%%%%%%%%%%%%%%%%%%%%%%%%%%%%%%%%%

Qiu et al. (2017; see also Bell 2003 and Wu et al. 2008) present a comprehensive analysis of the q$_{\rm IR}$ parameter for a sample of metal-poor (12+log(O/H) $\lesssim$ 8.1) galaxies, many of which are common to the present samples. They find that metal-poor galaxies generally have lower q$_{\rm IR}$ parameters than more metal-rich galaxies, with offsets increasing at longer IR wavelengths. The q$_{\rm IR}$ parameter is also found to generally decrease with metallicity, IR -- far-ultraviolet (FUV) ratio and IR color. Their result that the total SFR (FUV+24 $\mu$m) -- RC ratio is constant among metal-rich and metal-poor galaxies signals that the RC is still an effective tracer of the SFR at low metallicity. The proposed Qiu et al. (2017) mechanism at low metallicity is a combination of low obscured-to-total SFR fraction (which reduces the overall IR emission) and warmer dust temperatures (which moves the IR emission peak to lower wavelengths). The reduced dust absorption of the FUV radiation due to the low dust content results in an increase in the ionization of the gas, and, hence an increase in the thermal emission.

% metal-poor have lower qIR with larger offsets at longer IR wavelengths -0.06 dex at 24 micron and -0.6 at 160 micron.
% qIR at 160 micron shows positive trend with metallicity and IR/FUV ratio and negative with IR color
% weaker correlations at lower IR wavelengths

%model where low obscured SFR/total SFR fraction (reduces IR radiation) and warm dust T (reduces IR emission at even longer IR wavelengths) at low metals to explain qIR

% mean total SFR/radio ratio of metal poor and metal rich the same

% Radio still effective SFR at low metallicity

% q24 = 1.34 similar metal poor (8.1) to metal rich

% but Wu finds decreasing trend. below 7.6 dispersion increases

% q70 mean metal-poor is lower than rich by 0.26 dex. decrease with metallicity.
% q100 (0.44) and q160 (0.61 dex) even more decrease

% no qIR IR dependence = not due to lower dust content (lower IR emission) Also Bell 2003 did not find it 
% metal-poor lower IR/FUV indicated lower dust extinction, also decrease q with IR/FUV (this is the reason) - low obscured SFR/total SFR and warm dust color (increased q decrease with longer IR wavelengths) in metal-poor. reduced dust absorbed radiation and increased FUV radiation escape. This ionizes more gas leading to enhanced thermal emission.
% known that dust IR SED of metal poor different (Remy Ruyer et all 2013; Shi et all 2014).
% only q160 shows relation with IR color.
%At longer IR wavelengths from 24 μm to 160 μm, the offsets become larger and the trend of decreasing qIR with the IR/FUV ratio and IR color becomes stronger.

The present results are similar to the results presented in T17 for the KINGFISH galaxies (green crosses). The more massive galaxies generally show a slightly negative slope in the total q$_{\rm FIR}$ parameter versus SFR (Fig.~4a), driven by the slightly negative slope in the non-thermal 1.4 GHz RC emission (Fig.~4c). Because it is assumed that the IR emission traces the SFR in massive galaxies, this signals that the total and non-thermal RC emission increases faster with increasing SFR than the FIR emission. This is consistent with the supra-linear dependence of the non-thermal RC emission on SFR (see also Fig.~3b), as well as with the (F)IR -- RC relation (see also Fig.~2) for the KINGFISH galaxies. Increased star formation may also generate and amplify magnetic fields, and increase CR production, increasing the non-thermal, and, hence, the total RC component (e.g., T17). 

The mean value of the total and non-thermal q$_{\rm FIR}$ parameter (Fig.~4a and c) for the dwarfs and XMPs (black solid line) is similar to the mean value for the KINGFISH galaxies (black dotted line; see also Table~2). The value of the q$_{\rm FIR}$ parameter mean can be compared to the mean value for the IRAS 2 Jy sample, after converting to the present definition in Equation [3] (5.87; orange solid line; Yun et al. 2002). After converting to the present total q$_{\rm FIR}$ parameter definition (Eq.~[3]), the Qiu et al. (2017) mean total q$_{\rm FIR}$ parameter value for their metal-poor sample (12+log(O/H) $\lesssim$ 8.1) is consistent with the present dwarf and XMP value (6.0), and the mean total q$_{\rm FIR}$ parameter value for their metal-rich sample (12+log(O/H) $\gtrsim$ 8.1) is also consistent with the present KINGFISH value (6.2; see also Table~2).

The overall behaviour in the total and non-thermal q$_{\rm FIR}$ parameter for dwarfs (blue and purple symbols) and XMPs (red symbols) is mainly driven by the behaviour of the IR (see also Fig.~2 and 3a), thermal (e.g., Lee et al. 2009) and non-thermal (see also Fig.~3b) emission with SFR, with a breakdown of the IR emission towards lower SFR, and a breakdown of the thermal (generally, $\propto$ SFR$_{\rm H\alpha}$; Eq.~[2]; see also Sect.~2 and 3.1.1), non-thermal, and, hence, total RC emission below SFR $\simeq$ 0.01 M$_{\odot}$ yr$^{-1}$ as adequate star formation tracers. 

The dispersion, and total and non-thermal q$_{\rm FIR}$ parameter values (Fig.~4a and c) increase towards decreasing SFR. However, several low-SFR LITTLE THINGS sources (purple symbols) and XMPs possess particularly low values, consistent with the findings of Qiu et al. (2017), whereby the total q$_{\rm IR}$ parameter decreases with metallicity. The variation at low SFR, or, equivalently, at low luminosity (Eq.~[2]), signals that the (F)IR emission generally decreases faster with decreasing SFR than the total and non-thermal RC emission; a nearly-linear IR -- RC correlation (see also Fig.~4), and/or consistency with the 'conspiracy' theory (Bell 2003), down to extreme low luminosities would require that the (F)IR-to-total RC ratio remain roughly constant. 

Because the thermal 1.4 GHz RC continuum emission is generally derived from the H$\alpha$ (or H$\alpha$ plus 24 $\mu$m for the KINGFISH sources) SFR (Eq.~[1]; see also Sect.~2 and 3.1.1), q$_{\rm TIR}^{\rm thermal} \propto$ log$\Big(\frac{\rm SFR_{FIR}}{\rm SFR_{H\alpha}}\Big)$ (or q$_{\rm TIR}^{\rm thermal} \propto$ log$\Big(\frac{\rm SFR_{FIR}}{\rm SFR_{H\alpha+24 \, \mu m}}\Big)$ for the KINGFISH sources). If SFR$_{\rm H\alpha}$ (or SFR$_{\rm H\alpha+24 \, \mu m}$ for the KINGFISH sources) and SFR$_{\rm FIR}$ trace star formation in roughly the same manner, than the q$_{\rm TIR}^{\rm thermal}$ parameter (Fig.~4d) is expected to show a somewhat flat distribution. For the same galaxy population, significant deviations from a flat slope signals that the two indicators are not tracing the star formation in a similar manner.

The distribution of massive galaxies (green crosses) and high-SFR (SFR $\gtrsim$ 0.01 M$_{\odot}$ yr$^{-1}$) dwarfs (blue symbols) is relatively flat, as expected if the SFR$_{\rm FIR}$ and SFR$_{\rm H\alpha}$ (or SFR$_{\rm FIR}$ and SFR$_{\rm H\alpha+24 \, \mu m}$ for the KINGFISH sources) trace the star formation in the same manner (Fig.~4d). As IR emission is assumed to trace SFR in massive galaxies, this is also consistent with the findings of Lee et al. (2009); at high SFR, the H$\alpha$ emission robustly traces the SFR, as probed by the UV emission. Approaching lower SFRs (SFR $\simeq$ 0.01 M$_{\odot}$ yr$^{-1}$), coinciding with the SFR at which the H$\alpha$ emission ($\propto$ L$^{\rm thermal}_{\rm 1.4 \, GHz}$; Eq.~[1]) no longer adequately traces the SFR (Lee et al. 2009), the dispersion increases, and the thermal q$_{\rm FIR}$ parameter values decrease overall (Fig.~4d). It is noteworthy that low-SFR LITTLE THINGS sources (purple symbols) and XMPs (red symbols) show the lowest thermal q$_{\rm FIR}$ parameter values, with a mean of 5.87 for the XMPs (red solid line; see also Table~2), significantly offset from the high-SFR dwarfs and KINGFISH sources. The low thermal q$_{\rm FIR}$ parameter values reflect the fact that, in low-SFR dwarfs and XMPs, the FIR emission decreases more rapidly with decreasing SFR than the thermal RC emission (see also Sect.~3.1.1 and 3.1.2). 

%The mean value (7.0; see also Table~2) of the massive galaxies (black dotted line) is offset relative to the high-SFR dwarfs. There are several effects that can contribute to this offset: the IR deficit in dwarfs, which decreases the q$^{\rm thermal}_{\rm TIR}$ parameter values with respect to more massive galaxies (see also Fig.~2 and 3), the use of the TIR emission instead of the TIR emission in the KINGFISH galaxies (see also Fig. 2 and 3), and excess TIR emission in more massive galaxies from non-star-forming radiation of dust and hotter-than-normal dust. 

Bell (2003) demonstrated that if the total RC emission perfectly traces the SFR, then the total q$_{\rm (T)IR}$ parameter should decrease steeply with decreasing luminosity, as the result of the effects of dust optical depth. Figure~4b contains the Bell (2003) predicted non-linear SFR -- RC relation (black curved dashed line), and the L$_{\rm 1.4 \, GHz} \propto$ SFR relation (black curved dotted line). The data (orange open stars) of Qiu et al. (2017; their Fig.~3), which go down to L$_{\rm FIR} \simeq$ 10$^{33}$ W and includes a broader range in metallicity (12+log(O/H) $\lesssim$ 8.1), have also been included in Figure~4d.

The Qiu et al. (2017) data points appear to flatten at total q$_{\rm FIR}$ parameter values between 5.4 and 5.9 (Fig.~4b). However, there is some indication from the dwarfs (blue and purple symbols) and XMPs (red symbols) that the total q$_{\rm FIR}$ parameter does decrease steeply with decreasing FIR luminosity (black curved dotted line). This result signals that the FIR emission below L$_{\rm FIR} \simeq$ 10$^{36}$ W ceases to adequately trace,  while the total RC emission continues to adequately trace, the SFR. Below L$_{\rm FIR} \simeq$ 10$^{33}$ W, the few lower-limit data points (purple and red symbols) hint that the decrease may become shallower (black curved dashed line) at extreme luminosity, signaling that the neither the total RC nor FIR are adequate SFR tracers in this regime (see also Sect.~3.1.2). Although similar analyses in the literature do not probe such extreme regimes, this last result is consistent with findings that show a breakdown of the IR, and thermal, non-thermal, and, hence, total RC emission with SFR at low luminosities (e.g., Bell 2003; Lee et al. 2009); however, in contrast, it is generally assumed that the IR and RC breakdown occur at similar luminosities and with similar magnitudes, resulting in a nearly-linear IR -- RC relation ('conspiracy' theory; Bell 2003). 

\subsubsection{B -- SFR Relation}

In the following, a series of assumptions are made in order to estimate the magnetic field strength. It is, however, necessary to present the caveats related to the method. Dwarfs and XMPs undergoing a star formation burst are subject to strong winds and outflows due to their lower gravitational potential wells (e.g., Olmo-Garc\'\i a et al. 2017 and references therein), which can be responsible for increased CR advection and escape from the galaxy; CR losses may render the equipartition assumption invalid. In addition, depending on the mass, star formation rate, density and covering fraction of the gas, increased ionizing photon escape (e.g., Fernandez \& Shull 2011; Benson, Venkatesan \& Shull 2013 Leitherer et al. 2016) may invalidate the case B recombination assumption, which is associated with photon trapping in high density mediums, and undermines the effectiveness of using SFR tracers to estimate the free-free emission in dwarfs and XMPs. In addition, the SFR calibration (Eq.~[1] and [2]) may not be entirely appropriate, as it assumes solar metallicity and continuous star formation, while, in the case of dwarfs and XMPs, metallicities are lower and star formation is generally bursty and stochastic (e.g., Weisz et al. 2012). The choice of an appropriate reference galaxy is also important, since it establishes the zero-point of the magnetic field strength -- non-thermal RC emission relation (Eq.~[4]). Finally, in order to estimate the inclination angles for the DGS sources, it has been assumed that they are disks with a fixed intrinsic thickness (see also Sect.~2), an assumption which may not be valid.

The assumption of equipartition between the magnetic field and CR energy densities allows an estimate of the magnetic field strength (total magnetic field plus its line-of-sight perpendicular component plus the pathlength through the medium) from the non-thermal 1.4 GHz RC emission in star-forming regions dominated by turbulence (e.g., Beck \& Krause 2005). Geometries of the galaxies are taken into account, as well as the contribution of the ordered and random components to the total magnetic field strength.
 
The mean equipartition magnetic field strength, B, was derived from the non-thermal 1.4 GHz RC emission using Eq.~[25] of T17 (see also Beck \& Krause 2005)

\begin{equation}
{\rm B \, [\mu G] = B_0 \bigg( \frac{cos \, i}{cos \, i_0} \bigg)^{(-1/4)} \Bigg( \frac{L^{non-thermal}_{1.4\,GHz}}{L^{non-thermal_0}_{1.4\,GHz}} \Bigg)^{(1/4)}}
\end{equation}

\noindent where i is the inclination of the magnetic field with respect to the line-of-sight (assumed to be the inclination angle of the source), L$^{\rm non-thermal}_{\rm 1.4 \, GHz}$ is the non-thermal 1.4 GHz RC luminosity, and the subscript '0' refers to the reference galaxy. The ratio between the number density of CR protons and electrons was assumed to be 100, the pathlength through the synchrotron medium was assumed to be 1 kpc/cos i, and the non-thermal spectral index was assumed to be one (i.e., CR cooling is dominated by synchrotron emission).

Similarly to T17, NGC6946 has been used as the reference galaxy for the KINGFISH galaxies, dwarfs and XMPs with values B$_0$ = 16 $\mu$G, i$_0$ = 33$^{\circ}$, and F$^{{\rm non-thermal}_0}_{\rm 1.4\,GHz}$ = F$_{\rm 1.4\,GHz}$(1 - f$^{\rm thermal}_{\rm 1.4\,GHz}$) = 1.3 Jy, where F$_{\rm 1.4\,GHz}$ = 1440 mJy, and the thermal fraction at 1.4 GHz is f$^{\rm thermal}_{\rm 1.4\,GHz}$ = 0.10 (T17; their Table 6 and 7). M33 has also been used as the reference galaxy for the dwarfs and XMPs, with values B$_0$ = 6.5 $\mu$G (Tabatabaei et al. 2008), i$_0$ = 56$^{\circ}$ (Regan \& Vogel 1994), and F$^{{\rm non-thermal}_0}_{\rm 1.4\,GHz}$ = F$_{\rm 1.4\,GHz}$(1 - f$^{\rm thermal}_{\rm 1.4\,GHz}$) = 2.2 Jy, where F$_{\rm 1.4\,GHz}$ = 2722 mJy (Tabatabaei et al. 2007), and the thermal fraction at 1.4 GHz is f$^{\rm thermal}_{\rm 1.4\,GHz}$ = 0.18 (Tabatabaei et al. 2007).

% {\bf M33 is spiral SA(s)cd local group galaxy}

% F_nonthermal at 5 GHz = F_total @ 5GHz (1-f_thermal at 5 GHz) = 660 mJy * (1 - 0.24) = 0.5 Jy 
% F_nonthermal at 1.4 GHz = F_total @ 1.4GHz (1-f_thermal at 1.4 GHz) = 1440 mJy * (1 - 0.10) = 1.3 Jy 

Figure~5 contains the relation between the magnetic field strength and the SFR, when M33 is used as the reference galaxy for dwarfs and XMPs, and NGC6946 is used as the reference galaxy for KINGFISH galaxies (a and c), and when NGC6946 is used as the reference galaxy for dwarfs and XMPs, and KINGFISH galaxies (b). The SFR is from the H$\alpha$ plus 24 $\mu$m emission for the KINGFISH sources (a, b and c), from, generally, the H$\alpha$ emission for the dwarfs and XMPs (a and b; see also Sect.~2 and 3.1.1), from, generally, the H$\alpha$ emission for the dwarfs and XMPs with SFR $\gtrsim$ 0.01 M$_{\odot}$ yr$^{-1}$ (c; see also Sect.~2 and 3.1.1) and from the recalibration of the SFR using the UV SFR (Lee et al. 2009) for the dwarfs and XMPs with SFR $\lesssim$ 0.01 M$_{\odot}$ yr$^{-1}$ (c).

%%%%%%%%%%%%%%%%%%%%%%%%%%%%%%%%%%%%%%%%%%%%%%%%%%%%%%%%%%%%%%%%%%%%%%%%%%

% Figure 5 - B-SFR

\begin{figure*}
	\includegraphics[width=8cm]{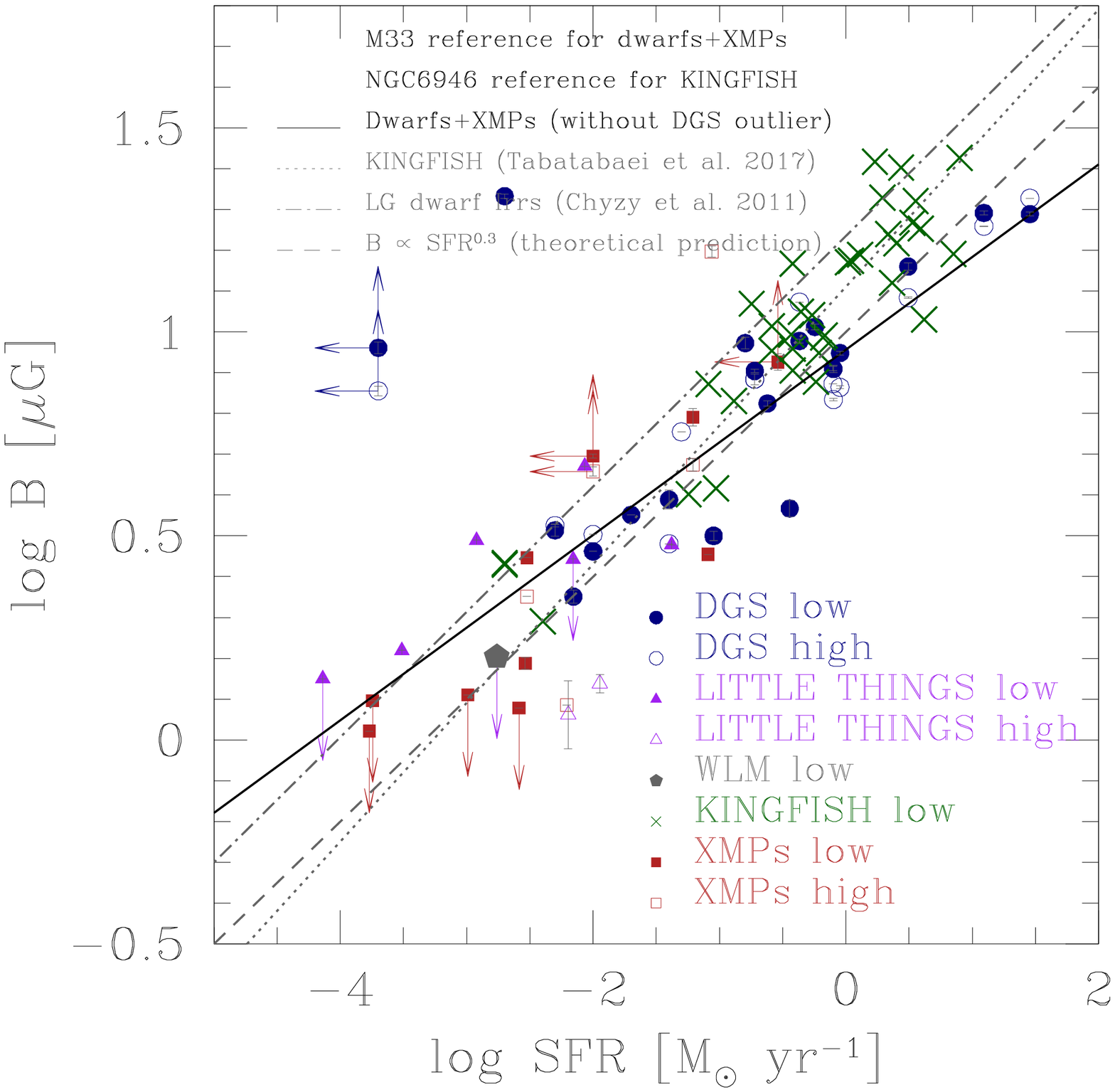}
	(a)
	\includegraphics[width=8cm]{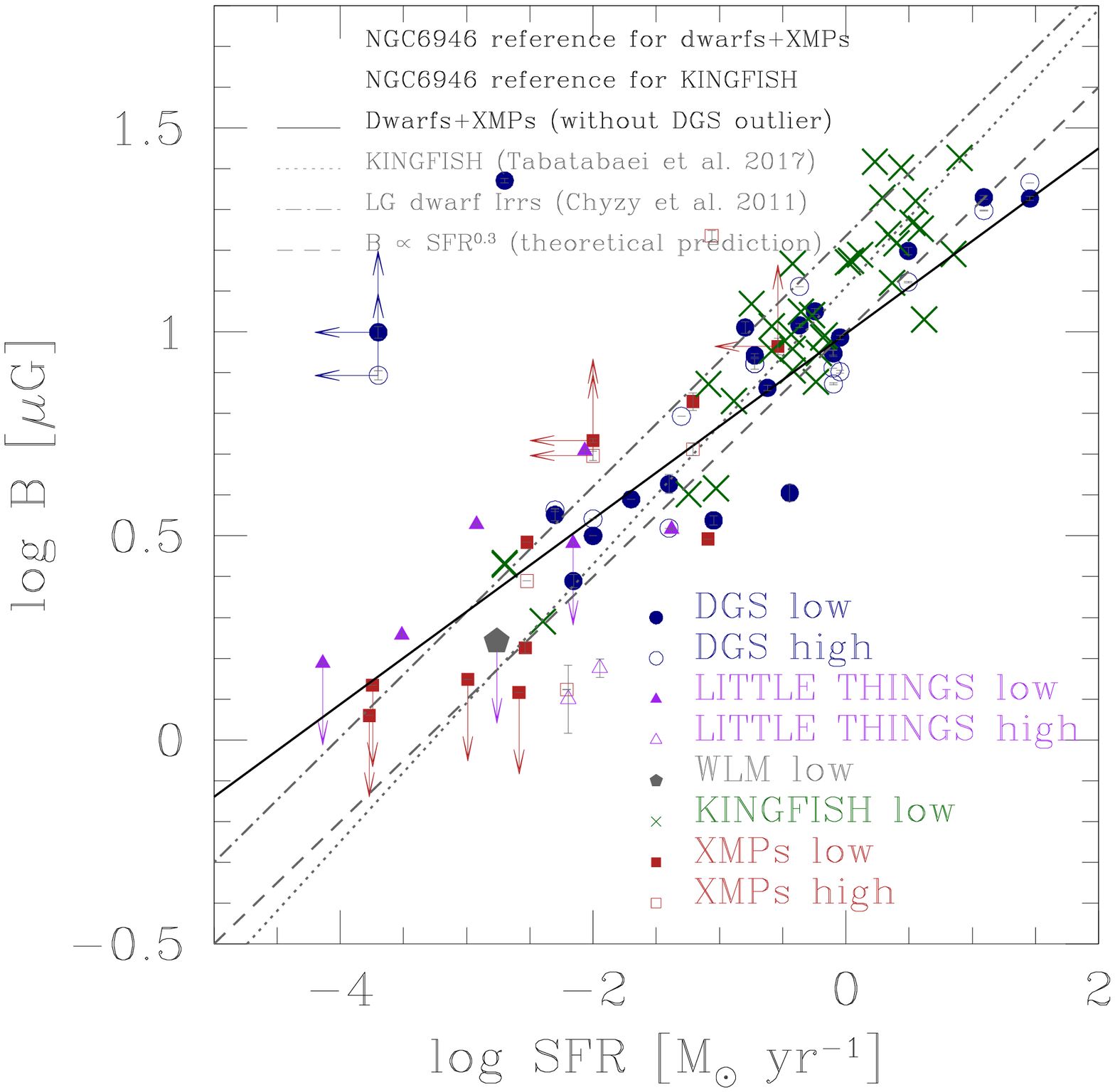}
	(b)
	\includegraphics[width=8cm]{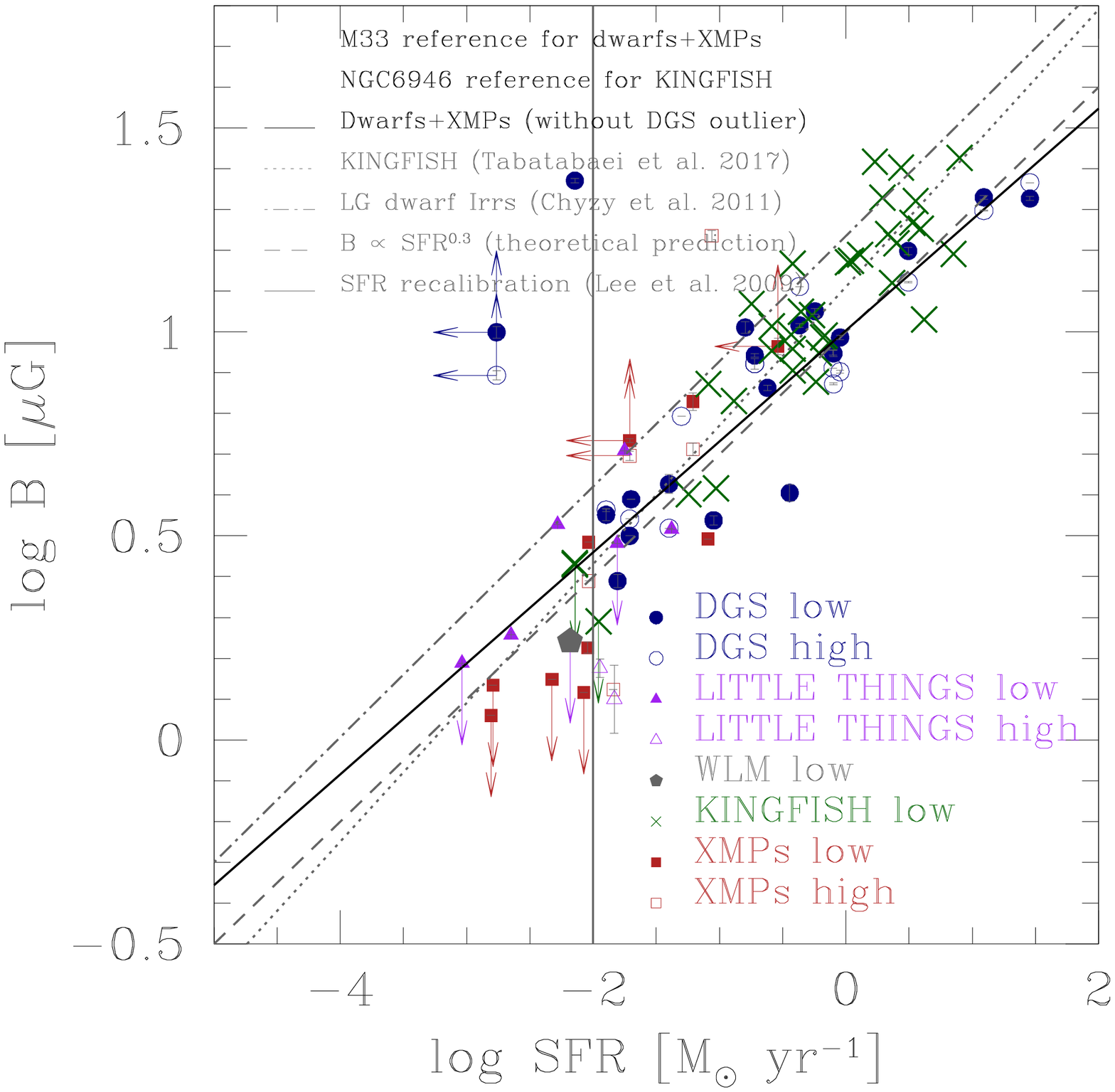}
	(c)
	\caption{{\it Top left (a):} Logarithm of the magnetic field strength as a function of the logarithm of the SFR, using M33 as the reference galaxy for dwarfs and XMPs, and NGC6946 for KINGFISH galaxies. {\it Top right (b):} Logarithm of the magnetic field strength as a function of the logarithm of the SFR, using NGC6946 as the reference galaxy for dwarfs and XMPs, and KINGFISH galaxies. {\it Bottom (c):} Logarithm of the magnetic field strength as a function of the logarithm of the SFR, using M33 as the reference galaxy for dwarfs and XMPs, and NGC6946 for KINGFISH galaxies. For the KINGFISH sources ({\it (a,b,c)}), SFR $\equiv$ SFR$_{\rm H\alpha+24 \, \mu m}$, for the LITTLE THINGS sources and XMPs, SFR $\equiv$ SFR$_{\rm H\alpha}$ and for the DGS sources, SFR $\equiv$ SFR$_{\rm TIR}$, SFR$_{\rm H\beta}$ or SFR$_{\rm H\alpha}$ ({\it (a,b)}). SFR $\equiv$ SFR$_{\rm UV}$, the empirical recalibration of Lee et al. (2009), when SFR $\lesssim$ 0.01 M$_{\odot}$ yr$^{-1}$ ({\it (c)}). Plotted also is the KINGFISH relation (grey dotted line; Tabatabaei et al. 2017), the (2.65 GHz) Local Group dwarf Irregulars relation (grey dotted -- dashed line; Chy\.{z}y et al. 2011), the expected B $\propto$ SFR$^{0.3}$ relation (with an arbitrary offset) if the F$^{\rm non-thermal}_{\rm 1.4 \, GHz} \propto$ SFR (grey dashed line), and the 'low' dwarfs and XMPs relation (black solid line; see also Table~1). Symbols are the same convention as Figure~1.}
\end{figure*}

% only dwarfs low resolution (no upper/lower limits) - log B = 0.02405288629 log SFR +0.299659729

%%%%%%%%%%%%%%%%%%%%%%%%%%%%%%%%%%%%%%%%%%%%%%%%%%%%%%%%%%%%%%%%%%%%%%%%%%%%%

Any discrepancy between the KINGFISH magnetic field strengths estimated with the present procedure (green crosses) and that of T17 stems from the different (more precise) method T17 used to estimate the thermal contribution to the RC emission (SED fitting) and, also, from the fact that T17 used the 4.8 GHz emission to calibrate the relation (instead of 1.4 GHz). Nonetheless, the present estimate of the KINGFISH magnetic field strengths are consistent with the estimations of T17. The slope of the present KINGFISH correlation is 0.28$\pm$0.02 (see also Table~1), while the slope of the T17 correlation is 0.34$\pm$0.08 (grey dotted line; Fig.~5).

There are 10 sources in common with the Chy\.{z}y et al. (2011) Local Group dwarf Irregular sample. There are a further two sources from Chy\.{z}y et al. (2000) and Kepley et al. (2010). For the common sources, the present estimated magnetic field strengths are generally lower than the Chy\.{z}y et al. (2011) measurements by 1 $\mu$G (IC1613) up to 5 $\mu$G (NGC1569), with the largest discrepancies in the sources with larger SFRs (Fig.~5). The difference persists irrespective of using M33 (Fig.~5a) or NGC6946 (Fig.~5b) as the dwarf and XMP reference galaxy, or if a recalibration of the low-SFRs is performed (Fig.~5c). The disagreement can be attributed to a combination of the following: the use of different SFRs, the different calibration, the different radio frequency (2.64 GHz) used by Chy\.{z}y et al. (2011) and the larger radio fluxes obtained by Chy\.{z}y et al. (2011) from using single-dish observations.

The largest magnetic field strengths were measured in the dwarf (DGS) galaxies NGC1140, IIZw40, Mrk153, SBS1533+574 and Haro2, with values between 11 (NGC1140) and 9 (Haro2) $\mu$G, comparable to the magnetic fields observed in some spiral galaxies (e.g., Beck 2016). Roychowdury \& Chengalur (2012) find a magnetic field strength for their faint dwarf Irregular galaxy sample of approximately 2 $\mu$G, consistent with values for the lower SFR dwarfs and XMPs (Fig.~5).

The dwarfs (blue and purple symbols) and XMPs (red symbols) show a shallower B -- SFR correlation (black solid line) than the KINGFISH sources (grey dotted line) and the Local Group dwarf Irregular sample (grey dotted -- dashed line; Fig.~5). Low SFR dwarfs and XMPs further suggest a steepening or downturn in the relation, more apparent when the SFR recalibration (Lee et al. 2009) is applied for SFR $\lesssim$ 0.01 M$_{\odot}$ yr$^{-1}$ (Fig.~5c). The overall behavior of the B -- SFR relation for dwarfs and XMPs mimics the behavior found with other quantities (see Sect.~3.1.1, 3.2 and 3.3).

The correlation slope for the KINGFISH sample (0.34$\pm$0.04; grey dotted line; T17), as well as for spiral galaxies (0.34$\pm$0.08; Niklas \& Beck 1997) are, in fact, very close to the theoretical prediction of B $\propto$ SFR$^{0.3}$ (grey dotted -- dashed line; Schleicher \& Beck 2013) for equipartition between the magnetic field and CR energy density (Fig.~5). Although this is also consistent with an increase of CR production and magnetic field amplification generated by the increased star formation rate, in massive galaxies, the generation and sustenance of magnetic fields is generally attributed to the large-scale dynamo action, where seed magnetic fields are amplified by large-scale flows and (shear-driven) turbulence, assisted by high rotation speeds (e.g., Beck et al. 1996).

%It is important, however, to stress that the Local Group dwarf Irregular sample (Chy\.{z}y et al. 2011) is quite small (12 plus 4 literature sources).} However, for the Chy\.{z}y et al. (2011) sample, it is important to stress that their fit is based on only 7 radio detections.

In dwarf galaxies, the strength of the total (turbulent) field is generally smaller than in spiral galaxies (Tabatabaei et al. 2016); the low mass, high velocity dispersion and slow differential rotation can further weaken or destroy the magnetic fields. The exception are low-mass galaxies undergoing vigorous star formation, where the operation of small-scale dynamos may be important (e.g., Chy\.{z}y 2008; Schleicher et al. 2013). In these sources, the magnetic field amplification by turbulence (via SN explosions or shocks) can be efficient, and occurs on short timescales. Because the local SFR determines the SN rate, which is the main source of turbulent energy needed for the small-scale dynamo effect, this creates a non-linear relation between the magnetic field strength and SFR (e.g., Chy\.{z}y 2008). Consistent with this scenario, a B -- SFR relation is observed for high-SFR dwarfs and XMPs (SFR $\gtrsim$ 0.1 M$_{\odot}$ yr$^{-1}$; Fig.~5c). The correlation slope (0.27$\pm$0.03; see also Table~1) can be compared to the correlation slope (0.21 -- 0.28) for a sample of low-mass dwarf and Magellanic-type galaxies, normal spirals and several massive starbursts (Chy\.{z}y, Sridhar \& Jurusik 2017). However, as signaled by the upper limit LITTLE things (purple symbols) and XMPs (red symbols) data points (Fig.~5c), the correlation is absent or distinct at low SFRs, suggesting a breakdown of equiparition and/or case B approximation assumptions (see also Hughes et al. 2006).

%However,no such relation is observed for the dwarfs and XMPs (Fig.~5), even when the SFR is recalibrated to the UV SFR (Lee et al. 2009) for low-SFR (SFR $\lesssim$ 0.01 M$_{\odot}$ yr$^{-1}$) dwarfs and XMPs (Fig.~5c). The lack of a clear correlation between the magnetic field strength and the SFR in dwarfs and XMPs can be attributed to the breakdown of equiparition and/or case B approximation assumptions (see also Hughes et al. 2006).

%%%%%%%%%%%%%%%%%%%%%%%%%%%%%%%%%%%%%%%%%%%%%%%%%%%%%%%%%%%%%%%%%%%%%%%%%%%%%%%%%%%%%%%%%%%%%%%%%%%%%%%%%%%%%%%%%%%%%%%%%%%%%%%%%%%%%%%%%%

\subsection{CO -- RC Relation}

It has been found that the correlation between the RC luminosity and the molecular gas (as traced by the CO emission) is as tight, if not more so, than the IR -- RC correlation (see also Sect.~3.1). A tight relation also exists between the CO and RC surface density (e.g., Murgia et al. 2002; Leroy et al. 2005; Murgia et al. 2005; Liu \& Gao 2010). 

The CO -- RC relation appears to be non-linear. Murgia et al. (2002) find a slope of 1.30 for the CO -- (1.4 GHz) RC relation in a sample of star-forming galaxies, while in Murgia et al. (2005), the slope is 1.28$\pm$0.13 for the BIMA sample (grey dotted line; Fig.~6). Normal spirals and LIRGs/ULIRGs (grey dotted -- dashed line; Fig.~6) show a slope of 1.31$\pm$0.09 for the CO -- (1.4 GHz) RC relation (Liu \& Gao 2010). The bright dwarf galaxies in Leroy et al. (2005) show a slope consistent with large spiral galaxies. Murgia et al. (2005) suggest that the correlation between CO and RC emission (on small and large scales) can be explained by a hydrostatic pressure regulating mechanism, and derive a relation with a slope of 1.4, consistent with the observations in the literature for bright galaxies.
  
% log L_1.4 GHz (Lsun) = 1.31 log CO (K km/s pc^2) - 6.20 (Liu & Gao 2010)

Figure~6 plots the 1.4 GHz RC luminosity as a function of the CO luminosity, for the total RC emission (a), as well as for the non-thermal (b) and thermal (c) contribution alone. 

%%%%%%%%%%%%%%%%%%%%%%%%%%%%%%%%%%%%%%%%%%%%%%%%%%%%%%%%%%%%%%%%%%%%%%%%%%

% Figure 6 - CO-RADIO

\begin{figure*} 
	\includegraphics[width=8cm]{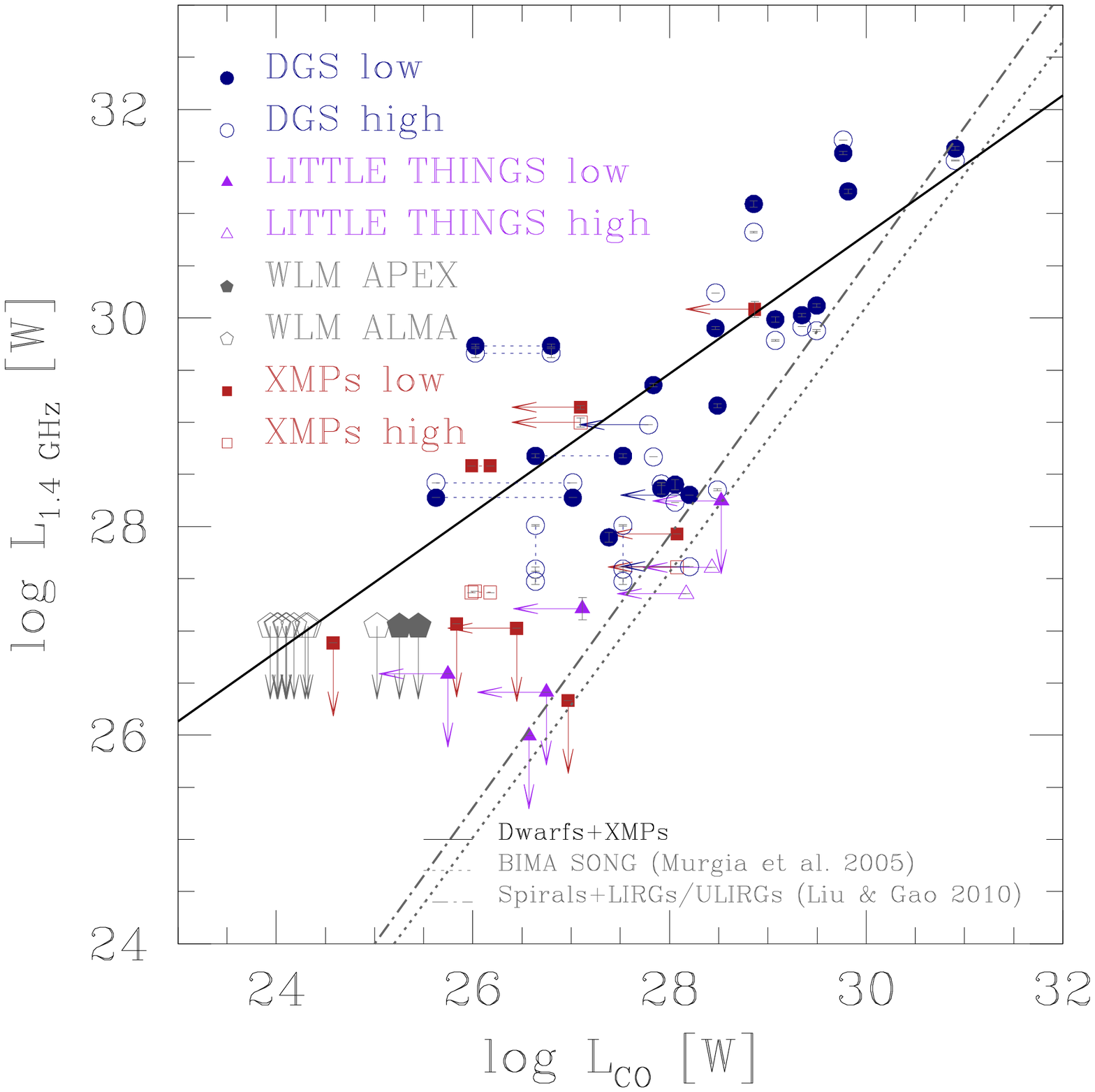}
	(a)
	\includegraphics[width=8cm]{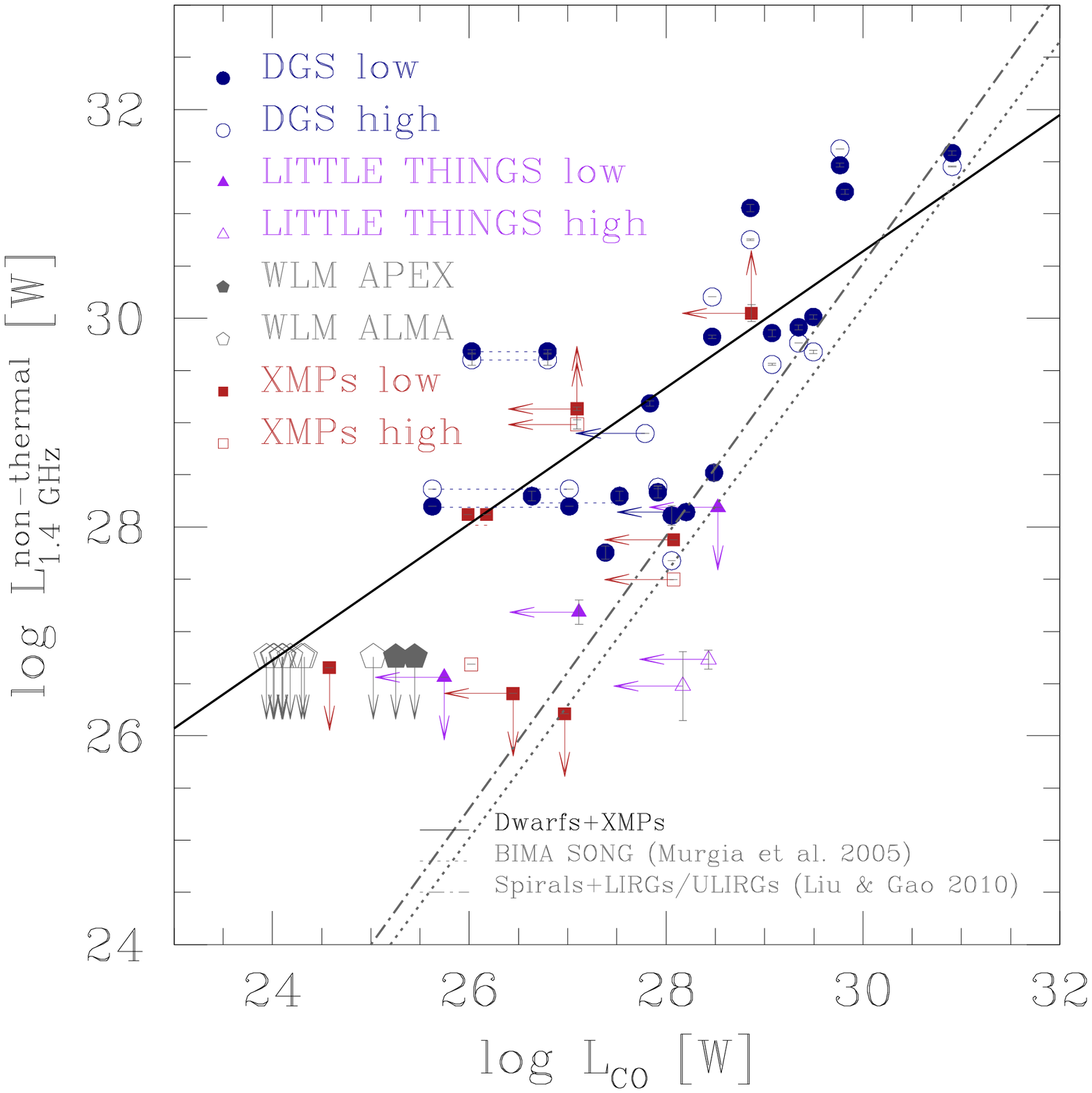}
	(b)
	\\
	\includegraphics[width=8cm]{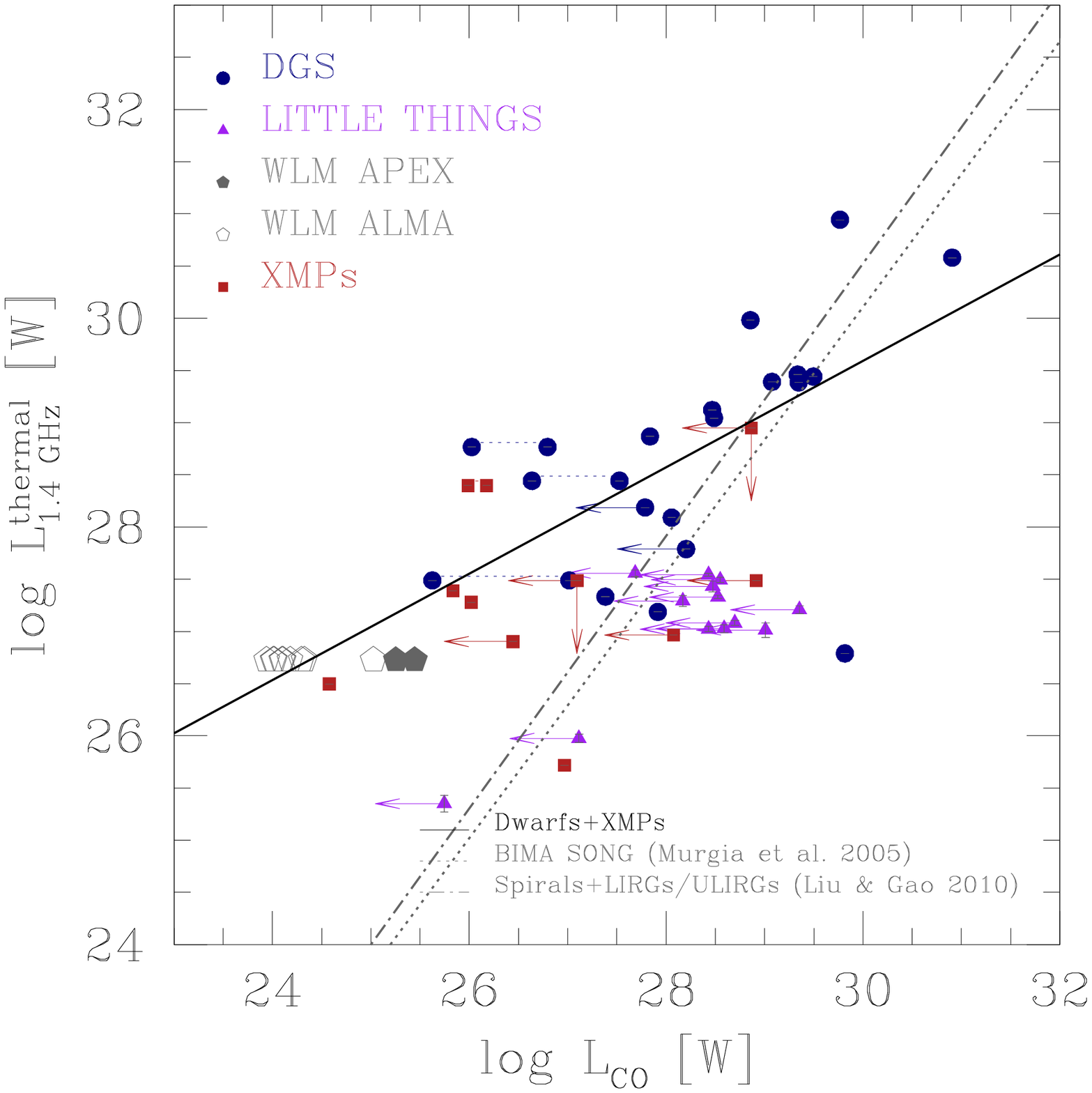}
	(c)

	\caption{{\it Top Left (a):} Logarithm of the total 1.4 GHz RC luminosity as a function of the logarithm of the CO luminosity. {\it Top Right (b):} Logarithm of the non-thermal 1.4 GHz RC luminosity as a function of the logarithm of the CO luminosity. {\it Bottom (c):} Logarithm of the thermal 1.4 GHz RC luminosity as a function of the logarithm of the CO luminosity. Data points with a range in CO luminosity are connected by a dotted line. The fit to the 'low' dwarfs and XMPs (only the faintest CO emission, in sources with multiple CO values) is plotted as the black solid line (see also Table~1). Plotted also is the total (1.4 GHz) relation (grey dotted -- dashed line) for nearby spirals from BIMA SONG (Murgia et al. 2005), and the (1.4 GHz) relation (grey dotted line) for normal spirals and LIRGs/ULIRGs (Liu \& Gao 2010). Symbols are the same convention as Figure~1.}
\end{figure*}

%%%%%%%%%%%%%%%%%%%%%%%%%%%%%%%%%%%%%%%%%%%%%%%%%%%%%%%%%%%%%%%%%%%%%%%%%%%%%

Although the scatter increases in the low-luminosity regime, the plot (Fig.~6a) demonstrates that the dwarf and XMP relation (black solid line) is not a simple extension of the CO -- RC relation for spiral galaxies (grey dotted line) and LIRGs/ULIRGs (grey dotted -- dashed line); the sub-linear slope for the dwarfs and XMPs (0.67$\pm$0.12; see also Table~1) is significantly shallower. This is in contrast to the results for the bright dwarfs in Leroy et al. (2005), and is inconsistent with the derived slope of 1.4 for the hydrostatic pressure regulating mechanism of Murgia et al. (2005). The CO emission increasingly weakens with decreasing luminosity, more than that predicted from the luminosity-scaling relation. In this regime, the CO is no longer an adequate indicator of the molecular gas content. However, there is some indication from the XMP (red symbols) and LITTLE THINGS (purple symbols) upper limits, for the total 1.4 GHz RC (Fig.~6a) and non-thermal (Fig.~6b) 1.4 GHz RC emission, that the relation may steepen below L$_{\rm 1.4 \, GHz} \simeq$ 10$^{27}$ W $\simeq$ L$^{\rm non-thermal}_{\rm 1.4 \, GHz} \simeq$ 10$^{27}$ W, coinciding with the luminosity at which the non-thermal (see also Fig.~3b), and, hence, total RC emission ceases to adequately trace the SFR. The overall behavior of the CO -- RC relation in dwarfs and XMPs (black solid line) is a combination of the following effects: the mass scaling, the cessation of the non-thermal RC emission as an adequate SFR tracer below L$_{\rm 1.4 \, GHz} \simeq$ 10$^{27}$ W $\simeq$ L$^{\rm non-thermal}_{\rm 1.4 \, GHz} \simeq$ 10$^{27}$ W (see also Fig.~3b), slower reaction rates and CO dissociation due to a hard ionizing field and low dust shielding (e.g., Richings \& Shaye 2016). 

%%%%%%%%%%%%%%%%%%%%%%%%%%%%%%%%%%%%%%%%%%%%%%%%%%%%%%%%%%%%%%%%%%%%%%%%%%%%%%%%%%%%%%%%%%%%%%%%%%%%%%%%%%%%%%%%%%%%%%%%%%%%%

\subsection{CO -- IR Relation}

An early analysis of the CO -- (F)IR relation is documented in Devereux \& Young (1990) and Young \& Scoville (1991), which show a relation with over an order of magnitude in scatter. Murgia et al. (2005) find a slope of 1.05$\pm$0.11 for the CO -- (60 $\mu$m) IR relation in the BIMA sample (black dashed line). A fairly tight CO -- (F)IR correlation, with a slope of 1.0$\pm$0.3 and increasing scatter at low luminosity, is present in the bright dwarf galaxy sample of Leroy et al. (2005). Shetty et al. (2016) find a slope of 1.1 -- 1.7 for the CO -- (T)IR relation in the Magellanic Clouds, with a higher intercept for the Small Magellanic Cloud (SMC), which they suggest could be due to its lower metallicity. Overall, it appears that the relation between the CO and the IR emission is linear, or close to, linear.

Figure~7 contains the relation between the CO and (70 $\mu$m) IR emission. 

%%%%%%%%%%%%%%%%%%%%%%%%%%%%%%%%%%%%%%%%%%%%%%%%%%%%%%%%%%%%%%%%%%%%%%%%%%

% Figure 7 - CO-IR

\begin{figure}
	\includegraphics[width=8cm]{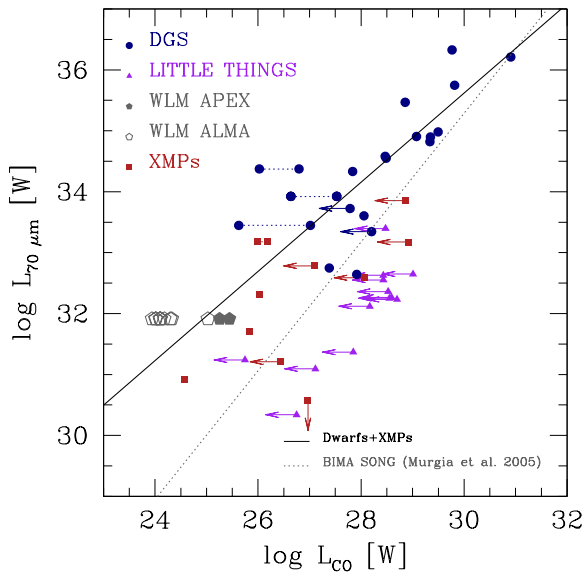}
	\caption{Logarithm of the 70 $\mu$m luminosity as a function of the logarithm of the CO luminosity. Data points with a range in CO luminosity are connected by a dotted line. The fit to the 'low' dwarfs and XMPs (only the faintest CO emission, in sources with multiple CO values) is plotted as the black solid line (see also Table~1). Plotted also is the total (1.4 GHz) relation (grey dotted line) for nearby spirals from BIMA SONG (Murgia et al. 2005). Symbols are the same convention as Figure~1.}
\end{figure}

% only dwarfs low resolution (no upper/lower limits) - log L_70 = 0.7317778468 log L_CO + 13.66476345

%%%%%%%%%%%%%%%%%%%%%%%%%%%%%%%%%%%%%%%%%%%%%%%%%%%%%%%%%%%%%%%%%%%%%%%%%%%%%

At low luminosities, a significant fraction of the galaxies possess only upper limits in CO emission, and the dispersion increases. The dwarf and XMP relation (black solid line) is not a simple extrapolation of the more massive galaxy relation (grey dotted line); the sub-linear slope for the dwarfs and XMPs (0.73$\pm$0.09; see also Table~1) is significantly shallower. This contrasts with the result of Leroy et al. (2005) for their sample of bright dwarf galaxies. The CO emission increasingly weakens with decreasing luminosity, more than that predicted from the luminosity-scaling relation. However, this is counterbalanced by the known increasing decrease of the IR emission per unit mass towards low luminosities (see also Sect.~3.1). The location of the XMP data points (red symbols) towards systematically lower CO luminosities relative to the massive galaxy relation (grey dotted line) is consistent with the metallicity effects observed by Shetty et al. (2016) for the SMC. The overall behavior of the CO -- IR relation in dwarfs and XMPs is a combination of the mass scaling, slower reaction rates and CO dissociation, and low dust content and decreased dust opacity (e.g., Richings \& Shaye 2016). 

% Excess TIR -- submillimeter (submm) emission (per unit dust mass) is observed in dwarf galaxies, particularly in low-metallicity systems, with increased dust temperatures with decreasing metallicity, and peaked emission at shorter wavelengths (50 -- 100 $\mu$m; e.g., R\'emy Ruyer et al. 2013; Shi et al. 2014). On the other hand, at low metallicities, and in the presence of a stronger radiation field and lower dust content, CO is dissociated (e.g., Richings \& Schaye 2016). 

%%%%%%%%%%%%%%%%%%%%%%%%%%%%%%%%%%%%%%%%%%%%%%%%%%%%%%%%%%%%%%%%%%%%%%%%%%%%%%%%%%%%

% TABLE 1 - LINEAR REGRESSIONS

\setcounter{table}{0}

\begin{table*}

\scriptsize

\begin{center}

\begin{minipage}{126mm}	

\caption{Least squares fits, and Pearson coefficient and Spearman rank, for the data.}

\begin{tabular}{l l l l l l l l}

\hline07

Sample 			&	Independent			& Dependent		& Slope		& Offset 	& r$_{\rm p}$	& r$_{\rm s}$ &	Fig. \\
				&   Variable			& Variable		& 			& 			&				&			   & \\
(1) & (2) & (3) & (4) & (5) & (6) & (7) & (8) \\
\hline
\hline
	
%Dwarfs+XMPs & log L$^{\rm thermal}_{\rm 1.4 \, GHz}$  & log L$^{\rm non-thermal}_{\rm 1.4 \, GHz}$ & 0.77$\pm$0.14 & 7.12$\pm$4.01 & 0.74 & 0.72 & 1  OLD \\

Dwarfs+XMPs & log L$^{\rm thermal}_{\rm 1.4 \, GHz}$  & log L$^{\rm non-thermal}_{\rm 1.4 \, GHz}$ & 0.78$\pm$0.14 & 7.04$\pm$3.90 & 0.75 & 0.73 & 1b \\

%Dwarfs$^a$+XMPs & log L$^{\rm thermal}_{\rm 1.4 \, GHz}$  & log L$^{\rm non-thermal}_{\rm 1.4 \, GHz}$ & 0.92$\pm$0.09 & 2.92$\pm$2.63 & 0.90 & 0.86 & 1 OLD\\

Dwarfs$^a$+XMPs & log L$^{\rm thermal}_{\rm 1.4 \, GHz}$  & log L$^{\rm non-thermal}_{\rm 1.4 \, GHz}$ & 0.92$\pm$0.09 & 2.84$\pm$2.46 & 0.91 & 0.88 & 1b \\

%KINGFISH & log L$^{\rm thermal}_{\rm 1.4 \, GHz}$  & log L$^{\rm non-thermal}_{\rm 1.4 \, GHz}$ & 1.23$\pm$0.11 & -5.92$\pm$3.21 & 0.90 & 0.87 & 1 OLD \\

KINGFISH & log L$^{\rm thermal}_{\rm 1.4 \, GHz}$  & log L$^{\rm non-thermal}_{\rm 1.4 \, GHz}$ & 1.22$\pm$0.11 & -5.63$\pm$3.13 & 0.90 & 0.87 & 1b \\

%All$^b$			& log L$^{\rm thermal}_{\rm 1.4 \, GHz}$  & log L$^{\rm non-thermal}_{\rm 1.4 \, GHz}$ & 0.94$\pm$0.08 & 2.61$\pm$2.41 & 0.83 & 0.86 & 1 OLD \\

All$^b$			& log L$^{\rm thermal}_{\rm 1.4 \, GHz}$  & log L$^{\rm non-thermal}_{\rm 1.4 \, GHz}$ & 0.94$\pm$0.08 & 2.72$\pm$2.34 & 0.84 & 0.86 & 1b \\

%All$^a$			& log L$^{\rm thermal}_{\rm 1.4 \, GHz}$  & log L$^{\rm non-thermal}_{\rm 1.4 \, GHz}$ & 1.06$\pm$0.06 & -0.95$\pm$1.79 & 0.92 & 0.93 & 1 OLD \\

All$^a$			& log L$^{\rm thermal}_{\rm 1.4 \, GHz}$  & log L$^{\rm non-thermal}_{\rm 1.4 \, GHz}$ & 1.06$\pm$0.06 & -0.83$\pm$1.70 & 0.92 & 0.93 & 1b \\

\hline

%Dwarfs+XMPs & log F$_{\rm 70 \mu m}$  & log F$_{\rm 1.4 \, GHz}$ & 0.54$\pm$0.05 & -1.98$\pm$0.05 & 0.88 & 0.83 & 2a \\

Dwarfs+XMPs & log L$_{\rm 70 \mu m}$  & log L$_{\rm 1.4 \, GHz}$ & 0.89$\pm$0.08 & -1.91$\pm$2.77 & 0.91 & 0.89 & 2a \\

%KINGFISH & log F$_{\rm 70 \mu m}$  & log F$_{\rm 1.4 \, GHz}$ & 1.00$\pm$0.08 & -2.29$\pm$0.12 & 0.91 & 0.88 & 2a \\

KINGFISH & log L$_{\rm 70 \mu m}$  & log L$_{\rm 1.4 \, GHz}$ & 1.07$\pm$0.07 & -8.19$\pm$2.39 & 0.94 & 0.92 & 2a \\

%All			& log F$_{\rm 70 \mu m}$  & log F$_{\rm 1.4 \, GHz}$ 	& 0.72$\pm$0.04 & -1.97$\pm$0.05 &  0.91 & 0.93 & 2a \\

All			& log L$_{\rm 70 \mu m}$  & log L$_{\rm 1.4 \, GHz}$ 	& 0.92$\pm$0.05 & -2.91$\pm$1.65 &  0.93 & 0.93 & 2a \\

\hline

%Dwarfs+XMPs & log F$_{\rm 100 \mu m}$  & log F$_{\rm 1.4 \, GHz}$ & 0.52$\pm$0.06 & -2.03$\pm$0.06 & 0.87 & 0.85 & 2b \\

Dwarfs+XMPs & log L$_{\rm 100 \mu m}$  & log L$_{\rm 1.4 \, GHz}$ & 0.95$\pm$0.11 & -4.00$\pm$3.75 & 0.88 & 0.85 & 2b \\

%KINGFISH & log F$_{\rm 100 \mu m}$  & log F$_{\rm 1.4 \, GHz}$ & 0.98$\pm$0.07 & -2.49$\pm$0.12 & 0.93 & 0.88 & 2b \\

KINGFISH & log L$_{\rm 100 \mu m}$  & log L$_{\rm 1.4 \, GHz}$ & 1.09$\pm$0.05 & -9.16$\pm$1.65 & 0.96 & 0.93 & 2b \\

%All			& log F$_{\rm 100 \mu m}$  & log F$_{\rm 1.4 \, GHz}$ 	& 0.70$\pm$0.04 & -2.07$\pm$0.06 & 0.91 & 0.93 & 2b  \\

All			& log L$_{\rm 100 \mu m}$  & log L$_{\rm 1.4 \, GHz}$ 	& 0.93$\pm$0.05 & -3.35$\pm$1.90 & 0.92 & 0.92 & 2b  \\

\hline

%Dwarfs+XMPs & log F$_{\rm 160 \mu m}$  & log F$_{\rm 1.4 \, GHz}$ & 0.47$\pm$0.05 & -1.92$\pm$0.06 & 0.86 & 0.82 & 2c \\

Dwarfs+XMPs & log L$_{\rm 160 \mu m}$  & log L$_{\rm 1.4 \, GHz}$ & 0.89$\pm$0.11 & -1.29$\pm$3.66 & 0.85 & 0.85 & 2c \\

%KINGFISH & log F$_{\rm 160 \mu m}$  & log F$_{\rm 1.4 \, GHz}$ & 0.94$\pm$0.07 & -2.49$\pm$0.13 & 0.92 & 0.87 & 2c \\

KINGFISH & log L$_{\rm 160 \mu m}$  & log L$_{\rm 1.4 \, GHz}$ & 1.11$\pm$0.06 & -9.61$\pm$2.28 & 0.95 & 0.92 & 2c \\

%All			& log F$_{\rm 160 \mu m}$  & log F$_{\rm 1.4 \, GHz}$ & 0.61$\pm$0.04 & -1.94$\pm$0.06 & 0.90 & 0.92 & 2c \\

All			& log L$_{\rm 160 \mu m}$  & log L$_{\rm 1.4 \, GHz}$ & 0.84$\pm$0.06 & 0.14$\pm$1.98 & 0.89 & 0.91 & 2c \\

\hline

%Dwarfs+XMPs$^c$		& log L$_{\rm FIR}$  & log L$_{\rm 1.4 \, GHz}$ & 0.91$\pm$0.07 & -2.38$\pm$2.47 & 0.93 & 0.91 & 2d \\

Dwarfs+XMPs$^c$		& log L$_{\rm FIR}$  & log L$_{\rm 1.4 \, GHz}$ & 0.89$\pm$0.07 & -1.97$\pm$2.57 & 0.93 & 0.91 & 2d \\

%KINGFISH & log L$_{\rm FIR}$  & log L$_{\rm 1.4 \, GHz}$ &  1.12$\pm$0.06 & -10.46$\pm$2.18 & 0.96 & 0.93 & 2d \\

KINGFISH & log L$_{\rm FIR}$ & log L$_{\rm 1.4 \, GHz}$ &  1.02$\pm$0.11 & -6.47$\pm$3.87 & 0.88 & 0.86 & 2d \\

%All$^c$			& log L$_{\rm FIR}$  & log L$_{\rm 1.4 \, GHz}$ & 0.85$\pm$0.04 & -0.68$\pm$1.51 & 0.93 & 0.94 & 2d \\

All$^c$			& log L$_{\rm FIR}$  & log L$_{\rm 1.4 \, GHz}$ & 0.94$\pm$0.05 & -3.80$\pm$1.82 & 0.93 & 0.93 & 2d \\

\hline

%Dwarfs+XMPs & log L$_{\rm TIR}$  & log L$^{\rm non-thermal}_{\rm 1.4 \, GHz}$ 	&  0.88$\pm$0.09 & -1.68$\pm$3.03 & 0.90 & 0.88 & 3a \\

Dwarfs+XMPs & log L$_{\rm FIR}$  & log L$^{\rm non-thermal}_{\rm 1.4 \, GHz}$ 	&  0.88$\pm$0.09 & -1.80$\pm$3.14 & 0.90 & 0.88 & 3a \\

%KINGFISH & log L$_{\rm TIR}$  & log L$^{\rm non-thermal}_{\rm 1.4 \, GHz}$  & 1.15$\pm$0.08 & -11.85$\pm$2.80 & 0.94 & 0.91 & 3a \\

KINGFISH & log L$_{\rm FIR}$  & log L$^{\rm non-thermal}_{\rm 1.4 \, GHz}$  & 1.01$\pm$0.13 & -6.14$\pm$4.87 & 0.84 & 0.83 & 3a \\

%All			& log L$_{\rm TIR}$  & log L$^{\rm non-thermal}_{\rm 1.4 \, GHz}$ 	& 0.88$\pm$0.05 & -1.66$\pm$1.70 & 0.92 & 0.94 & 3a \\

All			& log L$_{\rm FIR}$  & log L$^{\rm non-thermal}_{\rm 1.4 \, GHz}$ 	& 0.96$\pm$0.06 & -4.53$\pm$2.20 & 0.91 & 0.92 & 3a \\

\hline 

%Dwarfs$^a$+XMPs & log SFR & log L$_{\rm 1.4 \, GHz}^{\rm non-thermal}$ & 1.12$\pm$0.10 & 30.02$\pm$0.14 & 0.91 & 0.84 & 3b \\

Dwarfs$^a$+XMPs & log SFR & log L$_{\rm 1.4 \, GHz}^{\rm non-thermal}$ & 1.11$\pm$0.09 & 30.08$\pm$0.13 & 0.93 & 0.88 & 3b \\

%KINGFISH	& log SFR & log L$_{\rm 1.4 \, GHz}^{\rm non-thermal}$ & 1.23$\pm$0.11 & 30.42$\pm$0.06 & 0.90 & 0.87 & 3b \\

KINGFISH	& log SFR & log L$_{\rm 1.4 \, GHz}^{\rm non-thermal}$ & 1.22$\pm$0.11 & 30.42$\pm$0.06 & 0.90 & 0.87 & 3b \\

%All$^a$	& log SFR & log L$_{\rm 1.4 \, GHz}^{\rm non-thermal}$ & 1.24$\pm$0.07 & 30.29$\pm$0.07 & 0.93 & 0.93 & 3b \\

All$^a$	& log SFR & log L$_{\rm 1.4 \, GHz}^{\rm non-thermal}$ & 1.22$\pm$0.06 & 30.32$\pm$0.06 & 0.94 & 0.93 & 3b \\

\hline

%Dwarfs+XMPs 		& log SFR  & q$_{\rm TIR}$ 		&\ldots	 & \ldots & 0.20 & 0.22 & 4a \\

Dwarfs+XMPs 		& log SFR  & q$_{\rm FIR}$ 		&\ldots	 & \ldots & 0.20 & 0.22 & 4a \\

%KINGFISH 			& log SFR  & q$_{\rm TIR}$ 		&\ldots	 & \ldots & -0.34 &  -0.42 & 4a \\

KINGFISH 			& log SFR  & q$_{\rm FIR}$ 		&\ldots	 & \ldots & -0.27 & -0.17 & 4a \\

%All					& log SFR  & q$_{\rm TIR}$	 	& \ldots	& \ldots & 0.30 & 0.05 & 4a \\

All					& log SFR  & q$_{\rm FIR}$	 	& \ldots	& \ldots & -0.01 & -0.06 & 4a \\

\hline

%Dwarfs+XMPs	 	& log L$_{\rm TIR}$  & q$_{\rm TIR}$ 	&\ldots& \ldots & 0.38 & 0.30 & 4b \\

Dwarfs+XMPs	 	& log L$_{\rm FIR}$  & q$_{\rm FIR}$ 	&\ldots& \ldots & 0.38 & 0.30 & 4b \\

%KINGFISH			& log L$_{\rm TIR}$  & q$_{\rm TIR}$	& \ldots &\ldots & -0.28 & -0.40 & 4b \\

KINGFISH			& log L$_{\rm FIR}$  & q$_{\rm FIR}$	& \ldots &\ldots & -0.17 & -0.09 & 4b \\

%All			& log L$_{\rm TIR}$  & q$_{\rm TIR}$	& \ldots &\ldots & 0.49 & 0.20 & 4b \\

All			& log L$_{\rm TIR}$  & q$_{\rm TIR}$	& \ldots &\ldots & 0.26 & 0.04 & 4b \\

\hline

%Dwarfs+XMPs	 	& log SFR  & q$_{\rm TIR}^{\rm non-thermal}$ 	& \ldots& \ldots& 0.22 & 0.26 & 4c \\

Dwarfs+XMPs 		& log SFR  & q$_{\rm FIR}^{\rm non-thermal}$ 	&\ldots	 & \ldots & 0.22 & 0.24 & 4c \\

%KINGFISH	 	& log SFR  & q$_{\rm TIR}^{\rm non-thermal}$ 	& \ldots& \ldots& -0.36 &  -0.44 & 4c \\

KINGFISH	 	& log SFR  & q$_{\rm TIR}^{\rm non-thermal}$ 	& \ldots& \ldots& -0.27 &  -0.20 & 4c \\

%All			& log SFR  & q$_{\rm TIR}^{\rm non-thermal}$	& \ldots&\ldots & 0.26 & 0.03 & 4c \\

All			& log SFR  & q$_{\rm TIR}^{\rm non-thermal}$	& \ldots&\ldots & 0.03 & -0.06 & 4c \\

\hline

%Dwarfs+XMPs	 	& log SFR  & q$_{\rm TIR}^{\rm thermal}$ & \ldots& \ldots & -0.06 & 0.00 & 4d \\

Dwarfs+XMPs	 	& log SFR  & q$_{\rm FIR}^{\rm thermal}$ & \ldots& \ldots & -0.06 & 0.00 & 4d \\

%KINGFISH	 	& log SFR  & q$_{\rm TIR}^{\rm thermal}$ & \ldots& \ldots & 0.15 & 0.24 & 4d \\

KINGFISH	 	& log SFR  & q$_{\rm FIR}^{\rm thermal}$ & \ldots& \ldots & -0.04 & 0.15 & 4d \\

%All			& log SFR  & q$_{\rm TIR}^{\rm thermal}$ &	\ldots& \ldots& 0.22 & 0.40 & 4d \\

All			& log SFR  & q$_{\rm FIR}^{\rm thermal}$ &	\ldots& \ldots& 0.03 & 0.18 & 4d \\

\hline

Dwarfs+XMPs$^a$	& log SFR	& B		& 0.23$\pm$0.03 & 0.96$\pm$0.04 & 0.90 & 0.87 & 5a \\

KINGFISH		& log SFR	& B		& 0.28$\pm$0.02	& 1.11$\pm$0.02 &  0.91 & 0.84 & 5a \\

All$^a$			& log SFR	& B		& 0.27$\pm$0.02	& 1.06$\pm$0.02 & 0.91 & 0.91 & 5a \\

\hline

Dwarfs+XMPs$^a$	& log SFR	& B		& 0.23$\pm$0.02 & 1.00$\pm$0.04 & 0.90 & 0.87 & 5b \\

All$^a$			& log SFR	& B		& 0.26$\pm$0.02	& 1.07$\pm$0.02 & 0.90 & 0.90 & 5b \\

\hline

Dwarfs+XMPs$^a$	& log SFR	& B		& 0.27$\pm$0.03 & 1.00$\pm$0.04 & 0.91 & 0.87 & 5c \\

All$^a$			& log SFR	& B		& 0.30$\pm$0.02	& 1.08$\pm$0.02 & 0.91 & 0.90 & 5c \\

\hline

Dwarfs+XMPs 		& log L$_{\rm CO}$ & log L$_{\rm 1.4 \, GHz}$ & 0.67$\pm$0.12 & 10.79$\pm$3.31 & 0.82 & 0.87 & 6a \\

%Dwarfs+XMPs 		& log L$_{\rm CO}$ & log L$_{\rm 1.4 \, GHz}^{\rm non-thermal}$ & 0.66$\pm$0.15 & 10.82$\pm$4.16 & 0.76 & 0.85 & 6b \\

Dwarfs+XMPs 		& log L$_{\rm CO}$ & log L$_{\rm 1.4 \, GHz}^{\rm non-thermal}$ & 0.65$\pm$0.14 & 11.05$\pm$4.05 & 0.76 & 0.84 & 6b \\

%Dwarfs+XMPs 		& log L$_{\rm CO}$ & log L$_{\rm 1.4 \, GHz}^{\rm thermal}$ & 	0.54$\pm$0.13 & 13.54$\pm$3.52 & 0.68 & 0.68 & 6c \\

Dwarfs+XMPs 		& log L$_{\rm CO}$ & log L$_{\rm 1.4 \, GHz}^{\rm thermal}$ & 	0.50$\pm$0.11 & 14.29$\pm$3.07 & 0.65 & 0.66 & 6c \\

\hline

Dwarfs+XMPs 		& log L$_{\rm CO}$ 	& log L$_{\rm 70 \mu m}$ & 0.73$\pm$0.09 & 13.66$\pm$2.57 & 0.87 & 0.91 & 7 \\

\hline 

\end{tabular}

$^a$without the DGS Mrk1089 outlier \\
$^b$Dwarfs+XMPs+KINGFISH \\
$^c$without the XMP SBS0335-052 outlier \\

Column (1): Sample. Column (2): Independent variable. Column (3): Dependent variable. Column (4): Slope of the least squares fit. Column (5): Offset of the least squares fit. Column (6): Pearson coefficient. Column (7): Spearman rank. Column (8): Corresponding figure number.

\end{minipage}

\end{center}

\end{table*}

%%%%%%%%%%%%%%%%%%%%%%%%%%%%%%%%%%%%%%%%%%%%%%%%%%%%%%%%%%%%%%%%%

%%%%%%%%%%%%%%%%%%%%%%%%%%%%%%%%%%%%%%%%%%%%%%%%%%%%%%%%%%%%%%%%%%%%%%%%%%%%%%%%%%%%

% TABLE 2 - STATS q_TIR

\setcounter{table}{1}

\begin{table*}

\scriptsize

\begin{center}

\begin{minipage}{92mm}

\caption{Statistics for the q$_{\rm TIR}$ parameter.}

\begin{tabular}{l l l l l l l}

\hline

Sample 			&	Variable 		& Med. & Mean & Sigma & Kurt. & Fig.\\
(1) & (2) & (3) & (4) & (5) & (6) & (7) \\

\hline
\hline
		
%Dwarfs+XMPs			& q$_{\rm TIR}$				&  5.72   & 5.72 & 0.49	& -0.18 & 4a \\
Dwarfs+XMPs			& q$_{\rm FIR}$				&  6.00   & 5.72 & 0.49	& -0.18 & 4a \\

%KINGFISH		& q$_{\rm TIR}$				& 6.17		& 6.16 & 0.24 & 0.20 & 4a \\
KINGFISH		& q$_{\rm FIR}$				& 5.62		& 5.68 & 0.32 & -1.03 & 4a \\

%All				& q$_{\rm TIR}$				& 6.02		& 5.97	& 0.43	& 1.44 & 4a \\
All				& q$_{\rm FIR}$				& 5.66		& 5.70	& 0.41	& 0.09 & 4a \\

\hline

%Dwarfs+XMPs			& q$_{\rm TIR}^{\rm non-thermal}$	& 5.79 & 5.88 & 0.60 & 0.51 & 4c \\
Dwarfs+XMPs			& q$_{\rm FIR}^{\rm non-therml}$		&  5.78   & 5.85 & 0.57	& 0.22 & 4c \\

%KINGFISH		& q$_{\rm TIR}^{\rm non-thermal}$	& 6.23 & 6.24 & 0.29 &  0.59 & 4c \\
KINGFISH		& q$_{\rm FIR}^{\rm non-thermal}$	& 5.66 & 5.74 & 0.36 &  -0.71 & 4c \\

%All				& q$_{\rm TIR}^{\rm non-thermal}$	& 6.10 & 6.08 & 0.48 & 1.33 & 4c \\
All				& q$_{\rm FIR}^{\rm non-thermal}$	& 5.71 & 5.80 & 0.48 & 0.68 & 4c \\

\hline

%XMPs			& q$_{\rm TIR}^{\rm thermal}$		& 5.83 & 5.83 & 0.63 & -1.74 & 4d \\
XMPs			& q$_{\rm FIR}^{\rm thermal}$		& 5.87 & 5.87 & 0.63 & -1.74 & 4d \\

%Dwarfs+XMPs 			& q$_{\rm TIR}^{\rm thermal}$		& 6.46 & 6.44 & 0.69 & 12.47 & 4d \\
Dwarfs+XMPs 			& q$_{\rm FIR}^{\rm thermal}$		& 6.51 & 6.48 & 0.69 & 12.47 & 4d \\

%KINGFISH		& q$_{\rm TIR}^{\rm thermal}$		& 7.08 & 6.98 & 0.35 & 5.73 & 4d \\
KINGFISH		& q$_{\rm FIR}^{\rm thermal}$		& 6.66 & 6.61 & 0.27 & -0.27 & 4d \\

%All				& q$_{\rm TIR}^{\rm thermal}$		& 6.65 & 6.76 & 0.60 & 7.68 & 4d \\
All				& q$_{\rm FIR}^{\rm thermal}$		& 6.56 & 6.54 & 0.54 & 18.10 & 4d \\

\hline 

%Dwarfs+XMPs	   & log(SFR$_{\rm H\alpha}$/L$_{\rm 1.4 \, GHz}$)  & -30.18 & -30.24 & 0.92 & 7.37 & \ldots \\
%Dwarfs+XMPs	   & log(SFR$_{\rm UV}$/L$_{\rm 1.4 \, GHz}$)  & -30.07 & -30.07 & 0.83 & 7.09 & \ldots \\
%KINGFISH	    & log(SFR$_{\rm H\alpha}$/L$_{\rm 1.4 \, GHz}$) & -30.40 & -30.40 & 0.37 & -0.67 & \ldots \\

%Dwarfs$^a$+XMPs		&  log B	& 0.25 & 0.29 & 0.23 & -0.38 & 5a \\

%Dwarfs$^b$+XMPs		&  log B	& 0.79 & 0.83 & 0.23 & -0.38 & 5b \\

\hline 

\end{tabular}

$^a$M33 used as reference galaxy \\
$^b$NGC6946 used as reference galaxy\\

Column (1): Sample. Column (2): Variable. Column (3): Median of the distribution. Column (4): Mean of the distribution. Column (5): Sigma of the distribution. Column (6): Kurtosis of the distribution. Column (7): Corresponding figure number.

\end{minipage}

\end{center}

\end{table*}
	
%%%%%%%%%%%%%%%%%%%%%%%%%%%%%%%%%%%%%%%%%%%%%%%%%%%%%%%%%%%%%%%%%

\section{Summary of the Results and Conclusions}

The content and distribution of the ISM components (i.e., molecular, atomic, neutral and ionized gas, dust and CRs) are mainly determined by star formation and associated processes, such as accretion and feedback. It is therefore expected that star formation plays a significant role in linking these components together. Their inter-relations can be evaluated through the use of ISM component tracers, such as the RC (ionized gas and CRs), the IR (dust mixed with gas) and CO (molecular gas).

IR -- RC -- CO correlations have been extensively investigated for various types of galaxies, across several orders of magnitude, at several redshifts, and on different scales. The global IR -- RC correlation, as well as the global CO -- IR correlation, are shown to be nearly linear, while the global CO -- RC correlation is shown to be supra-linear. However, in the case of low-luminosity galaxies, these relations are less well-studied. It is uncertain whether the relations still hold at low luminosity, or if they are simple extrapolations of the relations observed in more massive galaxies (e.g., Leroy et al. 2005). There is some indication that the relations may show a dependency on luminosity (e.g., Bell 2003 and references therein), metallicity (e.g., Shetty et al. 2016), and redshift (e.g., Magnelli et al. 2015), and theoretical considerations predict a breakdown of the relations below a critical SFR surface density (e.g., Schleicher \& Beck 2016).

In order to probe the less well-known low-luminosity regime of several of these ISM relations, this paper contains a compilation and analysis of the molecular gas, dust, (H$\alpha$-based) thermal 1.4 GHz RC, total 1.4 GHz RC and 1.4 GHz synchrotron data for samples of dwarf galaxies, including extreme low-metallicity sources, as well as comparative samples and relations for more massive galaxies. The main advantage over previous studies is the sampling of the low-mass, low-luminosity and low-metallicity regime, and/or the inclusion of a larger number of sources in this regime. The main results of this analysis for the dwarfs and XMPs are summarized in the following:

\begin{itemize}

\item{Overall, the thermal 1.4 GHz RC fraction decreases with increasing total 1.4 GHz RC luminosity, as predicted by the standard model (Fig.~1a).}

\item{Even at low total 1.4 GHz RC luminosities, most dwarfs and XMPs still exhibit a non-thermal 1.4 GHz RC component in excess of 50\% (Fig.~1a), consistent with previous results for dwarfs in the literature, although the number of sources and probed luminosity range had been more restricted.}

\item{For dwarfs and XMPs, the thermal-to-non-thermal 1.4 GHz RC ratio decreases with decreasing thermal 1.4 GHz RC luminosity (Fig.~1b), consistent with a decrease in the thermal 1.4 GHz RC component, or, equivalently, to a decrease in the SFR, as traced by the H$\alpha$ emission (Eq.~[1] and [2]). This result is consistent with results from the literature for dwarfs, which find that the H$\alpha$ emission increasingly underestimates the true SFR with decreasing SFR.}

\item{In the IR -- RC relation, the dwarfs and XMPs tend to be underluminous in the IR for their luminosity, signaling that, below L$_{\rm FIR} \simeq$ 10$^{36}$ W, the IR emission ceases to adequately trace the SFR due to the low dust content and decreased dust opacity (Fig.~2). The IR depression is consistent with previous results in the literature for low-luminosity galaxies, although such low luminosities and large number of sources had not been probed.}

\item{There is some indication that the IR -- RC relation may be steeper below L$_{\rm 1.4 \, GHz} \simeq$ 10$^{27}$ W, demonstrating that the total 1.4 RC emission, alongside the IR emission, may no longer adequately trace the SFR in this low-luminosity regime (Fig.~2). The steepening is also hinted at in the thermal versus non-thermal 1.4 GHz RC relation (Fig.~1b), the non-thermal versus FIR relation (Fig.~3a) and the RC -- CO relation (Fig.~6). This result is consistent with published results and predictions for low-luminosity galaxies, although the extreme low-luminosity regime had not been adequately sampled.} 

\item{The sub-linear slope of the dwarf and XMP (F)IR -- non-thermal RC relation suggests that the non-thermal 1.4 GHz RC emission still traces the SFR down to L$_{\rm 1.4 \, GHz}^{\rm non-thermal} \simeq$ 10$^{27}$ W (Fig.~3a). Previous published results that failed to detect this flattening had not adequately sampled the lower-luminosity regime.}

\item{After recalibrating the SFR to the UV emission, the non-thermal RC -- SFR relation demonstrates that, below SFR $\simeq$ 0.01 M$_{\odot}$ yr$^{-1}$, or, equivalently, L$_{\rm 1.4 \, GHz}^{\rm thermal} \simeq$ L$_{\rm 1.4 \, GHz}^{\rm non-thermal} \simeq $ 10$^{27}$ W (Eq.~[1] and [2]), the non-thermal 1.4 GHz RC emission no longer adequately traces the SFR (Fig.~3b), driving the steepening of the relations at low luminosity. Previous published results that found a breakdown in the non-thermal RC versus SFR relation operating at higher luminosity had not adequately sampled the lower-luminosity regime.}

\item{Low-SFR dwarfs and XMPs exhibit the lowest total, non-thermal and thermal q$_{\rm FIR}$ parameters, signaling that the IR emission decreases more rapidly with decreasing SFR than the total, non-thermal and thermal 1.4 GHz RC emission (Fig.~4a, 4c and 4d). The effect is a combination of the IR deficit (Fig.~2) and the behavior of the non-thermal RC emission with SFR, namely the underestimation of the true SFR by the H$\alpha$ emission at low SFRs (Fig.~3b). The former is consistent with results for dwarfs in the literature, although the number of sources and probed luminosity range had been more restricted, while the latter is a documented result in the literature for dwarfs.}

\item{There is a suggestion that the total q$_{\rm FIR}$ parameter values decrease steeply with decreasing FIR luminosity, signaling that the total 1.4 GHz RC emission traces the SFR down to L$_{\rm 1.4 \, GHz} \simeq$ 10$^{27}$ W (Fig.~2 and 4c); this is supported by the shallow IR -- RC relation slope (Fig.~2 and Fig.~4a). Previous published results that had found a shallower decrease in the total q$_{\rm FIR}$ parameter, and/or failed to detect the  flattening in the IR -- RC relation, had not adequately sampled the lower-luminosity regime.}

%\item{{\bf From the fits to the IR -- RC correlation, and consistent with previous literature results for the DGS sources, there is indication that the IR flux relative to the RC decreases towards longer IR wavelengths, evidencing warmer dust temperatures and shorter wavelength IR SED peaks (Fig.~2).}}

\item{The lack of an apparent correlation between the magnetic field strength and the SFR in low-SFR dwarfs and XMPs may signal a breakdown of the equipartition and/or case B approximation assumption in this regime s(Fig.~5). This is in contrast with the results found for other dwarf samples in the literature; however, those results had been based on a small number of high-SFR dwarfs, and had not probed such extreme low luminosities.}

\item{In the CO -- RC and IR -- CO relation, the dwarfs and XMPs tend to be underluminous in CO for their luminosity (Fig.~6 and 7). The effect is a combination of slower reactions rates, CO dissociation due to the ionizing radiation field, low dust content and a decrease in dust opacity. The results are inconsistent with previous published results for dwarfs, although the number of sources had been smaller, and such extreme luminosities had not been probed.}

\end{itemize}

In summary, it is found that the IR, RC and CO properties of dwarfs and XMPs are not simple extrapolations of the properties of more massive galaxies. Previous studies that found a smooth transition did not include a large number of sources in the low-luminosity regime, or did not probe such extreme conditions (e.g., Bell 2003; Leroy et al. 2005). For luminosities above L$_{\rm 1.4 \, GHz} \simeq$ L$^{\rm non-thermal}_{\rm 1.4 \, GHz} \simeq$ L$^{\rm thermal}_{\rm 1.4 \, GHz} \simeq$ 10$^{27}$ W (where L$_{\rm 1.4 \, GHz} \approx$ 10$^{29}$ W for a Milky Way SFR of $\simeq$1 M$_{\odot}$ yr$^{-1}$), the dwarf and XMP IR -- RC -- CO relations are generally shallower than that found for more massive galaxies, signaling that the star formation conditions or coupling of the star formation to the observed quantities is modified. The dwarf and XMP relations demonstrate weaker CO and IR emission for their luminosity, consistent with CO dissociation and low dust content. The shallow relations also demonstrate that, below L$_{\rm FIR} \simeq$ 10$^{36}$ W (where L$_{\rm FIR} \simeq$ 10$^{36}$ W $\simeq$ 3 $\times$ 10$^9$ L$_{\odot}$), the IR ceases to adequately trace the SFR, while the total and non-thermal RC emission still trace the SFR down to L$_{\rm 1.4 \, GHz} \simeq$ L$^{\rm non-thermal}_{\rm 1.4 \, GHz} \simeq$ 10$^{27}$ W. Below L$_{\rm FIR} \simeq$ 10$^{33}$ W and L$_{\rm 1.4 \, GHz} \simeq$ L$^{\rm non-thermal}_{\rm 1.4 \, GHz} \simeq$ 10$^{27}$ W, the relations appear to show a steepening or a downturn, consistent with the H$\alpha$ ($\propto$ L$^{\rm thermal}_{\rm 1.4 \, GHz}$), non-thermal, and, hence, total 1.4 GHz RC emission ceasing to adequately trace the SFR. No clear correlation is found between the magnetic field strength and SFR in low-SFR dwarfs and XMPs, suggesting the breakdown of the equipartition and/or case B approximation assumption. Because this extreme low-luminosity and low-SFR regime is populated mainly by XMPs, the XMPs stand out in their IR -- RC -- CO properties.

\section*{Acknowledgements}

M. E. F. gratefully acknowledges the financial support of the ''Funda\c c\~ao para a Ci\^encia e Tecnologia'' (FCT -- Portugal), through the grant SFRH/BPD/107801/2015. This work has been partly funded by the Spanish Ministery of Economy and Competitiveness, project {\em Estallidos} AYA2013-47742-C04-02-P and AYA2016-79724-C04-2-P. F. S. T. acknowledges financial support from the Spanish Ministry of Economy and Competitiveness (MINECO) under the grant number AYA2016-76219-P. The authors would like to thank D. Dale for providing some KINGFISH data. The authors would also like to thank the referee for comments and suggestions that greatly improved this work.

%%%%%%%%%%%%%%%%%%%%%%%%%%%%%%%%%%%%%%%%%%%%%%%%%%%%%%%%%%%%%%%%%%%%%%%%%%%%%%%%%%%%%%%%%%%%%%%%%%%%%%%%%%%%%%%%%%%%%%%%%%%%%%%%%%%%%%%%%%%%%



%\bibliographystyle{mnras}

\section*{References}

\noindent Abazajian, K. N., Adelman-McCarthy, J. K., Ag\"ueros, M. A. 2009, ApJS, 182, 543

\noindent Amor\'\i n, R., Mu\~noz-Tu\~n\'on, C., Aguerri, J. A. L. \& Planesas, P. 2016, A\&A, 588, 23

\noindent Appleton, P. N., Fadda, D. T., Marleau, F. R. et al. 2004, ApJS, 154, 147

\noindent Beck, R., Brandenburg, A., Moss, D., Shukurov, A. \& Sokoloff, D. 1996, ARA\&A, 34, 155

\noindent Beck, R., \& Krause, M. 2005, Astronomische Nachrichten, 326, 414

\noindent Beck, R. 2016, A\&ARv, 24, 4

\noindent Bell, E. F.  2003, ApJ, 586, 794

\noindent Bennett, C. L., Larson, D., Weiland, J. L. \& Hinshaw, G. 2014, ApJ, 794, 135

\noindent Bettens, R. P. A., Brown, R. D., Cragg, D. M., Dickinson, C. J. \& Godfrey, P. D. 1993, MNRAS, 263, 93

\noindent Bicay, M. D., Kojoian, G., Seal, J., Dickinson, D. F. \& Malkan, M. A. 1995, ApJS, 98, 369

\noindent Bisbas, T. G., van Dishoeck, E. F., Papadopoulos, P. P. et al. 2017, ApJ, 839, 90

\noindent Bolatto, A. D., Wolfire, M. \& Leroy, A. K. 2013, ARA\&A, 51, 207

\noindent Bourne, N., Dunne, L., Ivison, R. J., Maddox, S. J., Dickinson, M. \& Frayer, D. T. 2011, MNRAS, 410, 1155

\noindent Calzetti, D. 2013, Proceedings of the XXIII Canary Islands Winter School of Astrophysics: Secular Evolution of Galaxies, edited by J. Falcon-Barroso and J.H. Knapen, 419

\noindent Cannon, J. M. \& Skillman, E. D. 2004, ApJ, 610, 772

\noindent Caplan, J. \& Deharveng, L. 1986, A\&A, 155, 297

\noindent Chy\.{z}y, K. T., Beck, R., Kohle, S., Klein, U. \& Urbanik, M. 2000, A\&A, 355, 128

\noindent Chy\.{z}y, K. T. 2008, A\&A, 482, 755

\noindent Chy\.{z}y, K. T., We\.{z}gowiec, M., Beck, R. \& Bomans, D. J. 2011, A\&A, 529, 94

\noindent Chy\.{z}y, K. T., Sridhar, S. S. \& Jurusik, W. 2017, A\&A, 603, 121

\noindent Condon, J. J., Anderson, M. L. \& Helou, G. 1991, ApJ, 376, 95

\noindent Condon, J. J. 1992, ARA\&A, 30, 575

\noindent Condon, J. J., Cotton, W. D., Greisen, E. W., Yin, Q. F., Perley, R. A., Taylor, G. B. \& Broderick, J. J. 1998, AJ, 115, 1693

\noindent Condon, J. J., Cotton, W. D. \& Broderick, J. J. 2002, AJ, 124, 675

\noindent Cormier, D., Madden, S. C., Lebouteiller, V. et al. 2014, A\&A, 564, 121 

\noindent Dale, D. A., Cohen, S. A., Johnson, L. C. et al. 2009, ApJ, 703, 517

\noindent Dale, D. A., Aniano, G., Engelbracht, C. W., Hinz, J. L., Krause, O. \& Montiel, E. J. 2012, ApJ, 745, 95

\noindent de Jong, T., Klein, U., Wielebinski, R. \& Wunderlich, E. 1985, A\&A, 147, 6

\noindent Devereux, N. A. \& Young, J. S. 1990, ApJ, 359, 42

\noindent Dickey, J. M. \& Salpeter, E. E. 1984, ApJ, 284, 461

\noindent Draine, B. T. 2011, EAS Publications Series, Vol, 46, 29

\noindent Dressel, L. L. \& Condon, J. J. 1978, ApJS, 36, 53

\noindent Engelbracht, C. W., Rieke, G. H., Gordon, K. D. et al. 2008, ApJ, 678, 804

\noindent Elmegreen, B. G., Rubio, M., Hunter, D. A., Verdugo, C., Brinks, E. \& Schruba, A. 2013, Nature, 495, 487

\noindent Fernandez, E. \& Shull, J. M. 2011, ApJ, 731, 20

\noindent Filho, M. E., Winkel, B., S\'anchez Almeida, J. et al. 2013, A\&A, 558, 18

\noindent Filho, M. E., S\'anchez Almeida, J., Amor\'\i n, R., Mu\~noz-Tu\~n\'on, C., Elmegreen, B. G. \& Elmegreen, D. M. 2016, ApJ, 820, 109 

\noindent Glover, S. C. O. \& Clark, P. C. 2012, MNRAS, 426, 377	

%\noindent Groves, B. A., Cho, J. Dopita, M. \& Lazarian, A. 2003, PASA, Volume 20, Issue 3, 252

\noindent Healey, S. E., Romani, R. W., Taylor, G. B. 2007, ApJS, 171, 61

\noindent Heesen, V., Rau, U., Rupen, M. P., Brinks, E. \& Hunter, D. A. 2011, ApJ, 739, 23

\noindent Helou, G., Soifer, B. T. \& Rowan-Robinson, M. 1985, ApJ, 298, 7

\noindent Helou, G. \& Bicay, M. D. 1993, ApJ, 415, 93

\noindent Hughes, A., Wong, T., Ekers, R. et al. 2006, MNRAS, 370, 363

\noindent Hunt, L. K., Dyer, K. K., Thuan, T. X. \& Ulvestad, J. S. 2004, ApJ, 606, 853

\noindent Hunter, D. A., Ficut-Vicas, D., Ashley, T. et al. 2012, AJ, 144, 134

\noindent Hunt, L. K., Garc\'\i a-Burillo, S., Casasola, V. et al. 2015, A\&A, 583, 114

\noindent Jarvis, M. J., Smith, D. J. B., Bonfield, D. G. et al. 2010, MNRAS, 409, 92

\noindent Katsianis, A., Blanc, G., Lagos, C. P. 2017, MNRAS, 472, 919

\noindent Kennicutt, R. C. 1989, ApJ, 344, 685

\noindent Kennicutt, R. C. 1998, ApJ, 498, 541

\noindent Kennicutt, R. C., Calzetti, D., Aniano, G. et al. 2011, PASP, 123, 1347

\noindent Kennicutt, R. C. \& Evans, N. J. 2012, ARA\&A, 50, 531 

\noindent Kepley, A. A., M\"uhle, S., Everett, J., Zweibel, E. G., Wilcots, E. M. \& Klein, U. 2010, ApJ, 712, 536

\noindent Klein, U. \& Graeve, R. 1986, A\&A, 161, 155

\noindent Klein, U., Weiland, H. \& Brinks, E.  1991, A\&A, 246, 323

\noindent Kistler, M. D., Stanek, K. Z., Kochanek, C. S., Prieto, J. L. \& Thompson, T. A. 2013, ApJ, 770, 88

\noindent Krumholz, M. R., McKee, C. F. \& Tumlinson, J. 2008, 689, 865

\noindent Krumholz, M. R., McKee, C. F. \& Tumlinson, J. 2009a, ApJ, 693, 216

\noindent Krumholz, M. R., McKee, C. F. \& Tumlinson, J. 2009b, ApJ, 699, 850

\noindent Krumholz, M. R. 2013, MNRAS, 436, 2747 

\noindent Lacki, B. C., Thompson, T. A. \& Quataert, E. 2010, ApJ, 717, 1

%\noindent Lacki, B. C. \& Thompson, T. A. 2010, ApJ, 717, 1

\noindent Lee, J. C., Gil de Paz, A., Tremonti, C. et al. 2009, ApJ, 706, 599 

\noindent Leitherer, C., Hernandez, S., Lee, J. \& Oey, S. 2016, ApJ, 823, 64

\noindent Lisenfeld, U., Verdes-Montenegro, L., Sulentic, J. 2007, A\&A, 462, 507

\noindent Lisenfeld, U., Wilding, T. W., Pooley, G. G. \& Alexander, P. 2004, MNRAS, 349, 1335

\noindent Leroy, A. K., Bolatto, A. D., Simon, J. D. \& Blitz, L. 2005, ApJ, 625, 763

\noindent Leroy, A. K., Bolatto, A. D., Gordon, K. et al. 2011, ApJ, 737, 12

\noindent Liu, F. \& Gao, Y. 2010, ApJ, 713, 524

\noindent Madden, S. C., R\'emy-Ruyer, A., Galametz, M. 2013, PASP, 125, 600

\noindent Magnelli, B., Ivison, R. J., Lutz, D. et al. 2015, A\&A, 573, 45

\noindent Morales-Luis, A. B., S\'anchez Almeida, J., Aguerri, J. A. L. \& Mu\~noz-Tu\~n\'on, C. 2011, ApJ, 743, 77

\noindent Murgia, M., Crapsi, A., Moscadelli, L. \& Gregorini, L. 2002, A\&A, 385, 412

\noindent Murgia, M., Helfer, T. T., Ekers, R., Blitz, L., Moscadelli, L., Wong, T. \& Paladino, R. 2005, A\&A, 437, 389

\noindent Murphy, E. J., Condon, J. J., Schinnerer, E., et al. 2011, ApJ, 737, 67

\noindent Nicholls, D. C., Dopita, M. A., Sutherland, R. S., Jerjen, H., \& Kewley, L. J. 2014, ApJ, 790, 75

\noindent Niklas, S. \& Beck, R. 1997, A\&A, 320, 54

\noindent Olmo-Garc\'\i a, A., S\'anchez Almeida, J., Mu\~noz-Tu\~n\'on, C. et al. 2017, ApJ, 834, 181

\noindent Osterbrock, D. E. \& Ferland, G. J. Astrophysics of Gaseous Nebulae and Active Galactic Nuclei, 2nd. ed. by D.E. Osterbrock and G.J. Ferland. Sausalito, CA: University Science Books, 2006

\noindent Pannella, M., Elbaz, D., Daddi, E. et al. 2015, ApJ, 807, 141 

\noindent Price, R. \& Duric, N. 1992, ApJ, 401, 81

\noindent Qiu, J., Shi, Y., Wang, J., Zhang, Z.-Y. \& Zhou, L. 2017, ApJ, 846, 68

\noindent Regan, M. W. \& Vogel, S. N. 1994, ApJ, 434, 536

\noindent R\'emy-Ruyer, A., Madden, S. C., Galliano, F. et al. 2013, 2013, A\&A, 557, 95

\noindent Rice, W., Lonsdale, C. J., Soifer, B. T. et al. 1988, ApJS, 68, 91

\noindent Richings, A. J. \& Schaye, J. 2016, MNRAS, 458, 270

\noindent R\"ollig, M., Ossenkopf, V., Jeyakumar, S., Stutzki, J. \& Sternberg, A. 2006, A\&A, 451, 917

\noindent R\"ollig, M. 2008, EAS Publications Series, Volume 31, 129

\noindent Rosa-Gonz\'alez, D., Burgarella, D., Nandra, K. 2007, MNRAS, 379, 357

\noindent Roy, A. L., Goss, W. M. \& Anantharamaiah, K. R. 2008, A\&A, 483, 79

\noindent Roychowdhury, S. \& Chengalur, J. N. 2012, MNRAS, 423, 127

\noindent Rutkowski, M. J., Scarlata, C., Haardt, F. et al. 2016, ApJ, 819, 81

\noindent S\'anchez-Janssen, R., M\'endez-Abreu, J., \& Aguerri, J. A. L. 2010, MNRAS, 406, 65

\noindent Sargent, M. T., Schinnerer, E., Murphy, E. et al. 2010, ApJ, 714, 190

\noindent Schleicher, D. R. G., \& Beck, R. 2013, A\&A, 556, A142

\noindent Schleicher, D. R. G., Schober, J., Federrath, C., Bovino, S. \& Schmidt, W. 2013, New Journal of Physics, Volume 15, Issue 2, 023017

\noindent Schleicher, D. R. G. \& Beck, R. 2016, A\&A, 593, 77

\noindent Schmidt, M. 1959, ApJ, 129, 243

\noindent Schmitt, H. R., Calzetti, D., Armus, L. et al. 2006, ApJS, 164, 52

\noindent Schober, J. Schleicher, D. R. G. \& Klessen, R. S. 2017, MNRAS, 468, 946

\noindent Schruba, A. Leroy, A. K., Walter, F. 2012, AJ, 143, 138

\noindent Seymour, N.;, Huynh, M., Dwelly, T., Symeonidis, M., Hopkins, A., McHardy, I. M., Page, M. J. \& Rieke, G. 2009, MNRAS, 398, 1573

\noindent Shetty, R. Glover, S. C., Dullemond, C. P. et al. 2011, MNRAS, 415, 3253

\noindent Shetty, R., Roman-Duval, J., Hony, S. et al. 2016, MNRAS, 460, 67

\noindent Shi, Y., Armus, L., Helou, G. et al. 2014, Nature, 514, 335

\noindent Shi, Y., Wang, J., Zhang, Z.-Y. et al. 2015, ApJ, 804, 11

\noindent Sparre, M., Hayward, C. C., Springer, V. 2015, MNRAS, 447, 3548

\noindent Suchkov, A., Allen, R. J. \& Heckman, T. M. 1993, ApJ, 413, 542

\noindent Tabatabaei, F. S., Beck, R., Kr\"ugel, E., et al. 2007, A\&A, 475, 133

\noindent Tabatabaei, F. S., Krause, M., Fletcher, A. \& Beck, R. 2008, A\&A, 490, 1005

\noindent Tabatabaei, F. S., Schinnerer, E., Murphy, E. J. et al. 2013a A\&A, 552, 19

\noindent Tabatabaei, F. S., Berkhuijsen, E. M., Frick, P., Beck, R. \& Schinnerer, E. 2013b, A\&A, 557, 129

\noindent Tabatabaei, F. S., Martinsson, T. P. K., Knapen, J. H., Beckman, J. E., Koribalski, B. \& Elmegreen, B. G. 2016, ApJ, 818, L10

\noindent Tabatabaei, F. S., Schinnerer, E., Krause, M. et al. 2017, ApJ, 836, 185

\noindent Thuan, T. X., Bauer, F. E., Papaderos, P. \& Izotov, Y. I. 2004, ApJ, 606, 213

\noindent van der Kruit, P. C., 1973a, A\&A, 29, 231

\noindent van der Kruit, P. C., 1973b, A\&A, 29, 249

\noindent van der Kruit, P. C., 1973c, A\&A, 29, 263

\noindent V\"olk, H. J. 1989, A\&A, 218, 67

\noindent Weisz, D. R., Johnson, B. D., Johnson, L. C., Skillman, E. D., Lee, J. C. et al. ApJ, 744, 44

\noindent White, R. L., Becker, R. H., Helfand, D. J. \& Gregg, M. D. 1997, ApJ, 475, 479

\noindent Wolfire, M. G., Hollenbach, D \& McKee, C. F. 2010, ApJ, 716, 1191

\noindent Wu, Y., Charmandaris, V., Houck, J. R., et al. 2008, ApJ, 676, 970

\noindent Young, J. S. \& Scoville, N. Z. 1991, ARA\&A, 29, 581 

\noindent Yun, M. S., Reddy, N. A. \& Condon, J. J. 2001, ApJ, 554, 803

\noindent Zijlstra, A., Pottasch, S. \& Bignell, C. 1990, A\&AS, 82, 273 

%%%%%%%%%%%%%%%%%%%%%%%%%%%%%%%%%%%%%%%%%%%%%%%%%%

\appendix

\section{Common Sources}

The XMP sample has 19 sources in common with the DGS: HS0017+1055, HS0822+3542, HS1222+3741, HS1236+3937, HS1319+3224, HS1442+4250, IZw18, SBS0335-052, SBS1159+545, SBS1211+540, SBS1249+493, SBS1415+437, Tol0618-402, Tol1214-277, UGC4483, UGC6456 (VIIZw403), UGCA20, UM133 and UM461 (J1151-0222). 

There are 11 sources in common between the XMP sample and LITTLE THINGS: UGCA292 (CVnIdwA), DDO50 (UGC4305, HolmbergII), DDO53, DDO69 (LeoA), DDO70 (SextansB), DDO75 (SextansA), DDO155 (GR8), DDO167, IC1613, SagDig and UGC6456 (VIIZw403). 

The XMP sample has 2 sources in common with KINGFISH: HolmbergII (DDO50, UGC4305) and DDO53. 

LITTLE THINGS and the DGS have 7 sources in common (one of which is an XMP): IC10, NGC1569, NGC2366, NGC4214, NGC6822, Mrk209 (Haro29) and VIIZw403 (UGC6456). 

KINGFISH and LITTLE THINGS have 5 sources in common (2 of which are XMPs): DDO50 (UGC4305, HolmbergII), DDO53, DDO63 (UGC5139, HolmbergI), DDO154 (UGC8024) and DDO165 (UGC8201). 

There are no sources in common between KINGFISH and the DGS.

%%%%%%%%%%%%%%%%%%%%%%%

\section{Data References}

This section contains a compilation of the data information sources for the different samples.

%%%%%%%%%%%%%%%%%%%%%%%%%%%%%%%%%%%%%%%%%%%%%%%%%%%%%%%%%%%%%%%%%%%%%%%%%%%%%%%%%%%%

% TABLE - Data

%\setcounter{table}{1}

\begin{table*}

\scriptsize

\begin{center}

\begin{minipage}{142mm}

\caption{Data references for the samples.}

\begin{tabular}{l l l l l l}

\hline

Sample 				& 	XMPs$^a$				& DGS			& LITTLE THINGS		& WLM 			& KINGFISH \\

\hline
\hline

Main Reference	& 	ML11, F13 				& M13 			& H12 				& H12 			& K11 \\	

Metallicity 	&	F13, M13, H12, K11		& M13 			& H12 				& H12			& K11 \\

Distance		&	F13, M13, H12, K11		& M13 			& H12 				& H12			& K11 \\

Star Formation Rate$^b$	&	F13, M13, H12, K11		& M13 				& H12 			& H12			& K11 \\

Inclination Angle & F13, H12, H15			& see Sect.~2 	& H12				& H12			& H15 \\

Infrared Data	& 	L07, E08, D09, D12, M13, R13	& M13, R13 			& E08, D09, D12			& R88, D09		& D12, IRAS \\

RC Data$^c$			&   see Sect.~2				& see Sect.~2	& see Sect.~2		& see Sect.~2	& T17 \\

CO Data$^c$			&   see Sect.~2				& see Sect.~2	& see Sect.~2		& see Sect.~2	& \ldots \\

\hline 

\end{tabular}

$^a$Data mainly taken from F13, or from the DGS (M13), LITTLE THINGS (H12) and KINGFISH (K11, H15) samples, when there are sources in common (see also Appendix A). \\
$^b$H$\alpha$ SFR for the XMPs, TIR SFR, or H$\alpha$ or H$\beta$ SFR when IR is unavailable, for the DGS sources, H$\alpha$ SFR for the LITTLE THINGS, and H$\alpha$+24$\mu$m SFR for the KINGFISH sources.\\
$^c$Data compilation from numerous sources. \\

Abbreviations: ML11 -- Morales-Luis et al. 2011, F13 -- Filho et al. 2013, M13 -- Madden et al. 2013, H12 -- Hunter et al. 2012, K11 -- Kennicutt et al. 2011, H15 -- Hunt et al. 2015, L07 -- Lisenfeld et al. 2007, E08 -- Engelbracht et al. 2008, D09 -- Dale et al. 2009, D12 -- Dale et al. 2012, R13 -- R\'emy-Ruyer et al. 2013, R88 -- Rice et al. 1988, T17 -- Tabatabaei et al. 2017. \\

\end{minipage}

\end{center}

\end{table*}

\bsp	% typesetting comment
\label{lastpage}
\end{document}